\documentclass[iop]{emulateapj}

\usepackage{epstopdf,hyperref,color}

\newcommand{\s}{\sigma}
\newcommand{\sn}{\times 10^}
\newcommand{\siii}{\ion{Si}{3}}
\newcommand{\siv}{\ion{Si}{4}}
\newcommand{\cii}{\ion{C}{2}}
\newcommand{\vsini}{$v\sin{i}$}

\newcommand{\dt}{\Delta t}
\newcommand{\al}{\alpha}
\newcommand{\lm}{\lambda}
\newcommand{\mf}{\left \langle \mathbf{F} \right \rangle}

\newcommand{\statstablefootnotes}{\tablenotetext{a}{Mean-normalized excess noise at 60~s cadence.}
					        \tablenotetext{b}{Radius of an occulting disk that would produce a transit depth equivalent to $\s_x$ projected to 3.5~h.}
					        \tablenotetext{c}{Number of flares identified.}
					        \tablenotetext{d}{Fraction of lightcurve points encompassed by flares.}
					        \tablenotetext{e}{Min photometric equivalent width of a detectable flare given the lightcurve scatter.}
					        \tablenotetext{f}{Data contain a periodic signal not fully suppressed by the high-pass filtering.}
					        \tablenotetext{g}{The literature did not provide an uncertainty on the stellar radius. The uncertainty on $R_{\s_x}$ accounts only for uncertainty in $\s_x$.}}
\newcommand{\nfe}{EW_p}

\shorttitle{Far-Ultraviolet Stellar Variability Measurements}
\shortauthors{Loyd and France}

\begin{document}

\title{Fluctuations and Flares in the Ultraviolet Line Emission of Cool Stars: Implications for Exoplanet Transit Observations}	

\author{R. O. Parke Loyd\altaffilmark{1} and Kevin France\altaffilmark{1,2}}
\altaffiltext{1}{Center for Astrophysics and Space Astronomy, Boulder, Colorado,  80303}
\altaffiltext{2}{NASA Nancy Grace Roman Fellow}

\email{robert.loyd@colorado.edu}

\begin{abstract}
Variations in stellar flux can potentially overwhelm the photometric signal of a transiting planet. Such variability has not previously been well-characterized in the ultraviolet lines used to probe the inflated atmospheres surrounding hot Jupiters. Therefore, we surveyed 38 F-M stars for intensity variations in four narrow spectroscopic bands: two enclosing strong lines from species known to inhabit hot Jupiter atmospheres, \ion{C}{2} $\lm\lm$1334,1335 and \ion{Si}{3} $\lm$1206; one enclosing \ion{Si}{4} $\lm\lm$1393,1402; and 36.5 \AA\ of interspersed continuum. For each star/band combination, we generated 60~s cadence lightcurves from archival {\it HST} COS and STIS time-tagged photon data. Within these lightcurves, we characterized flares and stochastic fluctuations as separate forms of variability. {\bf Flares:} We used a cross-correlation approach to detect 116 flares. These events occur in the time-series an average of once per 2.5 h, over 50\% last 4 min or less, and most produce the strongest response in \siv. If the flare occurred during a transit measurement integrated for 60 min, 90/116 would destroy the signal of an Earth, 27/116 Neptune, and 7/116 Jupiter, with the upward bias in flux ranging from 1-109\% of quiescent levels. {\bf Fluctuations:} Photon noise and underlying stellar fluctuations produce scatter in the quiescent data. We model the stellar fluctuations as Gaussian white noise with standard deviation $\s_x$. Maximum likelihood values of $\s_x$ range from 1-41\% for 60~s measurements. These values suggest that many cool stars will only permit a transit detection to high confidence in ultraviolet resonance lines if the radius of the occulting disk is $\gtrsim$ 1 R$_J$. However, for some M dwarfs this limit can be as low as several $R_\earth$.

\end{abstract}

\section{Introduction}
\label{sec:intro}

Transit observations in the far-ultraviolet (FUV, $1200\leq\lambda\leq1400$ \AA) have revealed the existence of inflated atmospheres surrounding the hot Jupiters HD209458b \citep{madjar03,madjar04,linsky10} and HD189733b \citep{lecavelier10,bourrier13,jaffel13}. During the transit of these planets, the UV resonance transitions of several species in their atmospheres -- \ion{H}{1}, \cii, \ion{O}{1}, and \siii\ -- produce detectable absorption against the background stellar emission. The depth of the absorption indicates that these species occupy a volume overfilling the planets' Roche lobes, suggesting atmospheric escape. Furthermore, such UV transit spectrophotometry can constrain the atmospheric mass loss rate (e.g. \citealt{madjar04,lecavelier10}) and characterize the atmospheric response to changes in the stellar radiation and particle flux \citep{lecavelier12,bourrier13}. 

However, variability in the background flux source itself --  the host star -- presents additional, instrument-independent challenges for all transit observations. In a lightcurve, the transit signal can be spuriously weakened, sometimes completely obliterated, by flares buried within it, or amplified by flares flanking it. Outside of flares, the transit signal can be obscured by stochastic fluctuations in stellar luminosity that act as additional noise, compounding that of photon statistics and instrumental sources. These forms of stellar variability fundamentally limit transit observations. Therefore, evaluating the possibilities for future transit work necessitates measurements of this variability for potential host stars.

The list of potential targets is diverse, encompassing stars beyond those just on the main sequence, such as HD209458b and HD189733b. Such candidate targets could include stars with transitional or debris disks (encompassing weak-line T-Tauri stars, WTTS) where planets might be in the process of coalescing from disk material. For example, \citet{eyken12} report on the transit signature of a super-Jupiter orbiting a WTTS star in the Orion-OB 1a/25-Ori region. High contrast imaging of stars with disks has also revealed (proto)planetary objects or evidence for these objects through disk gaps, such as LkCa15 \citep{kraus12}, PDS 70 \citep{hashimoto12}, RX J1633.9-2442 \citep{cieza12}, and TW Hya \citep{debes13}. Candidate targets also include post main-sequence stars. The evolved F5-F7 stars WASP-76, WASP-82, and WASP-90 host transiting hot Jupiters \citep{west13}, as does the G8III giant HIP 63242 \citep{jones13}. In addition, observations of remnant debris orbiting white dwarfs (e.g. \citealt{farihi13}) hint at the possibility of detecting extant or decomposing planets around stars nearing the end of their lives.

To begin characterizing the range of background fluctuations faced by FUV transit observations, we have conducted a survey of stellar flares and stochastic fluctuations for the largest possible sample of stars with archival FUV photon event data from {\it HST} covering \cii\ $\lm\lm$1334,1335 and \siii\ $\lambda$1206, two key lines for probing outer atmospheres of close-in exoplanets. We also include \siv\ $\lambda\lambda$1394,1403 and a composite of interspersed continuum bands. We did not attempt to include \ion{H}{1} $\lambda$1216 (Ly$\alpha$) or \ion{O}{1} $\lambda$1302 due to the correction for geocoronal emission that is required. This survey relies on UV data acquired by the two powerful UV spectrographs on board the {\it Hubble Space Telescope (HST)}: the Cosmic Origins Spectrograph (COS), with its G130M and G140L gratings, and the Space Telescope Imaging Spectrograph (STIS) with its E140M grating. To enable an analysis of temporal variability, we traded spectral resolution for temporal resolution. Thus, we employed short, 60~s time bins and summed photon counts over roughly the full width ($\sim$ 1-2 \AA) of each line and 36.5 \AA\ of interspersed continuum to create lightcurves. In comparison, transit observations have achieved resolutions of  $\sim0.1$ \AA\ (roughly 20-30 km s$^{-1}$) when coadding an entire transit dataset. However, these same observations commonly integrate over the full line-widths for increased signal to noise (e.g. \citealt{madjar03,lecavelier10}).

We present the results of an analysis of 153 lightcurves of 60~s cadence, including (1) 116 flares we identified by cross-correlating with a flare-like kernel and (2) estimates or upper limits for the standard deviation of stochastic fluctuations in the quiescent portions of the lightcurve. In the remainder of this introduction, we expand upon the implications and physical sources of stellar variability and provide pointers to previous variability surveys. In Section~\ref{sec:stars}, we describe the stellar sample and process of generating lightcurves from the FUV time-tagged photon data. In Section~\ref{sec:analysis} we outline the variability analysis, treating flares and stochastic fluctuations separately. In Section~\ref{sec:results}, we present the results, followed in Section~\ref{sec:discussion} by a discussion, including the implications for observing transits. We summarize the work in Section~\ref{sec:summary}.

Throughout this paper, we will treat 3.5~h as the typical transit timescale. This is near the average transit duration of 3.6 h (sample standard deviation 1.9 h) for exoplanets listed in the Exoplanet Data Explorer \citep{wright11} as of 2013 October.

\subsection{A Brief Discussion of Stellar Variability}
\label{sec:varintro}

Stellar variability can be divided into three categories based on the differing implications for transit observations:
\begin{enumerate}
	\item Periodicities: Oscillations of the stellar flux by phenomena that can be sufficiently characterized to predict the level of modulation over the course of a transit.
	\item Flares: Bursts of brightness isolated in time with respect to the cadence of the data, and well above the quiescent scatter in the lightcurve.
	\item Stochastic Fluctuations: Variations in the stellar flux that are too chaotic to be accurately predicted on transit timescales.
\end{enumerate}

Predictable periodic signals are surmountable obstacles. Any periodic signal strong enough to be even modestly detected would allow an accurate model fit, bootstrapped over many cycles, such that removing or accounting for the signal would not obscure an otherwise detectable transit.

Strong flares complicate the lightcurve analysis. While it is possible to describe the distribution of solar flares in both strength and frequency \citep{cassak08}, a statistical model cannot predict the onset, magnitude, and evolution of flares for any specific timeline, such as during a transit. Strong flares thus pose the risk of overwhelming, and weak flares of attenuating, a transit signal. Flares occurring near a transit could augment estimates of the out-of-transit flux, spuriously deepening the transit signal. Beyond interfering with transit observations, flares might also impact the atmospheres of exoplanets, such as the stripping of atomic hydrogen from HD189733b conjectured by \citet{bourrier13} to be caused by an observed host-star flare. 

Stochastic fluctuations, however, represent the greatest barrier to transit photometry. Because these fluctuations cannot be predicted deterministically over the course of a transit, they must be treated as noise. As with photon noise, they pose the risk of obscuring a true signal or causing a false one. This ``noise" can be overcome by averaging measurements until the uncertainty is within tolerable limits, but with a caveat: Unlike photon noise, stochastic fluctuations are probably not white noise. For example, in 60~s cadence, broadband optical photometry from {\it Kepler}, stochastic fluctuations of stellar flux (attributed to granulation and magnetic activity) have a power spectrum that, unlike white noise, is not flat \citep{gilliland10}. Although {\it Kepler} measures broadband optical flux, not the chromosphere and transition region FUV emission line flux we analyzed, the fluctuations of the chromospheric near UV flux from YZ CMi also show a frequency-dependent power spectrum \citep{robinson99}. Therefore, it seems probable that stochastic fluctuations in stellar flux will not behave as white noise in any band. What appears as true noise at one cadence would resolve into smooth variations at some faster cadence. Below this threshold cadence, lightcurve points will be highly correlated. Binning adjacent flux measurements in this regime will not average out the scatter.

The emission line flux we analyzed samples regions with temperatures of $10^4-10^5$ K in the outer atmospheres of stars. Specifically, the peak formation temperatures of the lines in solar conditions are estimated by \citet{dere09} to be $3.2\sn4$ K for \cii\ $\lm\lm$1334,1335, $5\sn4$ K for \siii\ $\lambda$1206, and $8\sn4$ K for \siv\ $\lambda\lambda$1393,1402. Reference models of the solar atmosphere place emission from these lines in the thin transition region between the chromosphere and the corona. In contrast to the lines, the solar FUV continuum (shortward of 1500 \AA) forms at slightly lower altitudes, primarily in a region of initial chromospheric temperature rise above the photospheric temperature minimum \citep{linsky12a}.

Variability in these regions of stellar atmospheres can result from several phenomena. Although drawing conclusions about these underlying physical phenomena is not our objective, the generally accepted origins of periodicities, flares, and stochastic fluctuations bear mentioning.

Periodic variability can be the result of a pulsational instability in the star \citep{gautschy95, gautschy96} and/or rotation of long-lived, localized brightness variations (starspots, faculae, etc.)  through the observer's field of view \citep{vaughan81}. A periodic signal can result from extrinsic phenomena as well. Gradual oscillations in flux are produced by phase changes of an orbiting planet (e.g. \citealt{borucki09}). Isolated, but nonetheless periodic, dips in flux occur when an orbiting object, such as an exoplanet or stellar companion, transits the host star (e.g. \citealt{wilson71}).

Flares are generally thought to be the result of magnetic reconnection events in the corona that abruptly convert magnetic energy into plasma kinetic energy. Some of this energy is deposited in the chromosphere and photosphere and radiated away \citep{haisch91,gershberg05}. Flares commonly produce a sharp rise in flux followed by an exponential decay lasting from hours to minutes, possibly even seconds \citep{pettersen89,gershberg05}. Within a single flare, multiple peaks and changes in the decay rate are possible. Some researchers have identified as flares events in which the stellar flux rises and fades more gradually \citep{houdebine03,tovmassian03}. 

The strength and frequency of flares typically exhibit an inverse power-law relationship (e.g. \citealt{shakhovskaia89,audard00} for stellar flares, \citealt{lin84,nita02} for solar flares). This implies that weaker flares are more prevalent than conspicuous events, such that many flares will occur in observations that cannot be clearly resolved as such. In fact, if the power law is steep enough, the lowest energy flares, often termed microflares, might inject enough heat into the corona of a star to explain the high temperatures present there \citep{hudson91,audard00}.

Low energy ``microflares'' or even ``regular'' flares, if the data is not of sufficient quality to resolve them, will contribute to the observed stochastic fluctuations of a target. For instance, \citet{robinson99} suggest microflaring as an explanation for the quiescent stochastic fluctuations in near UV flux that they observed from YZ CMi . They simulate the production of such stochastic fluctuations with a microflare model and find that it closely resembles the YZ CMi quiescent data. Ultimately, the extent of the contribution of flares to the observed stochastic fluctuations of any target is determined by the level of stellar magnetic activity, the photometric quality of the data, and the threshold set for identifying a lightcurve anomaly as a flare rather than fluctuation.

The remaining proportion of stochastic fluctuations in transition region line emission could be explained by several phenomena. Transition region explosions, smaller events possibly associated with magnetic restructuring at the edges of newly emerging flux loops \citep{gershberg05}, could introduce variability while also serving as a dominant heating source for the transition region. \citet{wood97} suggested such events might explain broad components of \siv\ and \ion{C}{4} emission in the FUV spectra of 11 late-type stars. However, \citet{peter06} suggests magnetic flux braiding and consequent Joule dissipation might be the dominant heat source for the transition region. Both braiding of surface field and the emergence of field loops produce pockets of rapid heating in the three-dimensional MHD models of \citet{hansteen10} that could explain much of the temporal variability of line emission originating in the transition region. In the \citet{hansteen10} model, the injected energy results from work done on the magnetic field by photospheric motions, tying transition region variability to the convective cells and p-mode oscillations within the star. These convective cells and p-mode oscillations also affect the transition region environment by initiating high altitude shock waves \citep{wedemeyer09}.

Variability of a planet-hosting star could be influenced by the planet itself. Planets orbiting close enough to a star will interact tidally and, possibly, magnetically with the host \citep{cuntz00}. Magnetic interactions could lead to flares from the reconnection of planetary and stellar fields \citep{rubenstein00,lanza08}, increased stochastic fluctuations from overall magnetic activity enhancements \citep{cuntz00}, or periodicities from enhanced plages and faculae surrounding the sub-planetary point on the star \citep{lanza08,cohen09,kopp11}. Tidal interactions could produce flows and turbulence associated with the tidal bulge \citep{cuntz00}. They could also spin up the star \citep{aigrain08}, indirectly increasing overall stellar magnetic activity. These interactions are supported by some evidence (beginning with \citealt{shkolnik03}), but more definitive conclusions require future, dedicated observations \citep{lanza11}.

\subsection{A Selection of Relevant Flare and Variability Studies}
\label{sec:context}
There is a long history of research into the frequency and intensity of flares on the Sun and other stars. Especially relevant is recent work by \citet{hilton10} and \citet{davenport12}, and references therein, examining large (several $10^4$) samples of M dwarf stars using multi-epoch data in the optical from the Sloan Digital Sky Survey (SDSS) and in the infrared from the Two Micron All Sky Survey (2MASS). \citet{tofflemire12} specifically assessed the impact of M dwarf flares on exoplanet observations in the infrared using three such stars. Recently, \citet{kowalski13} conducted a detailed spectrophotometric study in the near UV and optical of 20 M dwarf flares in order to probe the various mechanisms responsible for flare emission. Previous studies in the far and extreme UV are scarcer. \citet{welsh07} leveraged data from the {\it Galaxy Evolution Explorer} in the broadband near and far-ultraviolet to find 49 variable sources exhibiting 52 flares. The {\it Galaxy Evolution Explorer} FUV band data contain the \siv\ line we analyzed for variability. In addition, \citet{mullan06} examined 44 F-M stars in broadband extreme UV time-series data from the {\it Extreme Ultraviolet Explorer}. The band they utilized is dominated by emission lines of \ion{Fe}{18} -- \ion{Fe}{22} formed at coronal temperatures upwards of $10^7$ K, expected in magnetically active regions. 

Several previous studies have quantified the stochastic variability of large samples of stars in the optical, most notably employing {\it Kepler} results to place the Sun's well-characterized variability in the context of other stars (see \citealt{basri10} and \citealt{mcquillan12} for examples using {\it Kepler} data and \citealt{eyer97} for one using {\it Hipparcos} data). In addition, it is standard practice to quantify the variability of the exoplanet host star complimentary to radial-velocity or transit measurements, so many individual measurements of stellar variability exist (e.g. \citealt{dragomir12, kane11, berta11}). However, to the knowledge of the authors this paper presents the first analysis, focusing specifically on the implications for transit observations, of stellar variability in UV line emission flux. 


\section{Stellar Sample and Data Reduction}
\label{sec:stars}

\subsection{Sample Selection}
Because the motivation for this work is the characterization of stellar variability in all potential targets for FUV transit work (Section~\ref{sec:intro}), we constructed a stellar sample of \emph{all} F-M stellar targets with archival {\it HST} time-tagged photon data covering the wavelengths of the \ion{C}{2}, \ion{Si}{4}, and occasionally (27/42 datasets) \ion{Si}{3} lines. These wavelengths are observed with the STIS E140M, COS G130M, and COS G140L gratings. Thus, we retrieved all public time-tagged photon data for the sample acquired with these gratings from the Mikulski Archive for Space Telescopes (MAST). We also obtained some data still proprietary under program 12464 \citep{france13}. 

We culled datasets from target stars known to have circumstellar gas disks or outflows because: (1) the hot gas lines have a (sometimes large) contribution from accretion of circumstellar gas onto the star and (2) emission from photoexcited H$_2$ and CO can overwhelm the chromospheric signal \citep{france11, herczeg02, ardila13}. However, we retained many Weak-line T Tauri stars (WTTS) for which there did not seem to be significant contamination of the spectrum by disk or accretion-related emission because there is promise of finding transiting (proto)planets around such stars (Section~\ref{sec:intro}). We also culled datasets where line emission was very weak compared to the background plus continuum (where the ratio of fluxes was roughly less than half). Lastly, we discarded individual exposures (but not entire datasets) where the exposure contained some portion of a known planet's transit.

After culling, 42 datasets remained covering 38 stars. (Four stars have data from two different instrument/grating configurations that we keep separate.) For these stars, we retrieved fundamental properties from a wide range of catalogs and individual studies. These properties include spectral type; age; temperature, $T_\star$; mass, $M_\star$; surface gravity, $\log_{10}{g}$; luminosity, $L_\star$; radius, $R_\star$; rotation period $P_{rot}$; and projected equatorial velocity, $v \sin i$. Table~\ref{tbl:props} lists the 38 stars, together with an abridged summary of properties. (The online table lists all properties.\footnote{Online tables and figure sets available at http://iopscience.iop.org/0067-0049/211/1/9/}) The sample contains 5 F stars, 12 G stars, 8 K stars, and 13 M stars.  According to the SIMBAD database \citep{wenger00}, 3 stars are characterized as Cepheids, 5 as flare stars, and 12 as variable stars. Another 8 are designated WTTS by either \citet{herbig88}, \citet{alcala95}, \citet{sterzik99}, or \citet{neuhauser00}. The remaining 10 members have no unusual classifications. The sample is diverse, ranging in age from roughly 2 Myr to 10,500 Myr; mass from 0.08 $M_\sun$ to 7.7 $M_\sun$; $\log_{10}{g}$ from 1.6 to 5.3 (cgs units); luminosity from 0.0003 $L_\sun$ to 5300 $L_\sun$; radius from 0.1 $R_\sun$ to 71 $R_\sun$; effective temperature from 2564 K to 6959 K; $v \sin i$ from $<1$ km s$^{-1}$ to 163 km s$^{-1}$; and rotation period from 0.4 d to 286 d. The dataset(s) analyzed for each star are summarized in Table~\ref{tbl:obs} (the online table has information on each exposure). 

\begin{deluxetable*}{lllllrrlrlrlrlrlrl}
\tabletypesize{\scriptsize}
\tablecaption{Abridged properties of the stars in the sample\tablenotemark{a}\label{tbl:props}}
\tablehead{\colhead{Star} & \colhead{Spectral} &  \colhead{Ref} & \colhead{Other} & \colhead{Ref} & \colhead{$N_{p}$} & \colhead{Age} & \colhead{Ref} & \colhead{$P_{rot}$} & \colhead{Ref} & \colhead{$M$} & \colhead{Ref}\\
 & \colhead{Type\tablenotemark{b}} & & \colhead{Class.\tablenotemark{b}} & & & \colhead{(Myr)} & & \colhead{(d)} & & \colhead{$(M_\sun)$} & }
\startdata
$\beta$ Cas & F2IV &  & $\delta$ Sct Var & 1 &  & 1124$\pm$45\tablenotemark{c} & 2,3 & 0.89$\pm$0.03 & 2 & 1.91$\pm$0.02 & 2\\
$\delta$ Cep & F5Iab &  & Cepheid Var &  &  & 66 & 4 & $<$114\tablenotemark{d} & 5 & 4.82$\pm$0.26\tablenotemark{c} & 6\\
$\alpha$ Per & F5Iab &  & Var &  &  & 45.9$\pm$4.2 & 7 & 87.7 & 8 & 7.3$\pm$0.3 & 9\\
$\beta$ Dor & F6Ia &  & Cepheid Var &  &  & 42.5$\pm$2.7 & 7 & $<$286$\pm$98\tablenotemark{d} & 5 & 7.7$\pm$0.2 & 7\\
Polaris & F7Ib-IIv &  & Cepheid Var &  &  & 50$\pm$11 & 7 & $<$79\tablenotemark{d} & 5 & 6.9$\pm$0.5 & 7\\
HD25825 & G0 &  &  &  &  & 3700$\pm$1200\tablenotemark{c} & 10,11 & $\sim$6.5 & 12 & 1.055$\pm$0.024\tablenotemark{c} & 10,13\\
HD209458 & G0V & 14 &  &  & 1 & 2900$\pm$870\tablenotemark{c} & 10,3,11 & 11.4 & 15 & 1.128$\pm$0.018\tablenotemark{c} & 10,13,16\\
$\chi^1$ Ori & G0V & 17 & RS CVn Var &  &  & 2160$\pm$870\tablenotemark{c} & 10,11 & 5.1 & 18 & 0.90$\pm$0.01\tablenotemark{c} & 10,13\\
HII314 & G1-2V & 19 & BY Dra Var &  &  & 126 & 20 & 1.47851 & 21 & 1.1 & 20\\
EK Dra & G1.5V & 14 & BY Dra Var &  &  & 27.6$\pm$4.2 & 7 & 2.686\tablenotemark{c} & 22,23 & 1.044$^{+0.014}_{-0.02}$ & 10\\
$\pi^1$ UMa & G1.5Vb & 14 & BY Dra Var &  &  & 300 & 24 & 4.89 & 25 & 1 & 20\\
HD90508 & G1V & 26 &  &  &  & 10500$\pm$2000\tablenotemark{c} & 3,11 & $<$21\tablenotemark{d} & 5 & 1.02$\pm$0.13 & 13\\
HD199288 & G2V & 17 &  &  &  & 7700$\pm$2400\tablenotemark{c} & 3,11 & $<$12\tablenotemark{d} & 5 & 0.896$\pm$0.017\tablenotemark{c} & 13,27\\
18 Sco & G2Va & 28 & Var &  &  & 5500$\pm$1400\tablenotemark{c} & 29,10,11 & 22.7$\pm$0.5 & 30 & 1.008$\pm$0.024\tablenotemark{c} & 29,10,13\\
FK Com & G4III & 19 & Rot Var &  &  & \nodata &  & 2.40025 & 31 & 1.5 & 32\\
HD65583 & G8V &  &  &  &  & 5300$\pm$2600\tablenotemark{c} & 10,11 & 40 & 33 & 0.816$^{+0.02}_{-0.042}$ & 10\\
HD103095 & G8Vp &  &  &  &  & 8300$^{+3900}_{-3800}$ & 34 & 31 & 35 & 0.661$^{+0.028}_{-0.006}$ & 10\\
HD282630 & K0V &  & WTTS & 36 &  & 6.9$^{+6}_{-4}$ & 37 & 2.2321 & 38 & 1.35$^{+0.19}_{-0.16}$ & 37\\
HD189733 & K1V & 39 &  &  & 1 & 6800$^{+5200}_{-4400}$ & 16 & 11.95$\pm$0.01 & 40 & 0.816$\pm$0.025\tablenotemark{c} & 16,41\\
HD145417 & K3V & 17 &  &  &  & 7100$\pm$4700 & 11 & $<$6.9\tablenotemark{d} & 5 & 0.62 & 42\\
V410-$\tau$ & K4IV & 19 & WTTS & 36 &  & 2.0$\pm$0.4 & 7 & 1.872 & 43 & 1.2$\pm$0.2 & 7\\
EG Cha & K4Ve & 44 & WTTS & 45 &  & 5$\pm$2 & 46 & 4.5$\pm$0.04 & 47 & 1 & 48\\
HBC427\tablenotemark{e} & K5 &  & WTTS & 36 &  & 3.3 & 49 & 9.3898 & 38 & 1.4 & 50\\
61 Cyg A & K5V & 51 & BY Dra Var &  &  & 6000$\pm$1000 & 52 & 35.37 & 53 & 0.66$^{+0.01}_{-0.002}$ & 10\\
LkCa 4 & K7V & 54 & WTTS & 36 &  & 2.7$\pm$1.5 & 55 & 3.371 & 56 & 0.77$\pm$0.09 & 55\\
GJ832 & M1.5 & 57 &  &  & 1 & \nodata &  & \nodata &  & 0.45$\pm$0.05 & 58\\
TWA13B & M1Ve & 44 & WTTS & 59 &  & 8$\pm$2 & 60 & 5.35$\pm$0.03 & 48 & 0.68 & 61\\
TWA13A & M1Ve & 44 & WTTS & 59 &  & 8$\pm$2 & 60 & 5.56$\pm$0.03 & 48 & 0.7 & 61\\
AU Mic & M1Ve & 44 & BY Dra Var &  &  & 12$\pm$2 & 60 & 4.85$\pm$0.02 & 47 & 0.47$\pm$0.12\tablenotemark{c} & 60,48,62\\
CE Ant & M2Ve & 44 & WTTS & 63 &  & 5.3$\pm$1.9\tablenotemark{c} & 46,63 & 5$\pm$0.03 & 48 & 0.55$\pm$0.15 & 63\\
GJ436 & M3.5V & 64 &  &  & 1 & 6000$^{+4000}_{-5000}$ & 16 & 48 & 65 & 0.445$\pm$0.008\tablenotemark{c} & 66,67,68\\
EV Lac & M4.5V & 64 & Flare &  &  & 25 & 69 & 4.38 & 70 & 0.315$\pm$0.002 & 64\\
AD Leo & M4.5Ve & 14 & Flare &  &  & 25 & 69 & 2.6 & 70 & 0.390$\pm$0.032 & 64\\
IL Aqr & M5.0V & 64 & BY Dra Var &  & 4 & 2600$\pm$2500 & 71 & 96.7$\pm$1 & 72 & 0.33$\pm$0.01 & 68\\
HO Lib & M5.0V & 64 & BY Dra Var &  & 4 & 9000$\pm$2000 & 73 & 94.2$\pm$1 & 74 & 0.3087$\pm$0.0057\tablenotemark{c} & 64,68\\
Prox Cen & M6Ve & 44 & Flare &  &  & 5750$\pm$150 & 75 & 82.53 & 76 & 0.123$\pm$0.006 & 77\\
GJ3877 & M7.0V & 64 & Flare &  &  & 3100 & 78 & $<$1.2$\pm$0.5\tablenotemark{d} & 5 & 0.10$\pm$0.02\tablenotemark{f} & \\
GJ3517 & M9.0V & 64 & Flare &  &  & 3100 & 78 & $<$0.4\tablenotemark{d} & 5 & 0.08$\pm$0.02 & 79
\enddata
\tablenotetext{a}{Some stellar properties, uncertainty digits, planet references, and footnotes omitted for brevity. The full table is available at http://iopscience.iop.org/0067-0049/211/1/9/.}
\tablenotetext{b}{Reproduced from the SIMBAD database \citep{wenger00} with associated references when available. References for the WTTS classifications are not taken from SIMBAD.}
\tablenotetext{c}{Mean of multiple values found in the reference(s), using $1/\sigma^2$ weighting factors when possible. When uncertainties on a particular value were asymmetric, we used the average of the two uncertainties as $\sigma$. When the literature provided four or more values \emph{without uncertainties}, we estimated the uncertainty as the sample standard deviation of the values.}
\tablenotetext{d}{Represents the upper limit, $P/\sin i$, computed from the $v \sin{i}$ and $R$ values. Where possible, we used simple propagation of errors to estimate the uncertainty.}
\tablenotetext{e}{Unresolved binary system (see Section~\ref{sec:stars}).}
\tablenotetext{f}{Value assumed from stars of similar spectral type.}

\tablerefs{ (1) \citet{rodriguez00}; (2) \citet{che11}; (3) \citet{holmberg09}; (4) \citet{acharova12}; (5) this work; (6) \citet{caputo05}; (7) \citet{tetzlaff11}; (8) \citet{hatzes95}; (9) \citet{lyubimkov10}; (10) \citet{takeda07}; (11) \citet{casagrande11}; (12) \citet{linsky12}; (13) \citet{allende99}; (14) \citet{montes01mnras}; (15) \citet{silva08}; (16) \citet{torres08}; (17) \citet{gray06}; (18) \citet{telleschi05}; (19) \citet{strassmeier09}; (20) \citet{metchev09}; (21) \citet{hartman10}; (22) \citet{strassmeier98}; (23) \citet{konig05}; (24) \citet{montes01aap}; (25) \citet{gaidos00}; (26) \citet{cenarro07}; (27) \citet{sousa11}; (28) \citet{shenavrin11}; (29) \citet{valenti05}; (30) \citet{petit08}; (31) \citet{jetsu93}; (32) \citet{eggen89}; (33) \citet{isaacson10}; (34) \citet{mamajek08}; (35) \citet{baliunas96}; (36) \citet{herbig88}; (37) \citet{kraus09}; (38) \citet{watson06}; (39) \citet{belle09}; (40) \citet{henry08}; (41) \citet{bouchy05}; (42) \citet{santos05}; (43) \citet{pojmanski05}; (44) \citet{torres06}; (45) \citet{alcala95}; (46) \citet{weise10}; (47) \citet{messina11}; (48) \citet{messina10}; (49) \citet{palla02}; (50) \citet{kraus11}; (51) \citet{white07}; (52) \citet{robrade12}; (53) \citet{donahue96}; (54) \citet{riviere12}; (55) \citet{bertout07}; (56) \citet{xiao12}; (57) \citet{koen10}; (58) \citet{bailey09}; (59) \citet{sterzik99}; (60) \citet{plavchan09}; (61) \citet{manara13}; (62) \citet{kalas04}; (63) \citet{neuhauser00}; (64) \citet{jenkins09}; (65) \citet{demory07}; (66) \citet{braun12}; (67) \citet{torres07}; (68) \citet{onehag12}; (69) \citet{shkolnik09}; (70) \citet{hempelmann95}; (71) \citet{correia10}; (72) \citet{rivera05}; (73) \citet{selsis07}; (74) \citet{vogt10}; (75) \citet{yildiz07}; (76) \citet{kiraga07}; (77) \citet{demory09}; (78) \citet{reiners09}; (79) \citet{martin94}}
\end{deluxetable*}

\begin{deluxetable*}{rllllrrrrr}
\tabletypesize{\scriptsize}
\tablewidth{0pt}
\tablecaption{Abridged catalog of observations\label{tbl:obs}}
\tablehead{\colhead{Dataset\tablenotemark{a}} & \colhead{Star} & \colhead{Inst.} & \colhead{Grating} & \colhead{Start} & \colhead{$N_{exp}$} & \colhead{$T_{obs}$} & \colhead{$\sum{T_{exp}}$} & \colhead{Percent} & \colhead{$T_\textup{max}$}\tablenotemark{b}\\
\colhead{No.} & & & & \colhead{(UT)} & & \colhead{(h)} & \colhead{(h)} & \colhead{Observed} & (h)}
\startdata
1 & $\beta$ Cas & COS & G130M & 2010 Jun 07 18:00:37 UT & 2 & 0.39 & 0.36 & 91.63\% & 0.38\\
2 & $\delta$ Cep & COS & G130M & 2010 Oct 19 00:12:10 UT & 18 & 5705.01 & 2.38 & 0.04\% & 0.37\\
3 & $\alpha$ Per & COS & G130M & 2010 Jul 13 07:32:49 UT & 2 & 0.42 & 0.39 & 92.18\% & 0.41\\
4 & $\beta$ Dor & COS & G130M & 2010 Nov 14 01:53:04 UT & 14 & 6301.48 & 1.91 & 0.03\% & 0.36\\
5 & Polaris & COS & G130M & 2009 Dec 25 08:56:39 UT & 4 & 48.62 & 1.31 & 2.70\% & 0.70\\
6 & HD25825 & COS & G130M & 2010 Feb 20 23:15:33 UT & 2 & 0.36 & 0.32 & 90.06\% & 0.35\\
7 & HD209458 & COS & G130M & 2009 Sep 19 10:10:15 UT & 14 & 698.99 & 5.75 & 0.82\% & 0.65\\
8 & HD209458 & STIS & E140M & 2001 Aug 11 13:41:50 UT & 4 & 1925.79 & 2.11 & 0.11\% & 0.57\\
9 & $\chi^1$ Ori & COS & G130M & 2010 Mar 19 06:47:26 UT & 2 & 0.39 & 0.36 & 91.63\% & 0.38\\
10 & HII314 & COS & G130M & 2009 Dec 16 09:28:55 UT & 3 & 2.15 & 0.89 & 41.53\% & 0.38\\
11 & EK Dra & COS & G130M & 2010 Apr 22 09:18:42 UT & 18 & 16970.70 & 2.30 & 0.01\% & 0.74\\
12 & $\pi^1$ UMa & COS & G130M & 2010 Feb 28 22:31:16 UT & 2 & 0.39 & 0.36 & 91.63\% & 0.38\\
13 & HD90508 & COS & G140L & 2010 Jun 19 03:55:21 UT & 4 & 2.14 & 1.40 & 65.35\% & 0.85\\
14 & HD199288/G140L & COS & G140L & 2009 Nov 23 03:31:07 UT & 5 & 3.58 & 1.32 & 36.77\% & 0.71\\
15 & HD199288/G130M & COS & G130M & 2009 Nov 23 07:58:43 UT & 4 & 2.46 & 1.61 & 65.62\% & 0.40\\
16 & 18 Sco/G140L & COS & G140L & 2011 Feb 04 20:19:10 UT & 2 & 1.24 & 0.46 & 37.60\% & 0.23\\
17 & 18 Sco/G130M & COS & G130M & 2011 Feb 04 21:46:43 UT & 3 & 1.68 & 0.75 & 44.72\% & 0.54\\
18 & FK Com & COS & G130M & 2011 Mar 07 01:31:09 UT & 47 & 2079.45 & 9.19 & 0.44\% & 0.81\\
19 & HD65583 & COS & G140L & 2010 Mar 27 09:38:51 UT & 4 & 2.18 & 1.34 & 61.81\% & 0.81\\
20 & HD103095/G140L & COS & G140L & 2010 May 27 06:00:33 UT & 3 & 0.59 & 0.53 & 88.83\% & 0.58\\
21 & HD103095/G130M & COS & G130M & 2010 May 27 09:34:00 UT & 2 & 0.84 & 0.79 & 94.11\% & 0.38\\
22 & HD282630 & COS & G130M & 2011 Mar 30 23:21:36 UT & 4 & 1.94 & 0.54 & 27.90\% & 0.34\\
23 & HD189733 & COS & G130M & 2009 Sep 16 18:31:52 UT & 6 & 6.65 & 1.29 & 19.45\% & 0.82\\
24 & HD145417 & COS & G140L & 2010 Mar 11 21:57:16 UT & 4 & 2.22 & 1.49 & 66.82\% & 0.88\\
25 & V410-$\tau$ & STIS & E140M & 2001 Jan 30 10:35:33 UT & 4 & 5.25 & 2.66 & 50.70\% & 0.80\\
26 & EG Cha & COS & G130M & 2010 Jan 22 07:53:57 UT & 4 & 0.94 & 0.82 & 87.38\% & 0.44\\
27 & HBC427 & COS & G130M & 2011 Mar 29 23:42:24 UT & 4 & 1.83 & 0.59 & 32.27\% & 0.34\\
28 & 61 Cyg A & COS & G130M & 2010 Mar 28 23:08:26 UT & 2 & 0.39 & 0.36 & 91.62\% & 0.38\\
29 & LkCa 4 & COS & G130M & 2011 Mar 30 06:05:02 UT & 4 & 1.39 & 0.64 & 45.95\% & 0.34\\
30 & GJ832 & COS & G130M & 2012 Jul 28 22:12:56 UT & 2 & 1.26 & 0.60 & 47.68\% & 0.33\\
31 & TWA13B & COS & G130M & 2011 Apr 02 04:26:51 UT & 4 & 0.82 & 0.71 & 85.65\% & 0.38\\
32 & TWA13A & COS & G130M & 2011 Apr 02 01:26:37 UT & 4 & 0.68 & 0.55 & 81.49\% & 0.31\\
33 & AU Mic & STIS & E140M & 1998 Sep 06 12:17:14 UT & 4 & 5.35 & 2.81 & 52.45\% & 0.73\\
34 & CE Ant & COS & G130M & 2011 May 05 06:33:24 UT & 4 & 1.63 & 0.52 & 32.06\% & 0.28\\
35 & GJ436 & COS & G130M & 2012 Jun 23 07:22:56 UT & 3 & 1.74 & 0.94 & 53.82\% & 0.62\\
36 & EV Lac & STIS & E140M & 2001 Sep 20 16:45:48 UT & 4 & 5.30 & 3.03 & 57.21\% & 4.04\\
37 & AD Leo & STIS & E140M & 2000 Mar 10 03:28:05 UT & 26 & 19540.16 & 18.62 & 0.10\% & 4.01\\
38 & IL Aqr & COS & G130M & 2012 Jan 05 01:36:44 UT & 2 & 1.58 & 0.56 & 35.47\% & 0.33\\
39 & HO Lib & COS & G130M & 2011 Jul 20 13:47:36 UT & 3 & 1.36 & 0.41 & 30.20\% & 0.17\\
40 & Prox Cen & STIS & E140M & 2000 May 08 00:58:04 UT & 7 & 29.10 & 9.92 & 34.08\% & 6.01\\
41 & GJ3877 & COS & G140L & 2011 Jan 30 18:11:23 UT & 6 & 2.38 & 1.40 & 58.88\% & 0.82\\
42 & GJ3517 & COS & G140L & 2011 Feb 15 12:30:49 UT & 12 & 7.02 & 2.81 & 39.96\% & 0.81\\
\enddata
\tablenotetext{a}{Corresponds to the numbering of Figure Sets 4 and 9.}
\tablenotetext{b}{~Longest single block of \emph{quiescent} data with $\geq50$\% time exposed and constant central wavelength setting of the detector (i.e. the longest timescale of sampled variability).}
\end{deluxetable*}

\subsection{Lightcurve Extraction}
\label{sec:lcextract}
The data retrieved from MAST (``tag'' files for STIS and ``corrtag'' files for COS) consist of event lists of the time and detector coordinates for each recorded count. To process these data, we developed a customized IDL pipeline that constructs lightcurves by identifying a region corresponding to a chosen wavelength band (or regions if multiple orders in STIS/E140M exposures contain appropriate wavelengths) to extract signal counts, then bins these counts by the chosen cadence. It also identifies an adjacent background region, and subtracts the cadence-binned counts (adjusted according to difference in areas) from the signal. Each lightcurve point is assigned a photometric uncertainty equal to the sum, in quadrature, of the Poisson errors of the signal and background counts. 

\begin{figure}
\plotone{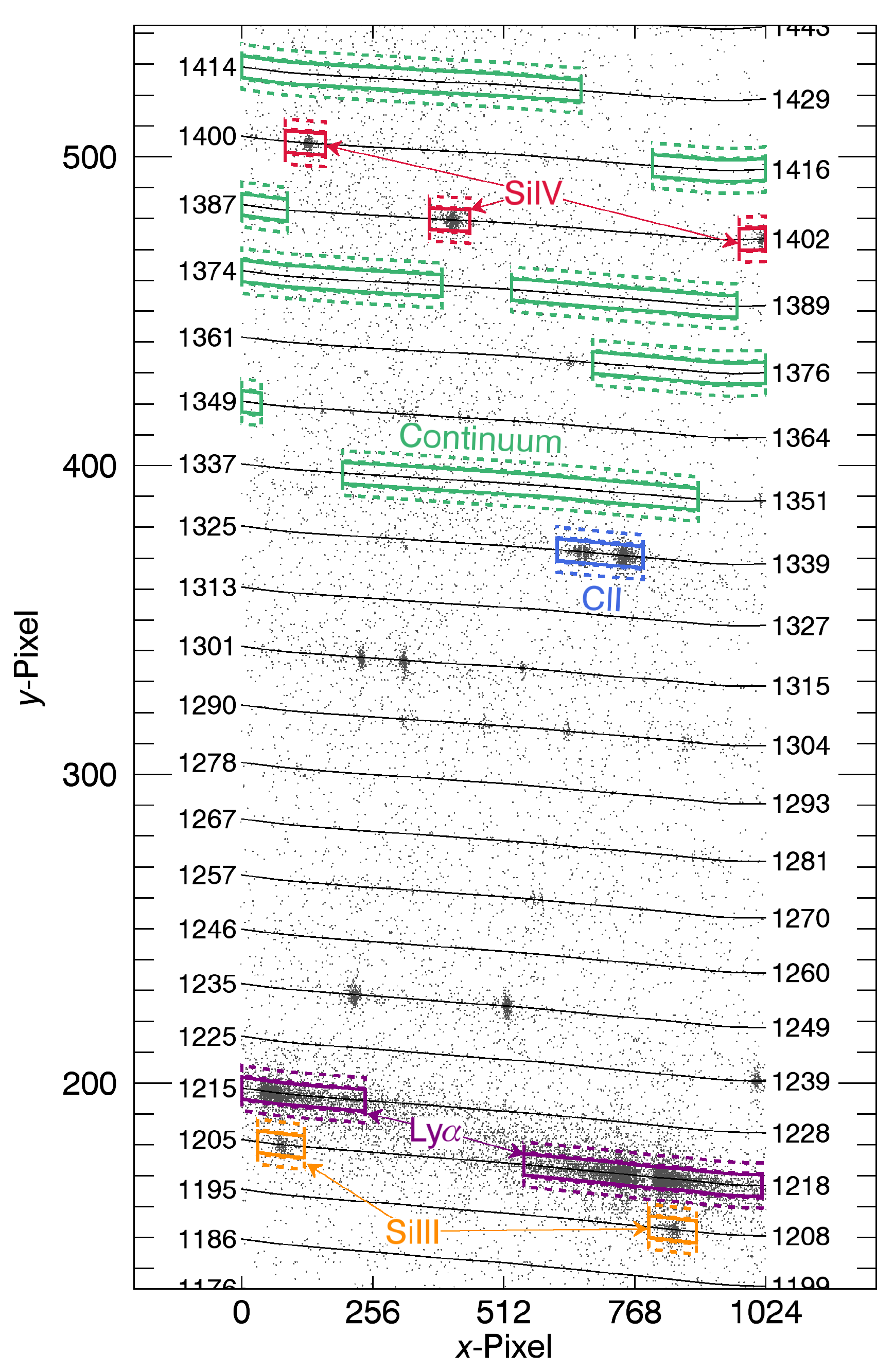}
\caption{Portion of exposure o61s01010 (MAST ID) from STIS of AD Leo. The figure has been cropped and stretched for display. Dots represent $\sim$30\% of the detected counts (many overlap). Spectral orders appear as black lines with labels showing the endpoint wavelengths in \AA\ of each order's dispersion axis. Signal and background count extraction areas are shown as solid and dashed lines respectively.}
\label{fig:stisimg}
\end{figure}

\begin{figure}
\plotone{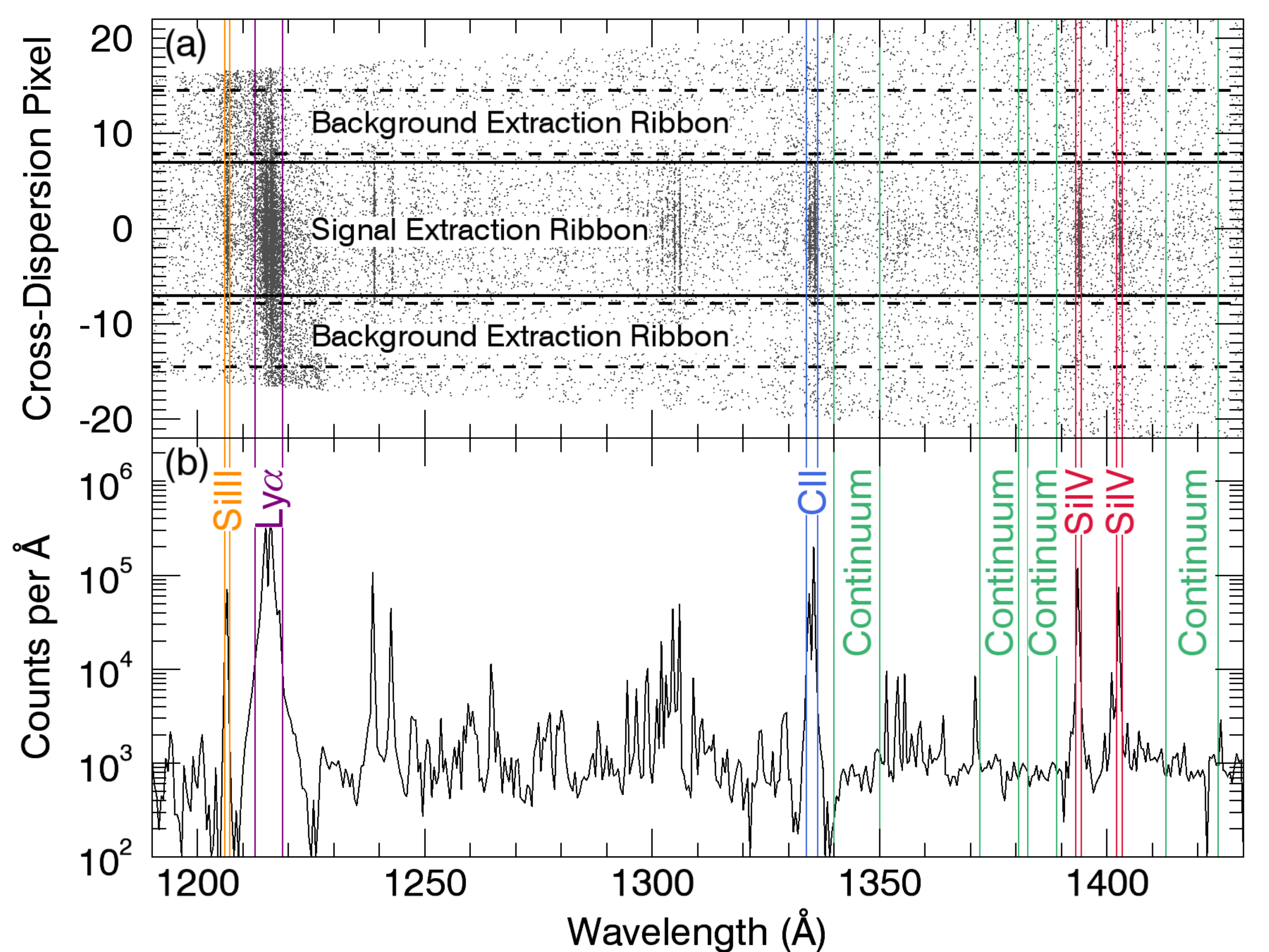}
\caption{(a) Example STIS counts from AD Leo after wavelength assignment (Figure~\ref{fig:stisimg}). Dots represent $\sim$1\% of the detected counts (many overlap). The upper and lower limits of the count cross-dispersion pixels increase with wavelength because the spectral orders are spaced farther apart (see Figure~\ref{fig:stisimg}). (b) Spectrum from the full set of background-subtracted signal counts histogrammed into 0.5 \AA\ bins. Vertical lines highlight the bands used to extract lightcurves. The dips to either side of strong lines are an artifact of light from bright lines scattered diagonally into the background ribbons and subtracted off (visible in Figure~\ref{fig:stisimg}).} 
\label{fig:stisspec}
\end{figure}

\begin{figure}
\plotone{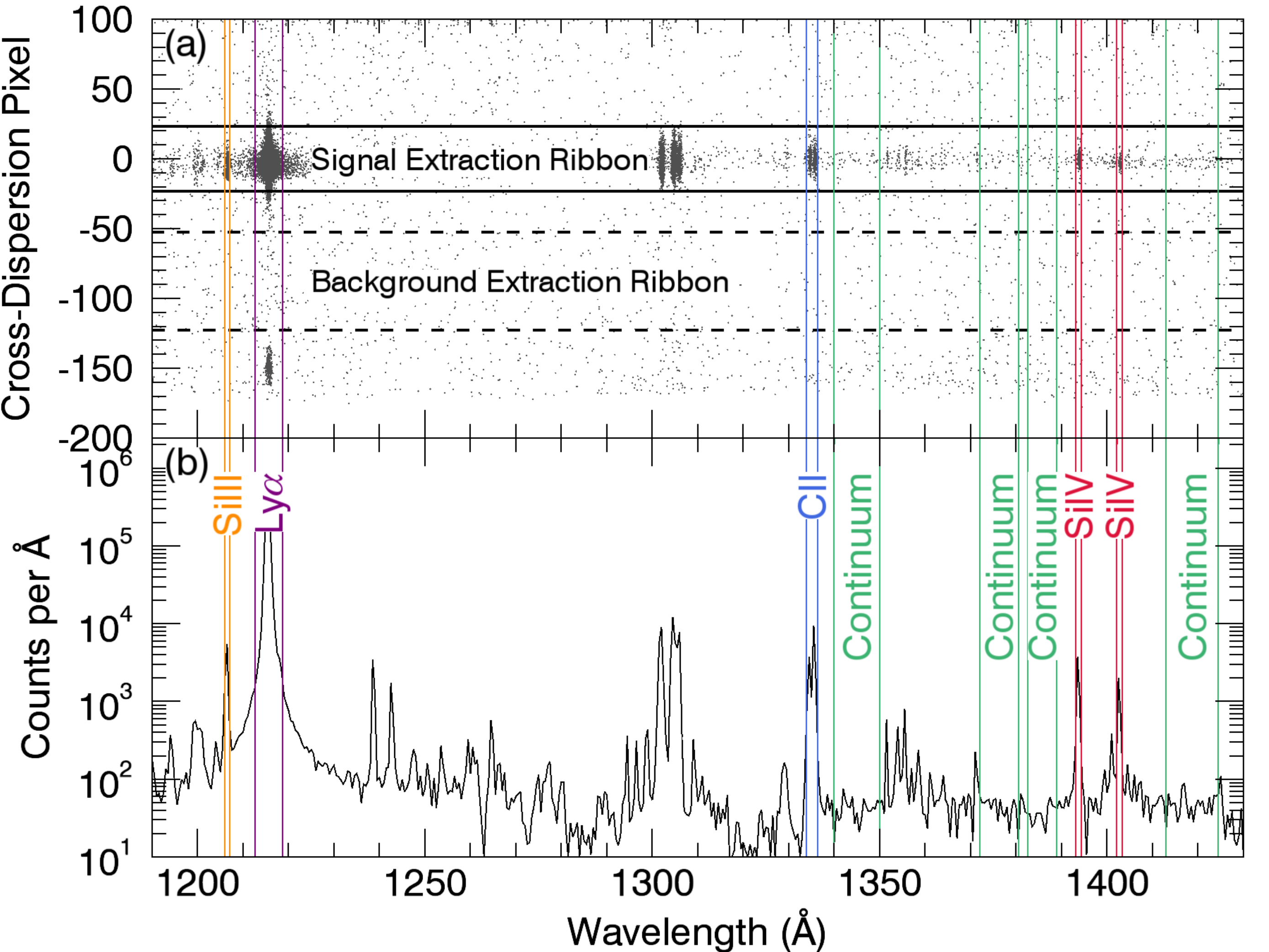}
\caption{Example COS G130M counts from CE Ant (Table~\ref{tbl:obs}), following the same format as Figure~\ref{fig:stisspec} and plotting $\sim$4\% of the counts.}
\label{fig:cosspec}
\end{figure}

A portion of the event-list data from a STIS exposure is displayed in Figure~\ref{fig:stisimg}, overplotted with the dispersion axis of each spectral order produced by the Echelle grating. From these data, assigning a wavelength to each count according to the nearest order and recording its cross-dispersion (vertical) distance from that order's dispersion axis produces Figure~\ref{fig:stisspec}a. Finally, accumulating all of the counts within the signal ribbon in Figure~\ref{fig:stisspec}a and subtracting the area-scaled background ribbon counts, results in the accumulated spectrum of Figure~\ref{fig:stisspec}b. Figure~\ref{fig:cosspec} is analogous to Figure~\ref{fig:stisspec} for example COS data. A plot similar to Figure~\ref{fig:stisimg} is unnecessary for COS exposures because they contain only a single spectral order.

\begin{deluxetable}{lcc}
\tablecaption{Observed line bands\label{tbl:bands}}
\tablewidth{\columnwidth}
\tablehead{\colhead{Ion/Line\tablenotemark{a}} & \colhead{$\lambda_\textup{rest}$\tablenotemark{b}} & \colhead{Band} \\
		 & (\AA) & (\AA)}
\startdata
 \ion{Si}{3} &1206.51 & 1205.92 -- 1207.12\\
\ion{Si}{4} & 1393.76 & 1393.16 -- 1394.36\\
 & 1402.77 & 1402.17 -- 1403.37\\
\ion{C}{2} & 1334.53 & 1333.90 -- 1336.30\\
 & 1335.71 &\\
FUV Continuum & & 1340.00 -- 1350.00\\
 & & 1372.00 -- 1380.50\\
 & & 1382.50 -- 1389.00\\
 & & 1413.00 -- 1424.50\\
Ly$\alpha$\tablenotemark{c} & 1215.67  & 1212.67 -- 1218.67\\
 & 1215.67  & \\
\enddata
\tablenotetext{a}{Throughout the paper we refer to these emission lines and bands simply by this identifier.}
\tablenotetext{b}{These values were retrieved from the National Institute of Standards and Technology Atomic Spectra Database at http://www.nist.gov/pml/data/asd.cfm \citep{kramida12}.}
\tablenotetext{c}{While not analyzed for variability, we included the Ly$\alpha$ line to compare to the lightcurves of other lines in search of contamination by geocoronal emission.}
\end{deluxetable}

We used the custom pipeline to create lightcurves with a uniform cadence of $\dt = 60$ s for the \ion{C}{2}, \ion{Si}{4}, FUV continuum, and where possible, \ion{Si}{3} bands outlined in Table~\ref{tbl:bands}. The OIV] $\lambda$1401 line falls between the two \ion{Si}{4} bands, but a visual inspection of the spectra for each star revealed no contamination. A $\dt = 60$ s cadence provided a balance between higher signal-to-noise, a greater number of lightcurve data points, and the ability to resolve flares. In addition, it nearly matches the $\dt=58.85$ s short cadence {\it Kepler} data \citep{koch10}, facilitating comparisons between the datasets. (Appendix~\ref{sec:sensitivity} provides an analysis of the effect of cadence on precision in estimating white-noise levels that also supports the use of a short cadence.) In Figures~\ref{fig:stisimg} and~\ref{fig:cosspec}a, solid lines outline signal extraction regions corresponding to the bands in Table~\ref{tbl:bands} and dotted lines outline background extraction regions. For each star, we concatenated all exposures from the same grating into a single temporal dataset. 

\begin{figure*}
\includegraphics[width=\textwidth]{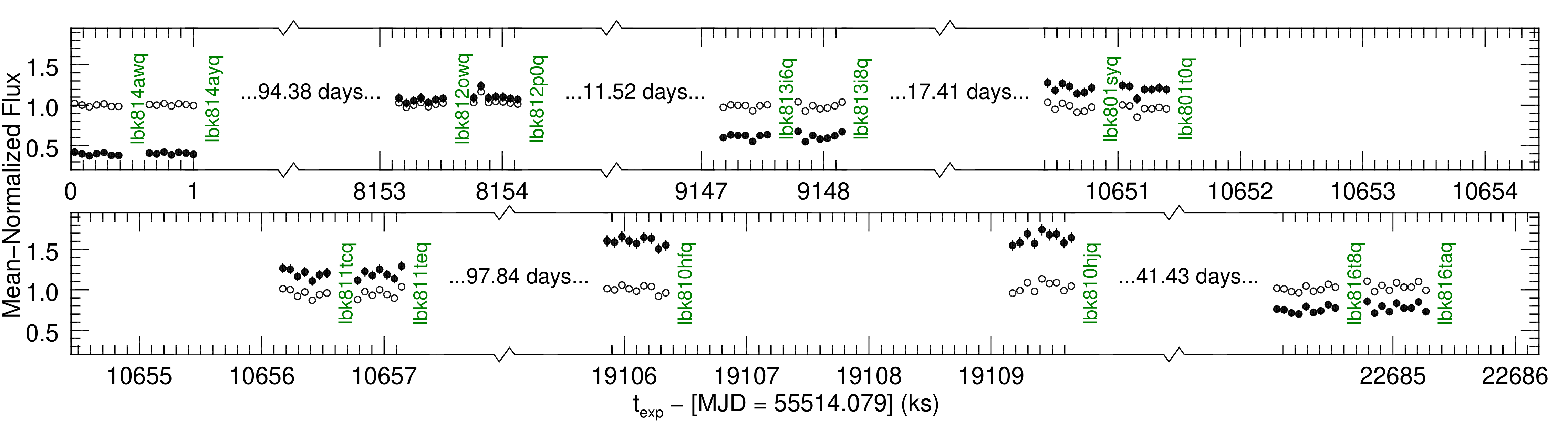}
\caption{Example lightcurve of a quiescent star created from 14 COS/G130M exposures of the Cepheid variable $\beta$ Dor in the \ion{C}{2} band. Solid points are the unfiltered values, while open points are high-pass filtered with cutoff frequency $(7\ \textup{h})^{-1}$. The lightcurve is normalized by the mean quiescent flux of $3.88\pm0.23\sn{-14}$ erg s$^{-1}$ cm$^{-2}$. A label of the MAST identifier (in green) marks the end of each exposure. Figure Set 4, available at http://iopscience.iop.org/0067-0049/211/1/9/, presents lightcurves like this for each star + band.}
\label{fig:LCquiet}
\end{figure*}

\begin{figure*}
\includegraphics[width=\textwidth]{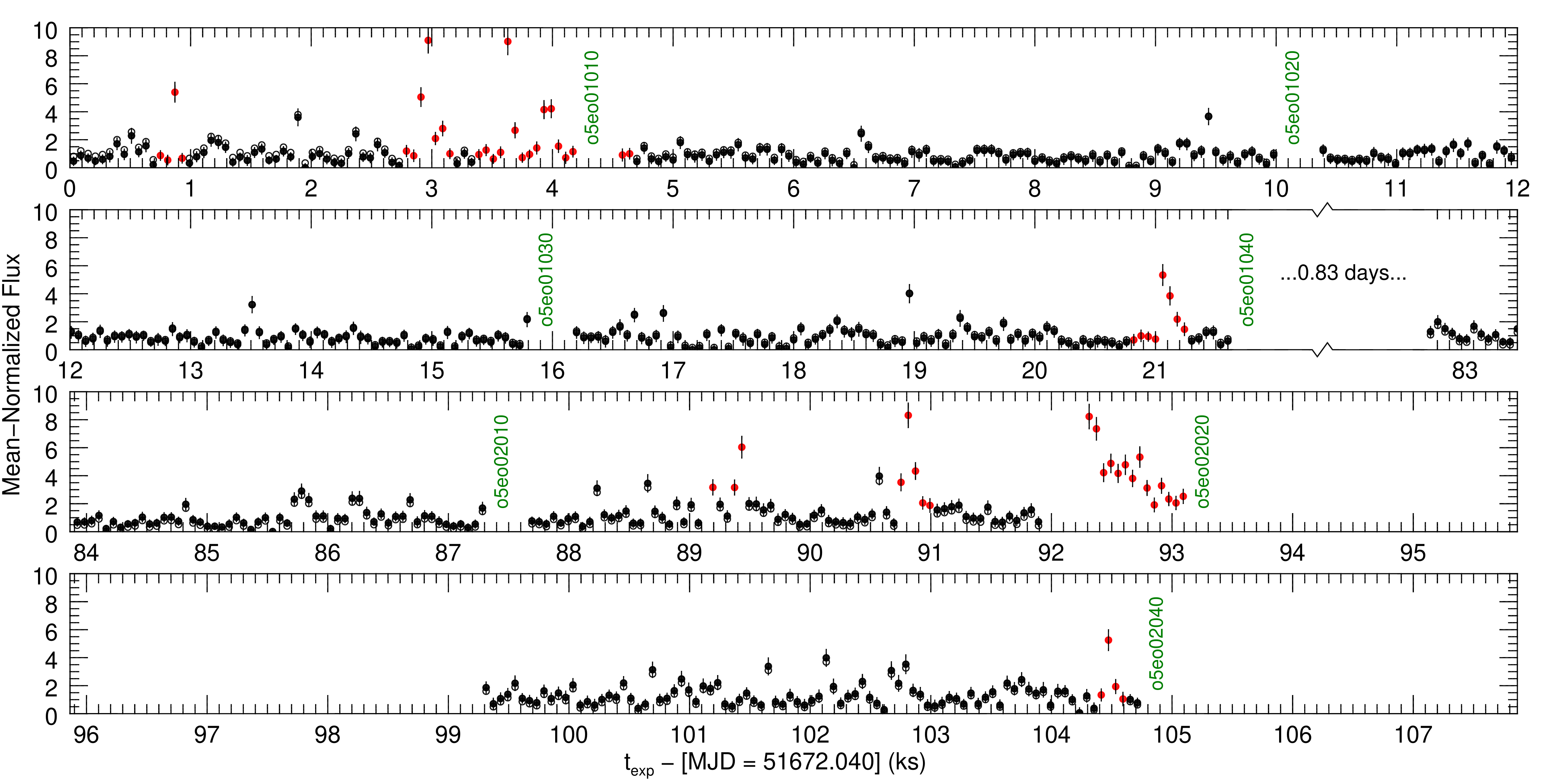}
\caption{Example lightcurve of a flaring star created from 7 STIS/E140M exposures of Prox Cen in the \ion{Si}{4} band following the same formatting as Figure~\ref{fig:LCquiet}, but with the $y$-axis clipped at 10. Points identified as flaring are red.  The mean quiescent flux is $5.24\pm0.42\sn{-15}$ erg s$^{-1}$ cm$^{-2}$. Some flare points exceed the $y$-axis maximum.}
\label{fig:LCflare}
\end{figure*}

Figures~\ref{fig:LCquiet} and~\ref{fig:LCflare} provide example lightcurves drawn from the full set of 153, viewable online as Figure Set 4. Both figures also exhibit high-pass filtered (Section~\ref{sec:filter}) versions of these data and highlight flare points (Section~\ref{sec:flaresweep}) in red. The full set of lightcurves exhibit a wide range of signal-to-noise; quoted as range (median) these are \cii: 0.8-150 (8.6), \siii: 0.8-22 (5.2), \siv: 0.5-115 (6.5), and FUV continuum: 0.8-222 (3.9).  The lightcurves also vary substantially in the number and spacing of points, containing from $\sim$20 to $\sim$10$^3$ (median $\sim$50) points. Consequently, some small datasets sample the stellar flux over less than 0.5 h, and some large datasets sample (exceedingly sparsely) the stellar flux over the course of several years.

\subsection{Continuum Subtraction}
In addition to the total flux within each line at 60~s time steps, we also computed the continuum subtracted flux in the lines. To this end, we totaled the photon counts in two bands near the line of interest. These were 1324.5-1328.3 \AA\ and 1338-1350.5 \AA\ for \cii, 1203-1205 \AA\ and 1208-1210 \AA\ for \siii, and 1382.5-1389 \AA\ and 1413-1424.5 \AA\ for \siv. The blueward \cii\ band fell on the detector gap for some COS datasets. In these instances, we halved the redward band and extrapolated from those two points. We assumed the counts in each pair of bands represented the integration of a linear continuum. With this assumption, we estimated the number of counts integrated over the emission line band attributable to the continuum. Subtracting this estimate from the line counts and augmenting the line flux error with the error in our continuum estimate completed the continuum subtraction. 

The continuum contributes significantly to the line flux of the F stars in the sample, with a median contribution of 27\% in \cii, 11\% in \siii, and 40\% in \siv. In the G, K, and M stars the median contribution is 2\% in \cii, 8\% in \siii, and 4\% in \siv. The contribution to \siii\ is similar in all the stars because the ``continuum'' is actually the blueward wing of the stellar Ly$\alpha$ that forms the base flux beneath that line in the spectra of all 38 stars. We use the continuum-subtracted data when identifying and characterizing flares but not when quantifying stochastic fluctuations, for reasons discussed in Section~\ref{sec:varanalysis}.

\section{Variability Analysis}
\label{sec:analysis}

\subsection{High-Pass Filtering}
\label{sec:filter}

\begin{figure}
\plotone{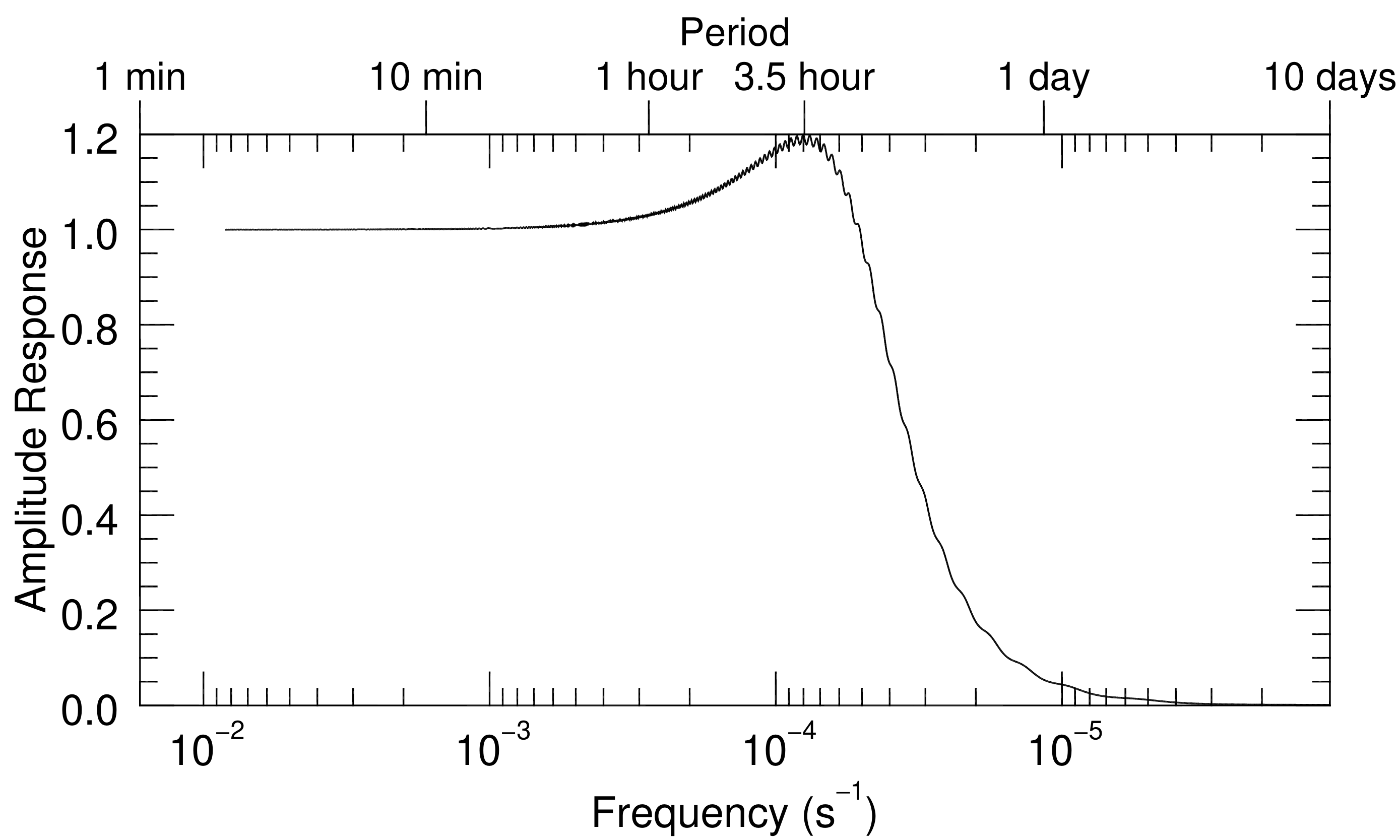}
\caption{Frequency response of the numerical technique we employed to high-pass filter the data with cutoff frequency $(7\ \textup{h})^{-1}$.}
\label{fig:filter}
\end{figure}

Filtering the lightcurves was necessary because low-frequency periodicities or overall drifts can sometimes dominate the signal. For example, the roughly two-day rotation of FK Com produces a periodic signal of much greater amplitude than rapid fluctuations visible in the lightcurve. Left unaddressed, such signals result in the misidentification of flares (Section~\ref{sec:flaresweep}) and excess noise estimates (Section~\ref{sec:varanalysis}) that overestimate the host-induced uncertainty in a transit depth measurement. However, precisely characterizing these signals is beyond the scope of this work because they do not seriously threaten transit observations -- they can generally be fitted and removed. Thus, rather than attempt to characterize low-frequency periodicities and drifts (difficult with highly clustered data), we mitigate their effects with a high-pass filter. The algorithm we employed is an exponential filter developed by \citet{rybicki95} for irregularly spaced data. As formulated by \citet{rybicki95}, the filter accepts as input a 3 dB cutoff frequency (in power). For this we chose 7 h, twice the typical transit duration. Figure~\ref{fig:filter} displays the full filter amplitude response.

Filtering the data has several side effects concerning the identification of flares and quantification of stochastic fluctuations. Most importantly, filtering shifts isolated clusters of points containing a large flare down due to their greater mean. This can significantly increase the scatter in the quiescent data. Consequently, after we filtered all of the data to identify flares, we then filtered only the quiescent data before quantifying stochastic fluctuations in the lightcurves. This and other effects of filtering are further addressed in the Sections~\ref{sec:flaresweep} and~\ref{sec:varanalysis}.

\subsection{Sweeping for Flares}
\label{sec:flaresweep}

We sought to identify flares both to characterize the risk they pose to transit observations and to prevent them from driving estimates of stochastic fluctuations.  When selecting a means of sweeping for flares, we made two important choices. First, we wished to treat the large volume of data consistently throughout. Thus, we avoided a by-eye approach. Second, we recognized that UV transit data will not always encompass the same set of chromospheric lines available in these data. Thus, we did not utilize a flare in one band to confirm a questionable, simultaneous event in another.

Initially, we hoped to identify flares by mimicking the methodology of one or more previous studies, thus facilitating comparison of the results of this analysis to the literature. However, we found no objective flare detection strategy well suited to this dataset. Therefore, we created a custom algorithm that identifies flares by cross correlating the lightcurve with a flare-shaped kernel.

In outline, the algorithm operates by first high-pass filtering the continuum-subtracted data, subtracting the mean, and normalizing by the sample standard deviation, $\s$, computed excluding outliers beyond $2.5 \s$. Once in $\s$ units, the algorithm correlates the lightcurve with the flare kernel (defined in $\s$ units as well, see below) and flags points where the correlation exceeds the value of the flare kernel correlated with itself. For each group of newly flagged points, the algorithm records the index of the peak point. The algorithm then repeats the filtering, mean-subtracting, and $\s$-normalizing steps with the original data but excluding the recorded points and again searches for flares. As this process is repeated, the algorithm adds and removes points from the list of recorded flare points one by one out to the first point at or below the quiescent mean. This slow growth of the identified regions avoids large changes between iterations that can prevent convergence. When points are no longer being added or removed, or when the same points are being  added and removed in repetition, the process is stopped.

Since the flare kernel used for correlating is in $\s$ units, the actual energy of a flare precisely matching the flare kernel is different for each lightcurve: The algorithm will identify flares only down to a minimum energy level appropriate to the noise in the given lightcurve. We chose a flare kernel that is an exponentially decaying curve with time constant of $2\dt$ (120 s) and lasting for six points, thus representing the canonical (see, e.g., \citealt{moffett72}) impulse-decay flare shape. This is near the lower bound of the decay times of flares observed by \citet{mullan06} in their extreme-UV data for 44 F-M stars. While shorter flares are possible, most observations have insufficient signal-to-noise to resolve them. 

After choosing the shape of the flare kernel, its amplitude remained an open parameter. This amplitude determines how conservative the algorithm will be when identifying flares. To set it, we ran the algorithm on simulated datasets of Gaussian white noise with flare kernels of varying amplitude until the algorithm made, on average, about one spurious detection. The simulated datasets had the same point spacing as the true datasets with 19,095 lightcurve points in total between all star and band combinations. Through these simulations, we found that a flare kernel with an amplitude of 3.5$\s$, i.e. the function  $3.5e^{-t/(120\textup{\scriptsize{ s}})}$ evaluated at the points $t=0,60,120,180,240,300$ s, produced about one false detection. More precisely, in $10^4$ simulations of the entire dataset (i.e., all stars, all bands), the algorithm with these definitions made on average 1.5 spurious flare detections, flagging 0.08\% of the simulated data. We used this same kernel for each lightcurve.

The results are moderately sensitive to the amplitude of the flare kernel. Decreasing the amplitude to 2.5$\s$ identified over twice as many events as flares (compared to a $3.5\s$-amplitude kernel) but also produced 43 false detections on average in simulated data. Consequently about a third of the additional detections in the actual data resulting from using a $2.5\s$-amplitude kernel were likely spurious. Alternatively, increasing the kernel amplitude to $4.5\s$ identified about 2/3 as many flares, but reduced spurious detections in simulated white-noise data to 0.03 events on average. The results are also sensitive to the shape of the flare kernel, particularly for events that are near the threshold of detection. For example, employing a boxcar kernel with the same area as the $3.5\s$ impulse-decay kernel (six points at $1.4\s$) identified about 50\% more events as flares and produced 23 spurious detections in simulated white-noise data.

Although the kernel shape affects identifications near the threshold, an event of any shape can trigger a detection if it causes enough of a flux boost. For example, a single point at 5.5$\s$ followed by five at the mean or six consecutive points at 2.3$\s$ above the mean would both result in a ``flare" detection. Figure~\ref{fig:LCflare} is an example of a star, Prox Cen, with multiple flares identified by the routine in the \ion{Si}{4} band, whereas the \ion{C}{2} band of $\beta$ Dor in Figure~\ref{fig:LCquiet} is an example where the algorithm identified no flares.

\begin{figure}
\plotone{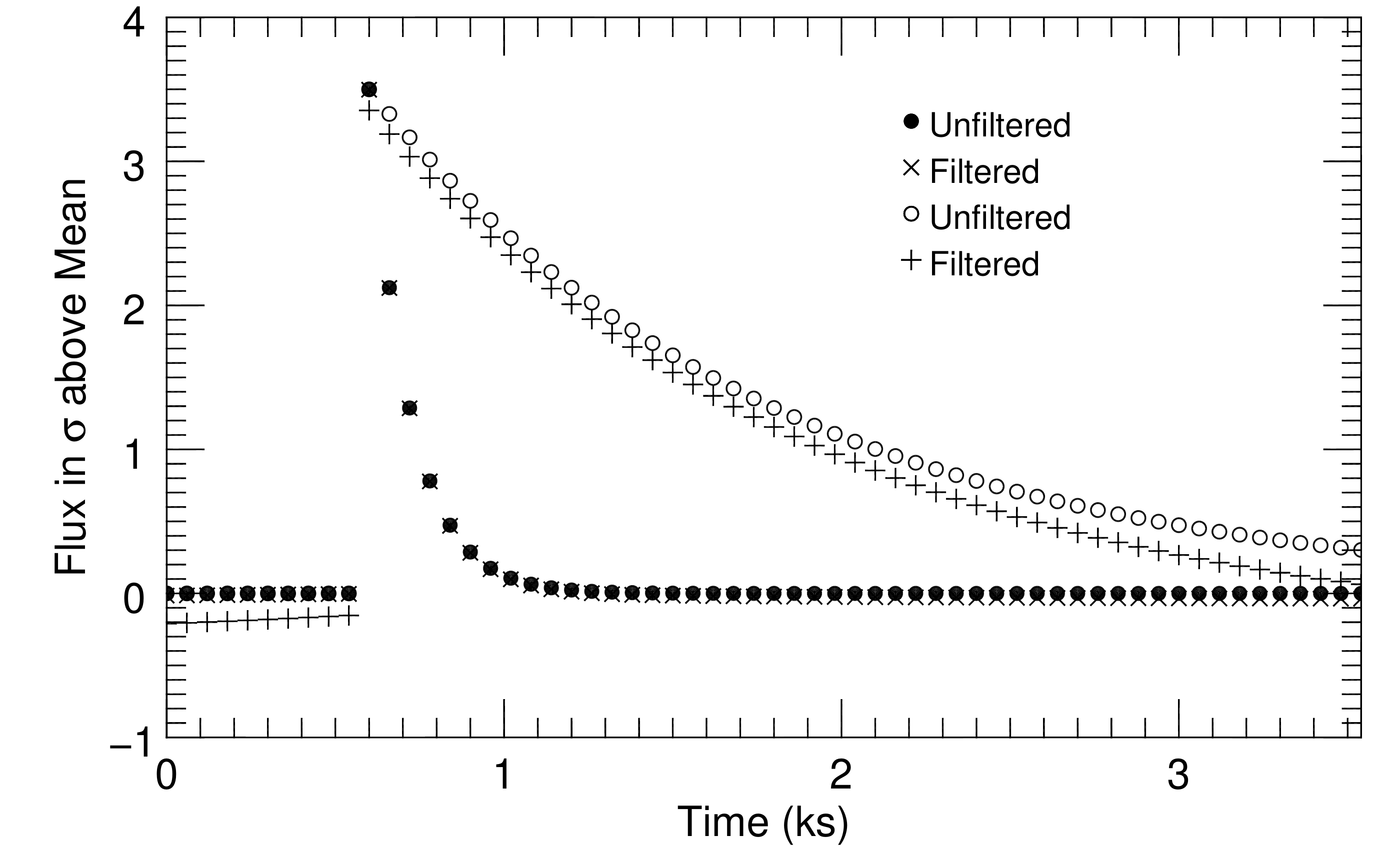}
\caption{The effect of high-pass filtering on canonical impulse-decay flare signals with 2 min and 20 min decay times. The short signal is identical to the flare kernel used to identify flares by cross-correlation, extended beyond the six points of the kernel.}
\label{fig:filteredkernel}
\end{figure}

High-pass filtering the lightcurves (Section~\ref{sec:filter}) when searching for flares strongly mitigated the false identification of more gradual but long-lived changes in stellar flux (e.g. pulsations of Cepheid variables). Although the filter will affect the flare signals, Figure~\ref{fig:filteredkernel} shows that the effect of filtering on the shape of flares lasting minutes is negligible and the effect on flares lasting tens of minutes (the longest detected, see Section~\ref{sec:flareresults}) is marginal. However, this is only true when the flares are surrounded by quiescent data: Without quiescent reference points flanking a flare, the filter would slide it down to the lightcurve mean, possibly preventing identification. Filtering also introduces slight changes in the amplitude of quiescent lightcurve scatter (Section~\ref{sec:varanalysis}), but these are well below the level of any detectable flare signal. 

\subsection{Quantification of Stochastic Fluctuations (Excess Noise)}
\label{sec:varanalysis}
\begin{figure}
\plotone{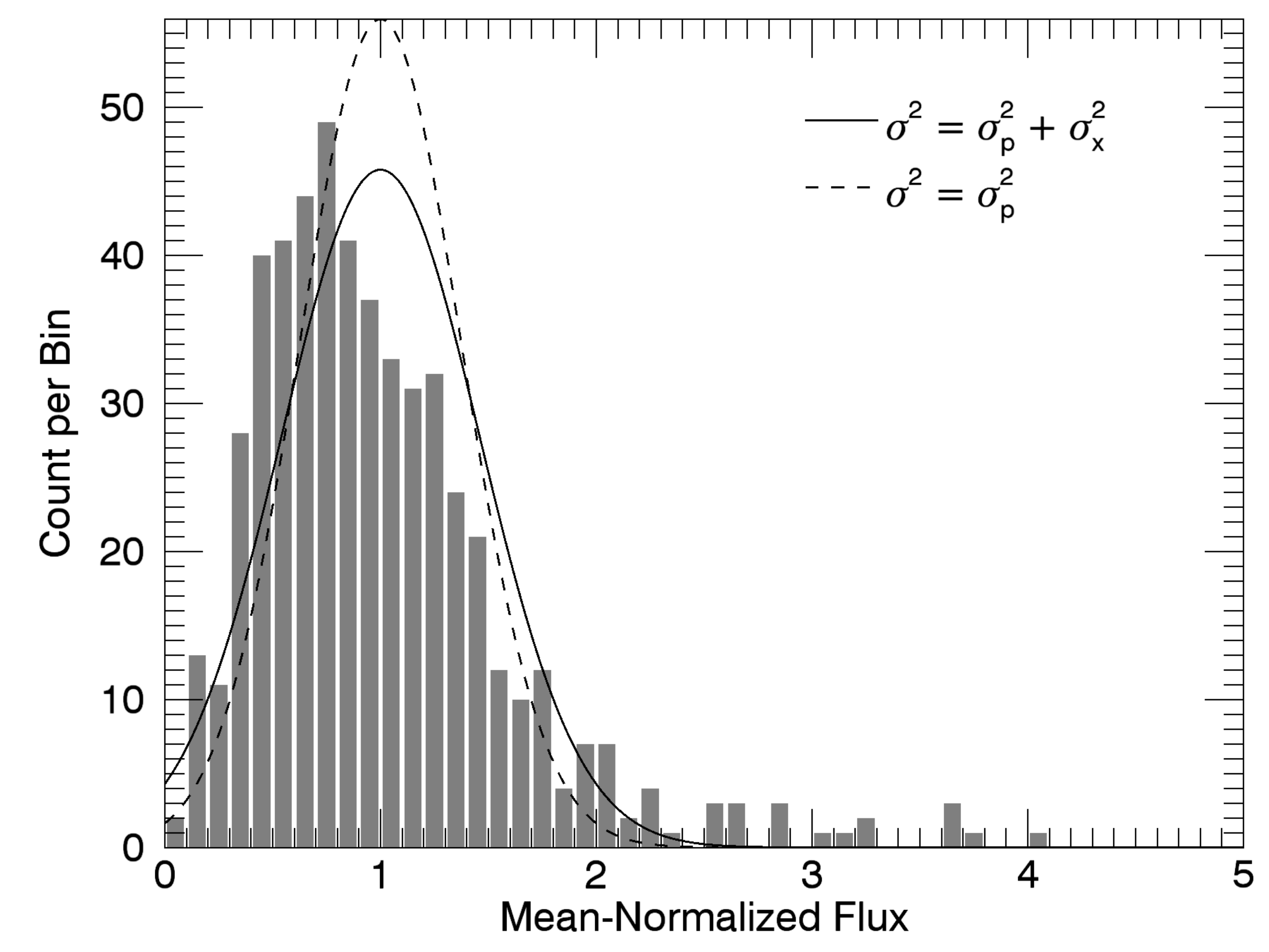}
\caption{A histogram of the high-pass filtered, quiescent lightcurve points from the Prox Cen \siv\ data (see Figure~\ref{fig:LCflare}). The lines represent normalized Gaussians with and without excess noise.}
\label{fig:hist}
\end{figure}

Stochastic fluctuations of stellar flux will produce scatter in lightcurves beyond that attributable to Poisson noise. It is simplest to treat these stochastic fluctuations as white noise. This is especially true given that these data have short temporal baselines and are highly clustered (see Table~\ref{tbl:obs}), unlike the lengthy sets of evenly spaced {\it Kepler} data that enable power spectrum analyses of stellar lightcurves for asteroseismology (e.g. \citealt{gilliland10}). In the white-noise model, the excess noise, with standard deviation $\s_x$, compounds the photometric noise, $\s_p$, in the data. The photometric noise is a combination of counting errors in the signal and background counts. Figure~\ref{fig:hist} illustrates the $\s_x$ parameter using the Prox Cen \siv\ data as an example, comparing Gaussians with $\s^2 = \left\langle\s_p^2\right\rangle$ and $\s^2 = \left\langle\s_p^2\right\rangle + \s_x^2$ to a histogram of the lightcurve points. We take care to use the average, $\left\langle\s_p^2\right\rangle$, because the photometric noise varies from point to point.

Several previous studies have estimated $\s_x$ (or the equivalent) by computing the sample standard deviation of the lightcurve, $\s$, and subtracting an estimate of $\s_p$ in quadrature (e.g. \citealt{gilliland11,jaffel13}). However, in the \citet{gilliland11} data the photometric errors are relatively constant, whereas in these data large variations in flux between points (sometimes factors of a few or more, as in Figure~\ref{fig:LCflare}) produce substantial variations in the photometric errors.

Therefore, rather than subtracting some representative value of $\s_p$ from $\s$ to estimate $\s_x$, we instead conducted a maximum likelihood analysis. For the analysis, we modeled each point in the high-pass filtered lightcurves as a random draw from its own Gaussian distribution. We let the distribution for point $i$ have variance $\s_i^2 = \s_x^2 + \s_{p,i}^2$ and mean equal to the quiescent lightcurve mean. Then we sampled the likelihood of the data for values of $\s_x$ ranging from zero to where the likelihood reached $10^{-5}$ times the maximum to generate a likelihood distribution for $\s_x$.

Intentionally, we do not utilize the continuum-subtracted data for this analysis. Both continuum and emission line photons will be indiscriminately absorbed by species in the atmosphere of a transiting exoplanet. As such, separating the relative contribution of each is irrelevant to this work; it is the variability of the two combined that will limit transit observations. With a few exceptions (most notably the F stars and HD103095) the continuum emission contributed $<10$\% to the stellar flux in the emission line bands.

Prior to the maximum likelihood analysis, we removed data points flagged as flaring (Section~\ref{sec:flaresweep}) to avoid contaminating our estimate of stochastic fluctuations. After removing the flares (for the reasons discussed in Section~\ref{sec:filter}), we high-pass filtered the remaining data. Filtering does not perfectly preserve the white noise in the data. Thus, the $\s_x$ values estimated before and after filtering would differ even in data exhibiting purely white noise and no periodic signals. Filtering thus introduces extra uncertainty in $\s_x$ because the effect of the filtering on white noise is not known (in some realizations of white noise filtering might increase scatter, while in others it might decrease it). We accounted for this uncertainty, to a reasonable approximation, through simulated white-noise data. This accounting process and the overall computation of maximum likelihoods are detailed in Appendix~\ref{sec:likelihood}.

\begin{figure}
\plotone{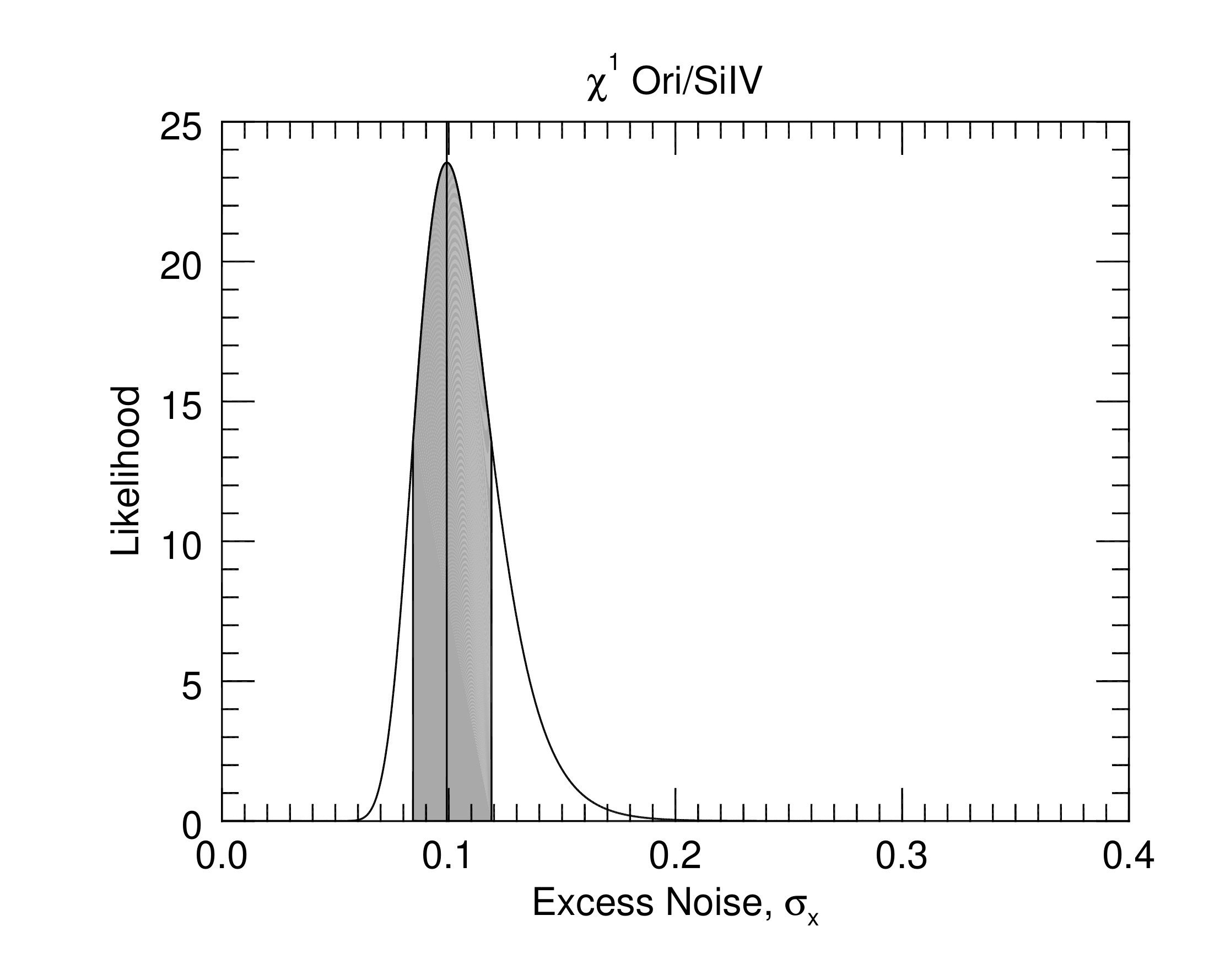}
\caption{Example likelihood distribution from Figure Set 9, available in the online at http://iopscience.iop.org/0067-0049/211/1/9/. The distribution shows a clear detection of excess noise for $\chi^1$ Ori in the \siv\ band. The 68.3\% confidence interval is shaded and the maximum marked by a vertical line.}
\label{fig:clearlike}
\end{figure}

\begin{figure}
\plotone{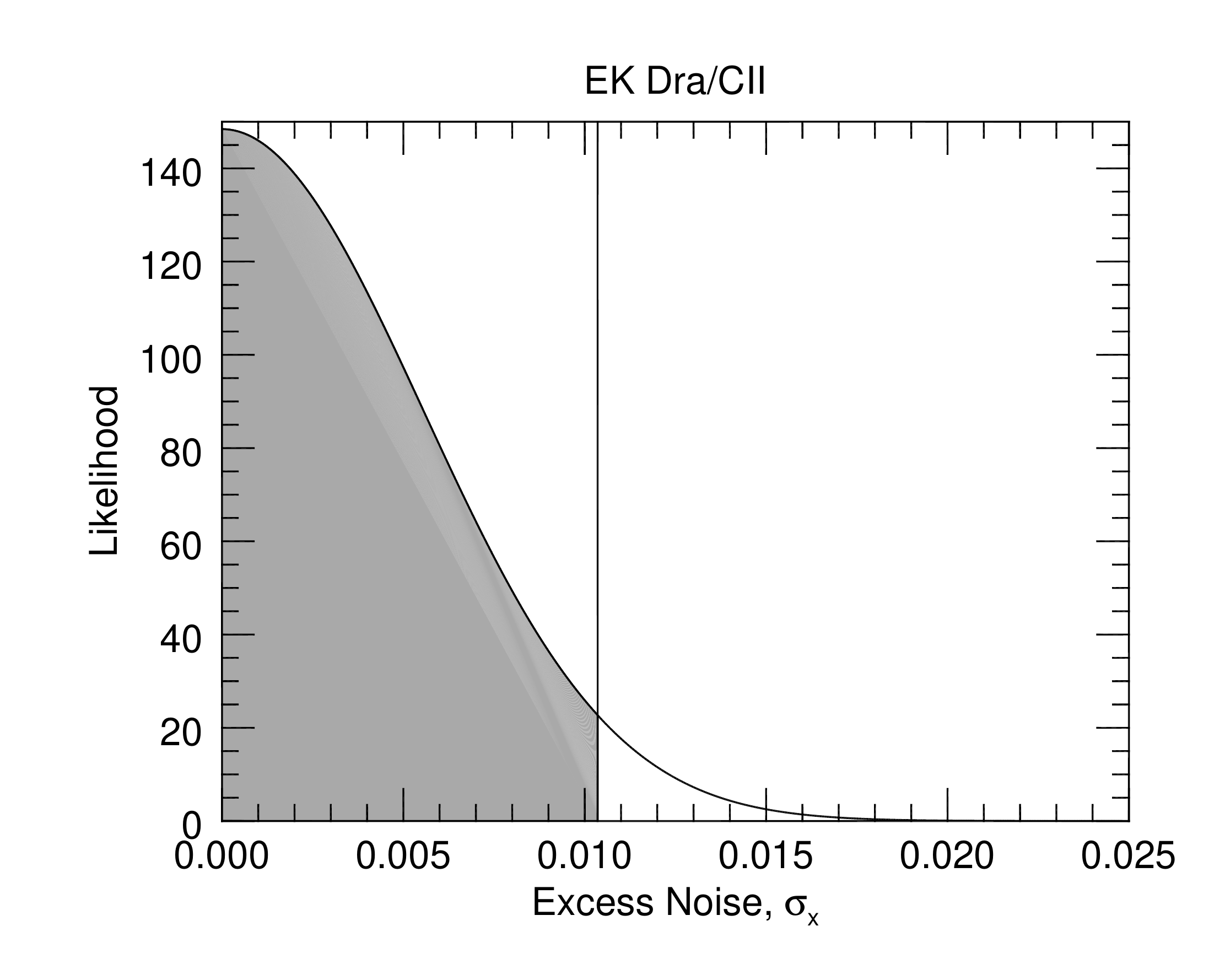}
\caption{Another example likelihood distribution from Figure Set 9, this time showing a non-detection of excess noise for EK Dra in the \cii\ band. The 95\% upper limit interval is shaded and the 95\% edge marked by a vertical line.}
\label{fig:limitlike}
\end{figure}

Once we specified a likelihood distribution for $\s_x$, we located the maximum and, as error bars, the equal-probability endpoints enclosing 68.3\% of the area under the curve. If the distribution was one sided, we instead set a 95\% upper limit on $\s_x$. Plots of the likelihood distribution of $\s_x$ for each lightcurve are available online as Figure Set 9.  Figure~\ref{fig:clearlike} is an example likelihood curve showing a clear detection of $\s_x$, and Figure~\ref{fig:limitlike} is an example likelihood curve that permits only an upper limit on $\s_x$.

\begin{samepage}
\subsubsection{Contamination in Excess Noise}
\label{sec:contamination}
The lightcurve scatter we quantified as excess noise could result from the star, another source in the aperture, or the instrument. To address additional sources, we searched the SIMBAD database for known objects within the view of the aperture. One star, HBC427 = V397 Aur, is a spectroscopic binary with angular separation 32.3$\pm$0.1 mas (as of 2008 December) and {\it Kp} band magnitude difference 0.87$\pm$0.01 \citep{kraus11}. For this system, the secondary likely contributes a significant portion of the flux in the observed spectral bands, as noted in Table~\ref{tbl:props}. Other targets with secondary, but insignificant, FUV sources within the instrument aperture are listed in Appendix~\ref{sec:doubles}.
\end{samepage}

Another non-stellar source is certain to be present in the aperture during each observation: geocoronal emission from \ion{H}{1} and \ion{O}{1}. This sky background is discussed in Section 7.4 of the {\it COS Instrument Handbook} \citep{cosinsthandbook}. To check for contamination from this source, we constructed lightcurves in a Ly$\alpha$ band (Table~\ref{tbl:bands}) for each star and visually compared the trends in these lightcurves to those of the other bands. We found no obvious contamination. Other non-instrumental sources of variability external to the star (e.g. planet phase changes, see Section~\ref{sec:varintro}) cannot be excluded.

Variability attributable to the instrument is a serious concern. Slow drifts in the instrument response are suppressed by high pass filtering. However, changes in the instrument configuration between exposures can alter its response on timescales too rapid for the effects to be filtered out.  Most notably, grid wires cast shadows of about 20\% depth over the COS detector at configuration-dependent locations (see Figure 5.10 of the COS Instrument Handbook, \citealt{cosinsthandbook}). Thus we separately normalize any exposures with different settings of the detector position before estimating $\s_x$. While this will exclude true changes in the stellar luminosity between such exposures from $\s_x$, it should strongly suppress variability resulting from different instrument configurations.

\section{Results}
\label{sec:results}

\subsection{Flares}
\label{sec:flareresults}
\begin{figure}
\plotone{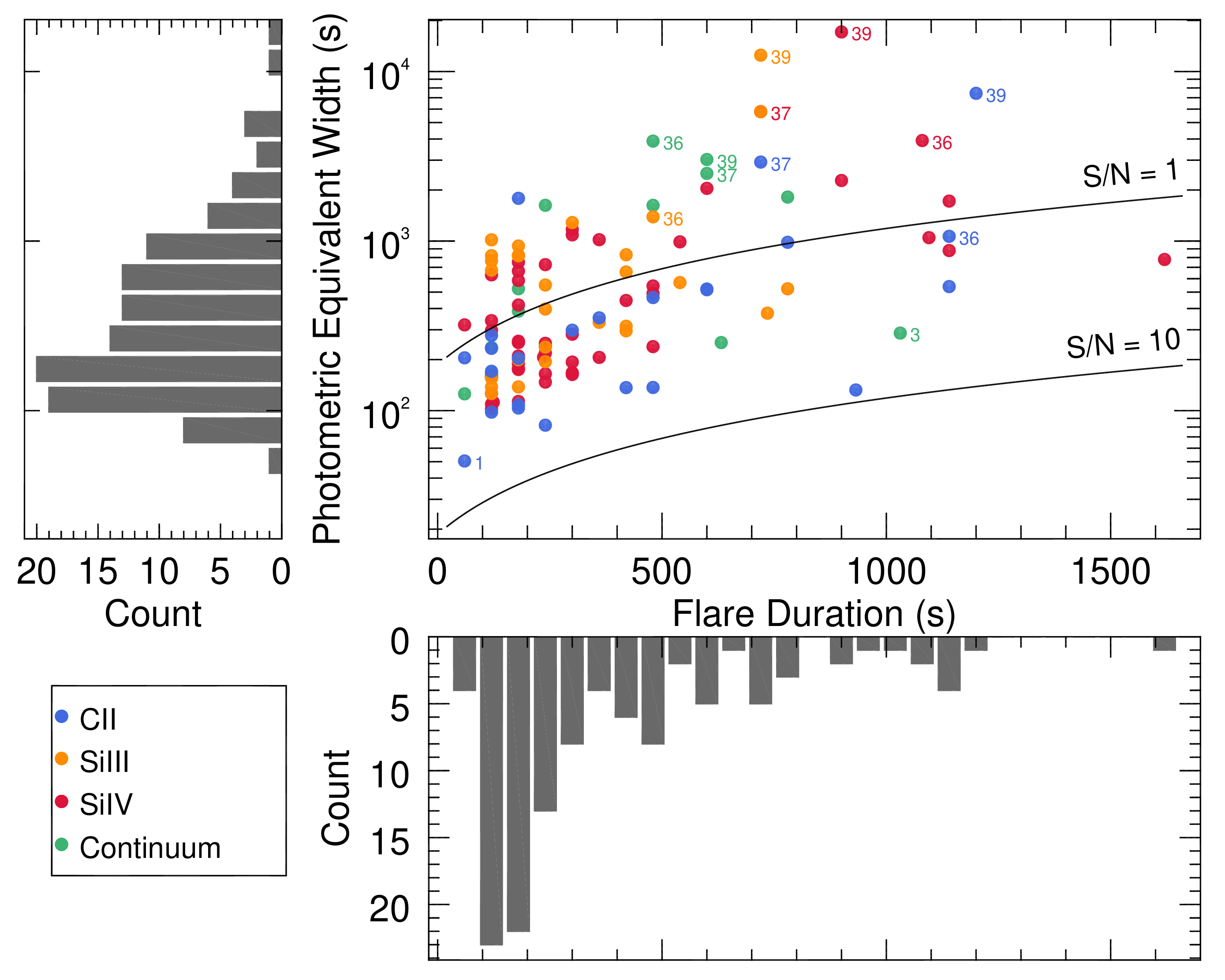}
\caption{The distribution of flares in duration and photometric equivalent width. See the text for definitions of these metrics and a discussion of expected biases. Solid black lines represent minimum detectable energies given a signal-to-noise of 1 or 10 in the quiescent 60~s lightcurve data. In this case alone, we use signal-to-noise to refer to all noise -- including stellar stochastic fluctuations -- rather than just photometric noise.}
\label{fig:flaredist}
\end{figure}

Both the number of flares identified in each star and the flare duty cycle (the fraction of lightcurve points flagged as flaring) are included in Tables~\ref{tbl:varcii}--\ref{tbl:varcont}. The lightcurve of each specific flare may be found within the applicable stellar lightcurves in Figure Set 4, available online. However, to avoid clutter, these lightcurves do not include the continuum-subtracted, filtered data used to compute flare properties. Figure~\ref{fig:flaredist} depicts the distribution of all detected flares in energy and duration (defined in the next paragraph). When flares were detected in multiple bands, the plot contains a separate point for the flare energy and duration as measured in each band. The plot also contains curves showing typical sensitivity limits for data with different ratios of mean signal to quiescent scatter. In the figure, some notable flares are labeled with the dataset number (from Table~\ref{tbl:obs}) of the star on which they occurred. These are discussed in Section~\ref{sec:flarebehave}. The figure omits two events flagged in all bands on FK Com, as these are likely a vestige of the strong stellar rotational signal that the high-pass filtering does not suppress below the level of quiescent scatter in that lightcurve (see Section~\ref{sec:flaresweep}). As such, they will be excluded from all further discussion.

We define duration as the time from the peak value to the first point below one sample standard deviation above the quiescent flux. As a metric of the total energy of a flare relative to the quiescent stellar emission, we use the photometric equivalent width (also termed equivalent duration), $\nfe$ \citep{gershberg72}. The photometric equivalent width is the integral of the mean-normalized, mean-subtracted flux over the flare duration and amounts to
\begin{equation}
\nfe = \int_{flare} \frac{F_f - \left \langle F_q \right \rangle}{\left \langle F_q \right \rangle}dt,
\end{equation}
for a star with continuously sampled flux $F_q$ during quiescent periods and flux $F_f$ during the flare. The result has units of time (we use s). By this definition, multiplying $\nfe$ by the quiescent stellar luminosity in the applicable band gives the absolute energy radiated by the flare. We computed these values from the continuum-subtracted, high-pass filtered data. Because filtering suppresses lower frequencies more (see Figure~\ref{fig:filteredkernel}), the energies of longer duration flares are systematically reduced. 

The flare sweeping algorithm flagged 116 flares (24 \cii, 28 \siii, 49 \siv, and 15 continuum), with median duration 6 min, median peak normalized flux 3.2, and median $\nfe$ 352 s. Of these flares, 62 were found in the AD Leo data, while 25 more were found in the Prox Cen data. Many of the 116 flares overlap with events flagged in other bands. Counting events that overlap as the same flare results in a tally of 58 separate events. The algorithm flagged data in only one band in 25 events, two bands in 17 events, three bands in 7 events, and all four bands in 9 events. The single band detections were dominated by events flagged in \siv\ data with 17, while \cii\ tallied 4, \siii\ tallied 2, and the continuum tallied 2. In the seven cases where a flare was detected in all bands but one, that missing detection was always the continuum.

The 9 flares detected in all four bands invite comparisons of the flare response of each band's flux. In 7 of these 9 the \siv\ band showed the greatest $\nfe$. Also in 7 of the 9, $\nfe$ was the lowest in the \cii\ data. Flare durations were mixed -- no band's response was typically longer or shorter than another. As for normalized peak flux, the \siv\ values were the highest in 6 of the 9 flares, while the continuum peaked the highest in the other three cases. The \cii\ data peaked lowest in 8 of the 9.

As expected, the histograms in Figure~\ref{fig:flaredist} show that the number of detected flares increases with decreasing energy and duration until the detection limits are approached. Figure~\ref{fig:flaredist} also shows a clear trend between the duration and radiated energy of a flare, confirmed to $>99.99$\% confidence by a Spearman Rank-Order test \citep{press02}. However, the trend could be explained by the sensitivity limits that increase with duration (example sensitivity curves are shown in the figure). For each lightcurve, the minimum $\nfe$ of a detectable flare depends on the scatter in the lightcurve. This minimum results when a single point occurs at 5.5 times the sample standard deviation of the quiescent lightcurve points with a subsequent point below the sample standard deviation. The minimum possible $\nfe$ value for each lightcurve is included in Tables~\ref{tbl:varcii} -~\ref{tbl:varcont}.

This work is the first to investigate flares as detected in the \ion{C}{2}, \ion{Si}{3}, and \ion{Si}{4} emission lines in more than a few stars. Even so, the sample of detected flares is too small to permit a detailed analysis of the distribution of events in energy and duration, or an analysis of the relationship between their frequency and stellar properties. Furthermore, weaker flares are not detectable in data where quiescent scatter is large. As the level of quiescent scatter relative to the mean varies by an order of magnitude or more  between lightcurves, flares detectable in some lightcurves are not detectable in others, biasing the population of low-energy flares. An additional bias results from gaps in the lightcurves. These restrict clusters of data points to shorter than an hour for most stars (see Table~\ref{tbl:obs}); even if a flare longer than a cluster were detected, the data provide no information on its true length. 

\subsection{Stochastic Fluctuations}
\label{sec:varresults}
The excess noise parameter used to quantify stochastic fluctuations represents the most probable standard deviation one would compute from 60~s cadence, mean-normalized flux data of the target in the absence of photometric noise. Tables~\ref{tbl:varcii} --~\ref{tbl:varcont} give the maximum-likelihood value or 95\% upper bound of $\s_x$ for all lightcurves. We detect excess noise in 19 \cii, 7 \siii, 17 \siv, and 6 FUV continuum lightcurves. In \siii\ fewer detections primarily result from lack of data for 15  stars. However, in the FUV continuum fewer detections could be a result of the lower integrated flux compared to \cii\ and \siv\ in about 2/3 of the datasets or lower excess noise. Quoted as range (median), the mean-normalized $\s_x$ estimates are 1-41\% (10\%) in \cii, 8-18\% (15\%) in \siii, 0.9-26\% (10\%) in \siv, and 1-22\% (5\%) in the FUV continuum. The remaining 101 lightcurves exhibited $\s_x$ below the sensitivity permitted by the quantity and photometric noise of the data (see Appendix~\ref{sec:sensitivity}). In these cases, we computed upper limits on $\s_x$, and many of these are below the typical values of the $\s_x$ detections. Some are under $\sim$1\%, making them valuable constraints on the stochastic fluctuations of the target stars.  

\begin{figure*}
\plottwo{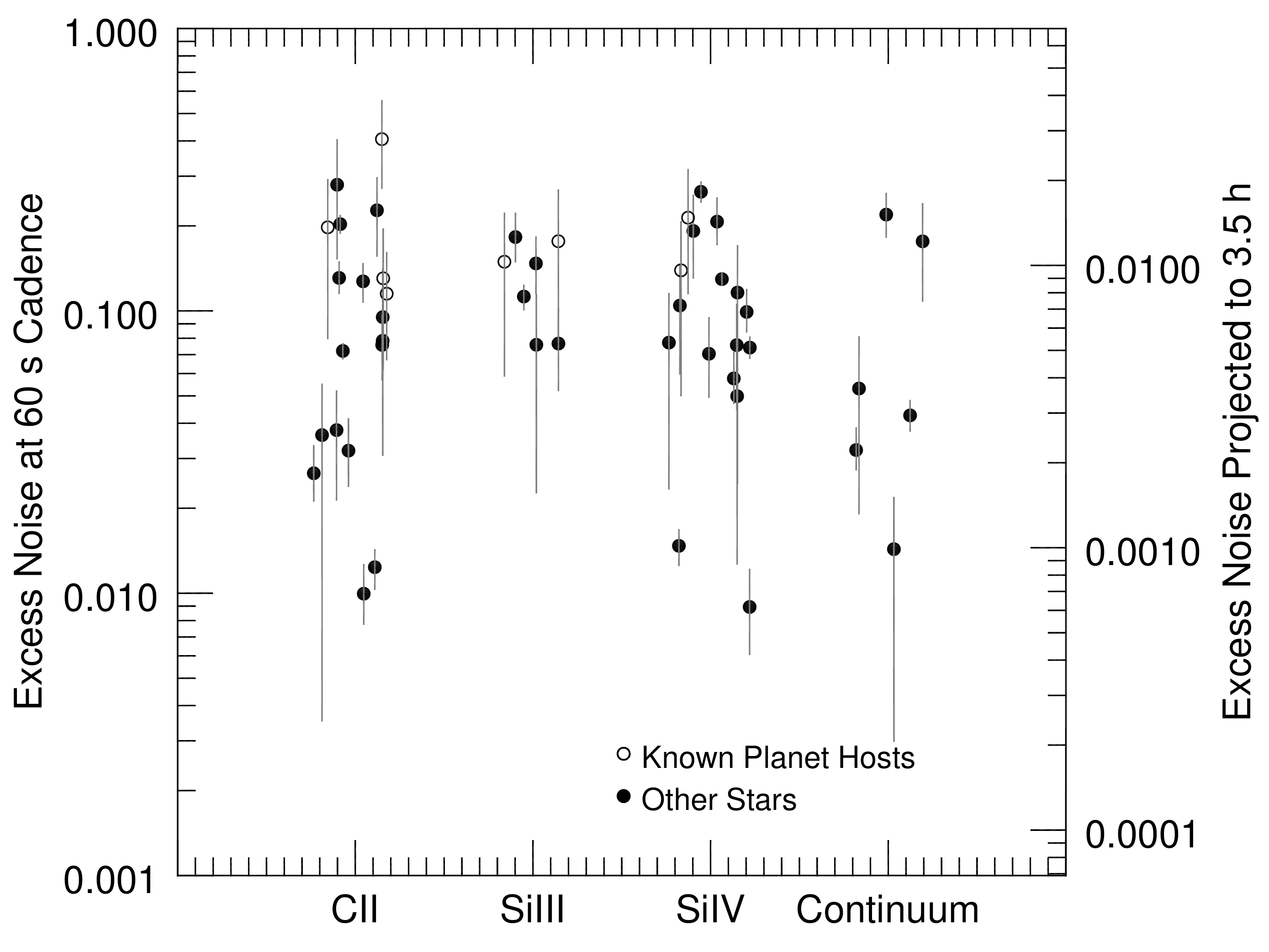}{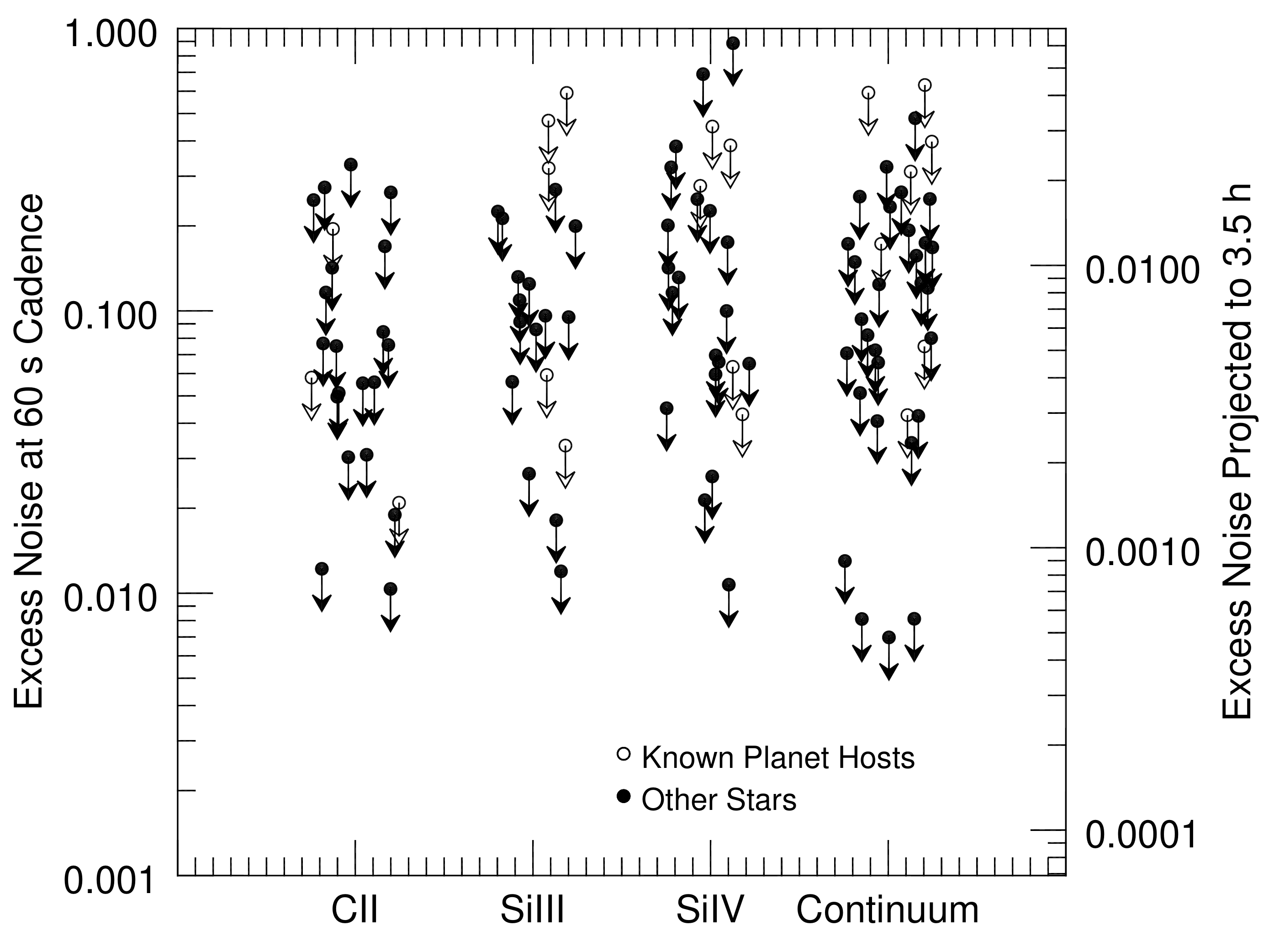}
\caption{Left: Reliable measurements of the excess noise, $\s_x$, at 60~s (this work) and projected to 3.5~h (typical transit duration). Gray bars give the 68.3\% confidence interval. Known planet hosts are GJ832, IL Aqr, and HO Lib in \cii; GJ832, IL Aqr, and GJ436 in \siii; and GJ832, IL Aqr, and HD209458 (E140M data) in \siv. Right: Upper limits on the excess noise for lightcurves where excess noise was not clearly detected for comparison to the detections (axes are identical). Known planet hosts are HD209458 (G130M and E140M data), HD189733, and GJ436 in \cii; HD209458 (G130M and E140M data), HD189733, and HO Lib in \siii; HD209458 (G130M data), HD189733, GJ436, and HO Lib in \siv; and all in the continuum. In both plots, random x-axis scatter was added for display.}
\label{fig:longvar}
\end{figure*}

Assuming that the stochastic fluctuations can be approximated as white noise , $\s_x$ will diminish with cadence length as $\dt^{-1/2}$. Without sufficient data for a detailed spectral analysis, this assumption is necessary. However, it is incorrect (see Section~\ref{sec:varintro}), and projecting as $\dt^{-1/2}$ will underestimate $\s_x$ at longer timescales. The severity of the underestimation will be dependent on the power spectrum of the star's stochastic fluctuations and will be different for different stars. Figures~\ref{fig:longvar} and~\ref{fig:longvar} illustrate the spread of mean-normalized excess noise estimates and upper limits projected by $\dt^{-1/2}$ to a timescale more meaningful to transit spectroscopy, the 3.5~h typical transit duration. These diagrams, in effect, illustrate the approximate error that would be associated with a 3.5~h integrated flux measurement due simply to stochastic fluctuations of the host star's emission lines. In actual data, these noise values would be compounded by additional photometric noise from photon statistics and instrumental sources.

For many lightcurves, the estimates of $\s_x$ at $\dt=$ 3.5~h are subject to one or more shortcomings not inherently evident from the information in Tables~\ref{tbl:varcii} --~\ref{tbl:varcont}. To begin with, as just mentioned above, $\s_x$ is unlikely to diminish by $\dt^{-1/2}$ for many, possibly all, stars in the sample. Additionally, most lightcurves do not contain any long enough blocks of closely spaced points to sample fluctuations over at least one full transit timescale (Table~\ref{tbl:obs}). Finally, lightcurves exhibiting low mean flux (generally $\lesssim5\sn{-16}$ erg s$^{-1}$ cm$^{-2}$) contain many time bins with zero counts after background subtraction. We model bins with even a single count as random draws from a Gaussian (not Poisson) distribution, though the subtraction of Poisson distributed background counts from Poisson distributed signal + background makes the distribution of the result very nearly Gaussian. To allow readers to form their own judgments of the quality of the excess noise measure for any given star, we have made all lightcurves and the associated $\s_x$ likelihood distributions available as Figure Sets 4 and 9.

A case of particular constancy in the sample is $\beta$ Cas, with estimated $\s_x$ values of under 1\% in the \cii\ and \siv\ bands, and about 3\% in the continuum band. Yet $\beta$ Cas is classified as a $\delta$ Scuti variable. Its pulsations have a period of $\sim$0.1 d \citep{riboni94}, so are not strongly suppressed by the high-pass filtering. However, the dataset for $\beta$ Cas spans 1419 s, only about 16\% of the total pulsation period. Were these sub-transit timescale pulsations observed in full, the value of $\s_x$ for $\beta$ Cas would likely be substantially higher.

\begin{deluxetable*}{lrrcrrr}
\tabletypesize{\scriptsize}
\tablecaption{Variability Statistics in \ion{C}{2} \label{tbl:varcii}}
\tablewidth{0pt}
\tablecolumns{7}
\tablehead{\colhead{Star} & \colhead{$\mf$} & \colhead{$\s_x$\tablenotemark{a}} & \colhead{$R_{\sigma_x}\tablenotemark{b}$} & \colhead{$N_{f}\tablenotemark{c}$} & \colhead{Duty\tablenotemark{d}} & \colhead{$\nfe\tablenotemark{e}$}\\
& \colhead{$10^{-17}$ erg s$^{-1}$ cm$^{-2}$} &  & \colhead{$R_J$} &  & \colhead{Cycle} & \colhead{s}}
\startdata
$\beta$ Cas\tablenotemark{f} & $249700\phd\phn\pm 4000\phd\phn$ & $\phm{<}\ 0.0100^{+0.0027}_{-0.0022}$ & $\phm{<}\ \phn0.88\phn\phn^{+0.12\phn\phn}_{-0.10\phn\phn\phn}\tablenotemark{g}$ & $\phn0$ & $0.000$ & $\phn\phn3.03$\\
$\delta$ Cep & $\phn\phn\phn280\phd\phn\pm \phn\phn53\phd\phn$ & $\phm{<}\ 0.038\phn^{+0.014\phn}_{-0.016\phn}$ & $\phm{<}\ 22.4\phn\phn\phn^{+5.9\phn\phn\phn}_{-6.4\phn\phn\phn\phn}\phm{\tablenotemark{a}}$ & $\phn1$ & $0.054$ & $\phn48.7\phn$\\
$\alpha$ Per & $\phn\phn1680\phd\phn\pm \phn300\phd\phn$ & $\phm{<}\ 0.036\phn^{+0.019\phn}_{-0.033\phn}$ & $\phm{<}\ 27\phd\phn\phn\phn\phn^{+8\phd\phn\phn\phn\phn}_{-13\phd\phn\phn\phn\phn}\phm{\tablenotemark{a}}$ & $\phn0$ & $0.000$ & $\phn22.7\phn$\\
$\beta$ Dor & $\phn\phn3880\phd\phn\pm \phn230\phd\phn$ & $<0.0122^{\phm{+}\phn\phd\phn\phn\phn\phn}_{\phm{-}\phn\phd\phn\phn\phn\phn}$ & $<20.1\phn\phn\phn^{\phm{+}\phn\phd\phn\phn\phn\phn}_{\phm{-}\phn\phn\phd\phn\phn\phn\phn}\phm{\tablenotemark{a}}$ & $\phn0$ & $0.000$ & $\phn14.0\phn$\\
Polaris & $\phn\phn2520\phd\phn\pm \phn190\phd\phn$ & $<0.0190^{\phm{+}\phn\phd\phn\phn\phn\phn}_{\phm{-}\phn\phd\phn\phn\phn\phn}$ & $<\phn7.04\phn\phn^{\phm{+}\phn\phd\phn\phn\phn\phn}_{\phm{-}\phn\phn\phd\phn\phn\phn\phn}\phm{\tablenotemark{a}}$ & $\phn0$ & $0.000$ & $\phn18.0\phn$\\
HD25825 & $\phn\phn1950\phd\phn\pm \phn350\phd\phn$ & $<0.0757^{\phm{+}\phn\phd\phn\phn\phn\phn}_{\phm{-}\phn\phd\phn\phn\phn\phn}$ & $<\phn0.767\phn^{\phm{+}\phn\phd\phn\phn\phn\phn}_{\phm{-}\phn\phn\phd\phn\phn\phn\phn}\phm{\tablenotemark{a}}$ & $\phn0$ & $0.000$ & $\phn18.6\phn$\\
HD209458/G130M & $\phn\phn\phn234\phd\phn\pm \phn\phn27\phd\phn$ & $<0.0580^{\phm{+}\phn\phd\phn\phn\phn\phn}_{\phm{-}\phn\phd\phn\phn\phn\phn}$ & $<\phn0.711\phn^{\phm{+}\phn\phd\phn\phn\phn\phn}_{\phm{-}\phn\phn\phd\phn\phn\phn\phn}\phm{\tablenotemark{a}}$ & $\phn0$ & $0.000$ & $\phn59.2\phn$\\
HD209458/E140M & $\phn\phn\phn\phn40\phd\phn\pm \phn\phn52\phd\phn$ & $<0.195\phn^{\phm{+}\phn\phd\phn\phn\phn\phn}_{\phm{-}\phn\phd\phn\phn\phn\phn}$ & $<\phn1.30\phn\phn^{\phm{+}\phn\phd\phn\phn\phn\phn}_{\phm{-}\phn\phn\phd\phn\phn\phn\phn}\phm{\tablenotemark{a}}$ & $\phn0$ & $0.000$ & $255\phd\phn\phn$\\
$\chi^1$ Ori & $\phn45800\phd\phn\pm 1700\phd\phn$ & $\phm{<}\ 0.0266^{+0.0067}_{-0.0054}$ & $\phm{<}\ \phn0.437\phn^{+0.056\phn}_{-0.045\phn\phn}\phm{\tablenotemark{a}}$ & $\phn0$ & $0.000$ & $\phn\phn7.82$\\
HII314 & $\phn\phn\phn830\phd\phn\pm \phn130\phd\phn$ & $<0.0559^{\phm{+}\phn\phd\phn\phn\phn\phn}_{\phm{-}\phn\phd\phn\phn\phn\phn}$ & $<\phn0.381\phn^{\phm{+}\phn\phd\phn\phn\phn\phn}_{\phm{-}\phn\phn\phd\phn\phn\phn\phn}\phm{\tablenotemark{a}}$ & $\phn0$ & $0.000$ & $\phn30.5\phn$\\
EK Dra & $\phn11630\phd\phn\pm \phn420\phd\phn$ & $<0.0104^{\phm{+}\phn\phd\phn\phn\phn\phn}_{\phm{-}\phn\phd\phn\phn\phn\phn}$ & $<\phn0.252\phn^{\phm{+}\phn\phd\phn\phn\phn\phn}_{\phm{-}\phn\phn\phd\phn\phn\phn\phn}\phm{\tablenotemark{a}}$ & $\phn1$ & $0.100$ & $\phn11.6\phn$\\
$\pi^1$ UMa & $\phn17300\phd\phn\pm 1000\phd\phn$ & $\phm{<}\ 0.0320^{+0.0095}_{-0.0080}$ & $\phm{<}\ \phn0.432\phn^{+0.065\phn}_{-0.055\phn\phn}\phm{\tablenotemark{a}}$ & $\phn0$ & $0.000$ & $\phn10.3\phn$\\
HD90508 & $\phn\phn\phn304\phd\phn\pm \phn\phn28\phd\phn$ & $<0.116\phn^{\phm{+}\phn\phd\phn\phn\phn\phn}_{\phm{-}\phn\phd\phn\phn\phn\phn}$ & $<\phn0.592\phn^{\phm{+}\phn\phd\phn\phn\phn\phn}_{\phm{-}\phn\phn\phd\phn\phn\phn\phn}\phm{\tablenotemark{a}}$ & $\phn0$ & $0.000$ & $\phn68.0\phn$\\
HD199288/G140L & $\phn\phn\phn210\phd\phn\pm \phn\phn23\phd\phn$ & $<0.0842^{\phm{+}\phn\phd\phn\phn\phn\phn}_{\phm{-}\phn\phd\phn\phn\phn\phn}$ & $<\phn0.719\phn^{\phm{+}\phn\phd\phn\phn\phn\phn}_{\phm{-}\phn\phn\phd\phn\phn\phn\phn}\phm{\tablenotemark{a}}$ & $\phn0$ & $0.000$ & $\phn65.2\phn$\\
HD199288/G130M & $\phn\phn\phn350\phd\phn\pm \phn\phn62\phd\phn$ & $<0.0554^{\phm{+}\phn\phd\phn\phn\phn\phn}_{\phm{-}\phn\phd\phn\phn\phn\phn}$ & $<\phn0.583\phn^{\phm{+}\phn\phd\phn\phn\phn\phn}_{\phm{-}\phn\phn\phd\phn\phn\phn\phn}\phm{\tablenotemark{a}}$ & $\phn0$ & $0.000$ & $\phn45.9\phn$\\
18 Sco/G140L & $\phn\phn1600\phd\phn\pm \phn120\phd\phn$ & $<0.0751^{\phm{+}\phn\phd\phn\phn\phn\phn}_{\phm{-}\phn\phd\phn\phn\phn\phn}$ & $<\phn0.728\phn^{\phm{+}\phn\phd\phn\phn\phn\phn}_{\phm{-}\phn\phn\phd\phn\phn\phn\phn}\phm{\tablenotemark{a}}$ & $\phn0$ & $0.000$ & $\phn25.4\phn$\\
18 Sco/G130M & $\phn\phn2010\phd\phn\pm \phn230\phd\phn$ & $<0.0496^{\phm{+}\phn\phd\phn\phn\phn\phn}_{\phm{-}\phn\phd\phn\phn\phn\phn}$ & $<\phn0.592\phn^{\phm{+}\phn\phd\phn\phn\phn\phn}_{\phm{-}\phn\phn\phd\phn\phn\phn\phn}\phm{\tablenotemark{a}}$ & $\phn0$ & $0.000$ & $\phn20.4\phn$\\
FK Com\tablenotemark{f} & $\phn18820\phd\phn\pm \phn240\phd\phn$ & $\phm{<}\ 0.0124^{+0.0019}_{-0.0020}$ & $\phm{<}\ \phn2.42\phn\phn^{+0.18\phn\phn}_{-0.20\phn\phn\phn}\tablenotemark{g}$ & $\phn2$ & $0.110$ & $\phn\phn9.26$\\
HD65583 & $\phn\phn\phn\phn33.7\pm \phn\phn\phn8.4$ & $\phm{<}\ 0.227\phn^{+0.069\phn}_{-0.071\phn}$ & $\phm{<}\ \phn0.94\phn\phn^{+0.14\phn\phn}_{-0.15\phn\phn\phn}\phm{\tablenotemark{a}}$ & $\phn0$ & $0.000$ & $123\phd\phn\phn$\\
HD103095/G140L & $\phn\phn\phn\phn33\phd\phn\pm \phn\phn14\phd\phn$ & $\phm{<}\ 0.28\phn\phn^{+0.12\phn\phn}_{-0.13\phn\phn}$ & $\phm{<}\ \phn0.89\phn\phn^{+0.20\phn\phn}_{-0.20\phn\phn\phn}\phm{\tablenotemark{a}}$ & $\phn0$ & $0.000$ & $138\phd\phn\phn$\\
HD103095/G130M & $\phn\phn\phn\phn48\phd\phn\pm \phn\phn32\phd\phn$ & $<0.247\phn^{\phm{+}\phn\phd\phn\phn\phn\phn}_{\phm{-}\phn\phd\phn\phn\phn\phn}$ & $<\phn0.838\phn^{\phm{+}\phn\phd\phn\phn\phn\phn}_{\phm{-}\phn\phn\phd\phn\phn\phn\phn}\phm{\tablenotemark{a}}$ & $\phn0$ & $0.000$ & $107\phd\phn\phn$\\
HD282630 & $\phn\phn1140\phd\phn\pm \phn200\phd\phn$ & $<0.0767^{\phm{+}\phn\phd\phn\phn\phn\phn}_{\phm{-}\phn\phd\phn\phn\phn\phn}$ & \nodata & $\phn0$ & $0.000$ & $\phn29.9\phn$\\
HD189733 & $\phn\phn2620\phd\phn\pm \phn200\phd\phn$ & $<0.0209^{\phm{+}\phn\phd\phn\phn\phn\phn}_{\phm{-}\phn\phd\phn\phn\phn\phn}$ & $<\phn0.280\phn^{\phm{+}\phn\phd\phn\phn\phn\phn}_{\phm{-}\phn\phn\phd\phn\phn\phn\phn}\phm{\tablenotemark{a}}$ & $\phn0$ & $0.000$ & $\phn15.9\phn$\\
HD145417 & $\phn\phn\phn\phn\phn7.2\pm \phn\phn\phn3.9$ & $<0.263\phn^{\phm{+}\phn\phd\phn\phn\phn\phn}_{\phm{-}\phn\phd\phn\phn\phn\phn}$ & $<\phn0.801\phn^{\phm{+}\phn\phd\phn\phn\phn\phn}_{\phm{-}\phn\phn\phd\phn\phn\phn\phn}\phm{\tablenotemark{a}}$ & $\phn0$ & $0.000$ & $226\phd\phn\phn$\\
V410-$\tau$ & $\phn\phn\phn420\phd\phn\pm \phn180\phd\phn$ & $<0.169\phn^{\phm{+}\phn\phd\phn\phn\phn\phn}_{\phm{-}\phn\phd\phn\phn\phn\phn}$ & $<\phn2.84\phn\phn^{\phm{+}\phn\phd\phn\phn\phn\phn}_{\phm{-}\phn\phn\phd\phn\phn\phn\phn}\phm{\tablenotemark{a}}$ & $\phn0$ & $0.000$ & $141\phd\phn\phn$\\
EG Cha & $\phn\phn2680\phd\phn\pm \phn250\phd\phn$ & $\phm{<}\ 0.131\phn^{+0.018\phn}_{-0.015\phn}$ & $\phm{<}\ \phn1.66\phn\phn^{+0.12\phn\phn}_{-0.10\phn\phn\phn}\tablenotemark{g}$ & $\phn0$ & $0.000$ & $\phn45.0\phn$\\
HBC427 & $\phn\phn\phn370\phd\phn\pm \phn110\phd\phn$ & $<0.142\phn^{\phm{+}\phn\phd\phn\phn\phn\phn}_{\phm{-}\phn\phd\phn\phn\phn\phn}$ & $<\phn1.77\phn\phn^{\phm{+}\phn\phd\phn\phn\phn\phn}_{\phm{-}\phn\phn\phd\phn\phn\phn\phn}\phm{\tablenotemark{a}}$ & $\phn0$ & $0.000$ & $\phn54.1\phn$\\
61 Cyg A & $\phn10350\phd\phn\pm \phn790\phd\phn$ & $<0.0309^{\phm{+}\phn\phd\phn\phn\phn\phn}_{\phm{-}\phn\phd\phn\phn\phn\phn}$ & $<\phn0.122\phn^{\phm{+}\phn\phd\phn\phn\phn\phn}_{\phm{-}\phn\phn\phd\phn\phn\phn\phn}\phm{\tablenotemark{a}}$ & $\phn0$ & $0.000$ & $\phn\phn8.22$\\
LkCa 4 & $\phn\phn\phn300\phd\phn\pm \phn100\phd\phn$ & $\phm{<}\ 0.095\phn^{+0.045\phn}_{-0.064\phn}$ & $\phm{<}\ \phn1.47\phn\phn^{+0.35\phn\phn}_{-0.50\phn\phn\phn}\tablenotemark{g}$ & $\phn0$ & $0.000$ & $\phn55.2\phn$\\
GJ832 & $\phn\phn\phn420\phd\phn\pm \phn120\phd\phn$ & $\phm{<}\ 0.115\phn^{+0.046\phn}_{-0.048\phn}$ & $\phm{<}\ \phn0.433\phn^{+0.087\phn}_{-0.091\phn\phn}\phm{\tablenotemark{a}}$ & $\phn0$ & $0.000$ & $\phn56.9\phn$\\
TWA13B & $\phn\phn1580\phd\phn\pm \phn220\phd\phn$ & $\phm{<}\ 0.078\phn^{+0.021\phn}_{-0.019\phn}$ & $\phm{<}\ \phn0.715\phn^{+0.095\phn}_{-0.089\phn\phn}\tablenotemark{g}$ & $\phn0$ & $0.000$ & $\phn30.3\phn$\\
TWA13A & $\phn\phn1660\phd\phn\pm \phn250\phd\phn$ & $<0.0512^{\phm{+}\phn\phd\phn\phn\phn\phn}_{\phm{-}\phn\phd\phn\phn\phn\phn}$ & $<\phn0.636\phn^{\phm{+}\phn\phd\phn\phn\phn\phn}_{\phm{-}\phn\phn\phd\phn\phn\phn\phn}\phm{\tablenotemark{a}}$ & $\phn0$ & $0.000$ & $\phn23.6\phn$\\
AU Mic & $\phn26000\phd\phn\pm 1200\phd\phn$ & $<0.0303^{\phm{+}\phn\phd\phn\phn\phn\phn}_{\phm{-}\phn\phd\phn\phn\phn\phn}$ & $<\phn0.373\phn^{\phm{+}\phn\phd\phn\phn\phn\phn}_{\phm{-}\phn\phn\phd\phn\phn\phn\phn}\phm{\tablenotemark{a}}$ & $\phn0$ & $0.000$ & $\phn28.5\phn$\\
CE Ant & $\phn\phn3020\phd\phn\pm \phn380\phd\phn$ & $\phm{<}\ 0.076\phn^{+0.021\phn}_{-0.019\phn}$ & $\phm{<}\ \phn0.703\phn^{+0.095\phn}_{-0.087\phn\phn}\tablenotemark{g}$ & $\phn0$ & $0.000$ & $\phn27.9\phn$\\
GJ436 & $\phn\phn\phn\phn11\phd\phn\pm \phn\phn13\phd\phn$ & $\phm{<}\ 0.20\phn\phn^{+0.09\phn\phn}_{-0.12\phn\phn}$ & $\phm{<}\ \phn0.50\phn\phn^{+0.12\phn\phn}_{-0.15\phn\phn\phn}\phm{\tablenotemark{a}}$ & $\phn1$ & $0.073$ & $132\phd\phn\phn$\\
EV Lac & $\phn\phn6220\phd\phn\pm \phn570\phd\phn$ & $\phm{<}\ 0.127\phn^{+0.020\phn}_{-0.020\phn}$ & $\phm{<}\ \phn0.337\phn^{+0.026\phn}_{-0.026\phn\phn}\tablenotemark{g}$ & $\phn0$ & $0.000$ & $\phn75.0\phn$\\
AD Leo & $\phn25530\phd\phn\pm \phn490\phd\phn$ & $\phm{<}\ 0.0721^{+0.0044}_{-0.0045}$ & $\phm{<}\ \phn0.3111^{+0.0094}_{-0.0097\phn}\tablenotemark{g}$ & $11$ & $0.150$ & $\phn35.1\phn$\\
IL Aqr & $\phn\phn\phn270\phd\phn\pm \phn140\phd\phn$ & $\phm{<}\ 0.130\phn^{+0.064\phn}_{-0.068\phn}$ & $\phm{<}\ \phn0.345\phn^{+0.085\phn}_{-0.090\phn\phn}\tablenotemark{g}$ & $\phn1$ & $0.375$ & $\phn53.3\phn$\\
HO Lib & $\phn\phn\phn\phn\phn7\phd\phn\pm \phn\phn14\phd\phn$ & $\phm{<}\ 0.41\phn\phn^{+0.15\phn\phn}_{-0.13\phn\phn}$ & $\phm{<}\ \phn0.487\phn^{+0.091\phn}_{-0.082\phn\phn}\phm{\tablenotemark{a}}$ & $\phn0$ & $0.000$ & $159\phd\phn\phn$\\
Prox Cen & $\phn\phn2910\phd\phn\pm \phn180\phd\phn$ & $\phm{<}\ 0.203\phn^{+0.015\phn}_{-0.014\phn}$ & $\phm{<}\ \phn0.1623^{+0.0099}_{-0.0099\phn}\phm{\tablenotemark{a}}$ & $\phn8$ & $0.176$ & $101\phd\phn\phn$\\
GJ3877 & $\phn\phn\phn\phn\phn6.9\pm \phn\phn\phn4.0$ & $<0.274\phn^{\phm{+}\phn\phd\phn\phn\phn\phn}_{\phm{-}\phn\phd\phn\phn\phn\phn}$ & $<\phn0.160\phn^{\phm{+}\phn\phd\phn\phn\phn\phn}_{\phm{-}\phn\phn\phd\phn\phn\phn\phn}\phm{\tablenotemark{a}}$ & $\phn0$ & $0.000$ & $234\phd\phn\phn$\\
GJ3517 & $\phn\phn\phn\phn\phn2.7\pm \phn\phn\phn2.2$ & $<0.330\phn^{\phm{+}\phn\phd\phn\phn\phn\phn}_{\phm{-}\phn\phd\phn\phn\phn\phn}$ & $<\phn0.148\phn^{\phm{+}\phn\phd\phn\phn\phn\phn}_{\phm{-}\phn\phn\phd\phn\phn\phn\phn}\phm{\tablenotemark{a}}$ & $\phn1$ & $0.030$ & $478\phd\phn\phn$\\
\enddata

\statstablefootnotes
\end{deluxetable*}

\begin{deluxetable*}{lrrcrrr}
\tabletypesize{\scriptsize}
\tablecaption{Variability Statistics in \ion{Si}{3} \label{tbl:varsiii}}
\tablewidth{0pt}
\tablecolumns{7}
\tablehead{\colhead{Star} & \colhead{$\mf$} & \colhead{$\s_x$\tablenotemark{a}} & \colhead{$R_{\sigma_x}\tablenotemark{b}$} & \colhead{$N_{f}\tablenotemark{c}$} & \colhead{Duty\tablenotemark{d}} & \colhead{$\nfe\tablenotemark{e}$}\\
& \colhead{$10^{-17}$ erg s$^{-1}$ cm$^{-2}$} &  & \colhead{$R_J$} &  & \colhead{Cycle} & \colhead{s}}
\startdata
$\beta$ Cas\tablenotemark{f} & \nodata & \nodata & \nodata & \nodata & \nodata & \nodata\\
$\delta$ Cep & $\phn\phn312\phd\phn\pm \phn\phn57\phd\phn$ & $<0.0265^{\phm{+}\phn\phd\phn\phn\phn}_{\phm{-}\phn\phd\phn\phn\phn}$ & $<18.7\phn\phn^{\phm{+}\phn\phd\phn\phn\phn}_{\phm{-}\phn\phd\phn\phn\phn}\phm{\tablenotemark{a}}$ & $\phn0$ & $0.000$ & $\phn41.5$\\
$\alpha$ Per & \nodata & \nodata & \nodata & \nodata & \nodata & \nodata\\
$\beta$ Dor & $\phn4940\phd\phn\pm \phn260\phd\phn$ & $<0.0120^{\phm{+}\phn\phd\phn\phn\phn}_{\phm{-}\phn\phd\phn\phn\phn}$ & $<19.9\phn\phn^{\phm{+}\phn\phd\phn\phn\phn}_{\phm{-}\phn\phd\phn\phn\phn}\phm{\tablenotemark{a}}$ & $\phn0$ & $0.000$ & $\phn11.9$\\
Polaris & $\phn2720\phd\phn\pm \phn200\phd\phn$ & $<0.0181^{\phm{+}\phn\phd\phn\phn\phn}_{\phm{-}\phn\phd\phn\phn\phn}$ & $<\phn6.88\phn^{\phm{+}\phn\phd\phn\phn\phn}_{\phm{-}\phn\phd\phn\phn\phn}\phm{\tablenotemark{a}}$ & $\phn0$ & $0.000$ & $\phn16.3$\\
HD25825 & \nodata & \nodata & \nodata & \nodata & \nodata & \nodata\\
HD209458/G130M & $\phn\phn189\phd\phn\pm \phn\phn24\phd\phn$ & $<0.0593^{\phm{+}\phn\phd\phn\phn\phn}_{\phm{-}\phn\phd\phn\phn\phn}$ & $<\phn0.718^{\phm{+}\phn\phd\phn\phn\phn}_{\phm{-}\phn\phd\phn\phn\phn}\phm{\tablenotemark{a}}$ & $\phn0$ & $0.000$ & $\phn64.2$\\
HD209458/E140M & $\phn\phn\phn13\phd\phn\pm \phn\phn57\phd\phn$ & $<0.320\phn^{\phm{+}\phn\phd\phn\phn\phn}_{\phm{-}\phn\phd\phn\phn\phn}$ & $<\phn1.67\phn^{\phm{+}\phn\phd\phn\phn\phn}_{\phm{-}\phn\phd\phn\phn\phn}\phm{\tablenotemark{a}}$ & $\phn0$ & $0.000$ & $425\phd\phn$\\
$\chi^1$ Ori & \nodata & \nodata & \nodata & \nodata & \nodata & \nodata\\
HII314 & $\phn\phn510\phd\phn\pm \phn110\phd\phn$ & $<0.0961^{\phm{+}\phn\phd\phn\phn\phn}_{\phm{-}\phn\phd\phn\phn\phn}$ & $<\phn0.499^{\phm{+}\phn\phd\phn\phn\phn}_{\phm{-}\phn\phd\phn\phn\phn}\phm{\tablenotemark{a}}$ & $\phn0$ & $0.000$ & $\phn48.1$\\
EK Dra & \nodata & \nodata & \nodata & \nodata & \nodata & \nodata\\
$\pi^1$ UMa & \nodata & \nodata & \nodata & \nodata & \nodata & \nodata\\
HD90508 & \nodata & \nodata & \nodata & \nodata & \nodata & \nodata\\
HD199288/G140L & \nodata & \nodata & \nodata & \nodata & \nodata & \nodata\\
HD199288/G130M & $\phn\phn197\phd\phn\pm \phn\phn47\phd\phn$ & $<0.0950^{\phm{+}\phn\phd\phn\phn\phn}_{\phm{-}\phn\phd\phn\phn\phn}$ & $<\phn0.763^{\phm{+}\phn\phd\phn\phn\phn}_{\phm{-}\phn\phd\phn\phn\phn}\phm{\tablenotemark{a}}$ & $\phn1$ & $0.052$ & $\phn59.7$\\
18 Sco/G140L & $\phn\phn825\phd\phn\pm \phn\phn86\phd\phn$ & $\phm{<}\ 0.076\phn^{+0.038}_{-0.053}$ & $\phm{<}\ \phn0.73\phn^{+0.19\phn}_{-0.26\phn}\phm{\tablenotemark{a}}$ & $\phn0$ & $0.000$ & $\phn42.1$\\
18 Sco/G130M & $\phn1250\phd\phn\pm \phn180\phd\phn$ & $<0.0561^{\phm{+}\phn\phd\phn\phn\phn}_{\phm{-}\phn\phd\phn\phn\phn}$ & $<\phn0.629^{\phm{+}\phn\phd\phn\phn\phn}_{\phm{-}\phn\phd\phn\phn\phn}\phm{\tablenotemark{a}}$ & $\phn0$ & $0.000$ & $\phn23.9$\\
FK Com\tablenotemark{f} & \nodata & \nodata & \nodata & \nodata & \nodata & \nodata\\
HD65583 & \nodata & \nodata & \nodata & \nodata & \nodata & \nodata\\
HD103095/G140L & \nodata & \nodata & \nodata & \nodata & \nodata & \nodata\\
HD103095/G130M & $\phn\phn113\phd\phn\pm \phn\phn51\phd\phn$ & $<0.200\phn^{\phm{+}\phn\phd\phn\phn\phn}_{\phm{-}\phn\phd\phn\phn\phn}$ & $<\phn0.754^{\phm{+}\phn\phd\phn\phn\phn}_{\phm{-}\phn\phd\phn\phn\phn}\phm{\tablenotemark{a}}$ & $\phn0$ & $0.000$ & $\phn75.9$\\
HD282630 & $\phn\phn540\phd\phn\pm \phn140\phd\phn$ & $<0.125\phn^{\phm{+}\phn\phd\phn\phn\phn}_{\phm{-}\phn\phd\phn\phn\phn}$ & \nodata & $\phn0$ & $0.000$ & $\phn45.3$\\
HD189733 & $\phn1130\phd\phn\pm \phn140\phd\phn$ & $<0.0333^{\phm{+}\phn\phd\phn\phn\phn}_{\phm{-}\phn\phd\phn\phn\phn}$ & $<\phn0.353^{\phm{+}\phn\phd\phn\phn\phn}_{\phm{-}\phn\phd\phn\phn\phn}\phm{\tablenotemark{a}}$ & $\phn0$ & $0.000$ & $\phn25.8$\\
HD145417 & \nodata & \nodata & \nodata & \nodata & \nodata & \nodata\\
V410-$\tau$ & $\phn\phn\phn\phn3\phd\phn\pm \phn\phn31\phd\phn$ & $<0.269\phn^{\phm{+}\phn\phd\phn\phn\phn}_{\phm{-}\phn\phd\phn\phn\phn}$ & $<\phn3.58\phn^{\phm{+}\phn\phd\phn\phn\phn}_{\phm{-}\phn\phd\phn\phn\phn}\phm{\tablenotemark{a}}$ & $\phn0$ & $0.000$ & $397\phd\phn$\\
EG Cha & $\phn1230\phd\phn\pm \phn180\phd\phn$ & $\phm{<}\ 0.077\phn^{+0.025}_{-0.024}$ & $\phm{<}\ \phn1.27\phn^{+0.21\phn}_{-0.20\phn}\tablenotemark{g}$ & $\phn1$ & $0.250$ & $\phn30.9$\\
HBC427 & $\phn\phn132\phd\phn\pm \phn\phn67\phd\phn$ & $<0.213\phn^{\phm{+}\phn\phd\phn\phn\phn}_{\phm{-}\phn\phd\phn\phn\phn}$ & $<\phn2.16\phn^{\phm{+}\phn\phd\phn\phn\phn}_{\phm{-}\phn\phd\phn\phn\phn}\phm{\tablenotemark{a}}$ & $\phn0$ & $0.000$ & $\phn72.5$\\
61 Cyg A & \nodata & \nodata & \nodata & \nodata & \nodata & \nodata\\
LkCa 4 & $\phn\phn110\phd\phn\pm \phn\phn59\phd\phn$ & $<0.225\phn^{\phm{+}\phn\phd\phn\phn\phn}_{\phm{-}\phn\phd\phn\phn\phn}$ & $<\phn2.26\phn^{\phm{+}\phn\phd\phn\phn\phn}_{\phm{-}\phn\phd\phn\phn\phn}\phm{\tablenotemark{a}}$ & $\phn0$ & $0.000$ & $\phn79.5$\\
GJ832 & $\phn\phn126\phd\phn\pm \phn\phn60\phd\phn$ & $\phm{<}\ 0.149\phn^{+0.072}_{-0.090}$ & $\phm{<}\ \phn0.49\phn^{+0.12\phn}_{-0.15\phn}\phm{\tablenotemark{a}}$ & $\phn0$ & $0.000$ & $\phn93.8$\\
TWA13B & $\phn\phn600\phd\phn\pm \phn130\phd\phn$ & $\phm{<}\ 0.147\phn^{+0.035}_{-0.032}$ & $\phm{<}\ \phn0.98\phn^{+0.12\phn}_{-0.11\phn}\tablenotemark{g}$ & $\phn0$ & $0.000$ & $\phn55.1$\\
TWA13A & $\phn\phn700\phd\phn\pm \phn160\phd\phn$ & $<0.0860^{\phm{+}\phn\phd\phn\phn\phn}_{\phm{-}\phn\phd\phn\phn\phn}$ & $<\phn0.824^{\phm{+}\phn\phd\phn\phn\phn}_{\phm{-}\phn\phd\phn\phn\phn}\phm{\tablenotemark{a}}$ & $\phn0$ & $0.000$ & $\phn32.5$\\
AU Mic & $\phn4400\phd\phn\pm 1100\phd\phn$ & $<0.0915^{\phm{+}\phn\phd\phn\phn\phn}_{\phm{-}\phn\phd\phn\phn\phn}$ & $<\phn0.648^{\phm{+}\phn\phd\phn\phn\phn}_{\phm{-}\phn\phd\phn\phn\phn}\phm{\tablenotemark{a}}$ & $\phn1$ & $0.144$ & $\phn83.5$\\
CE Ant & $\phn1420\phd\phn\pm \phn260\phd\phn$ & $\phm{<}\ 0.183\phn^{+0.039}_{-0.033}$ & $\phm{<}\ \phn1.09\phn^{+0.12\phn}_{-0.10\phn}\tablenotemark{g}$ & $\phn0$ & $0.000$ & $\phn58.1$\\
GJ436 & $\phn\phn\phn\phn1.1\pm \phn\phn\phn4.5$ & $<0.472\phn^{\phm{+}\phn\phd\phn\phn\phn}_{\phm{-}\phn\phd\phn\phn\phn}$ & $<\phn0.778^{\phm{+}\phn\phd\phn\phn\phn}_{\phm{-}\phn\phd\phn\phn\phn}\phm{\tablenotemark{a}}$ & $\phn1$ & $0.164$ & $323\phd\phn$\\
EV Lac & $\phn\phn820\phd\phn\pm \phn400\phd\phn$ & $<0.109\phn^{\phm{+}\phn\phd\phn\phn\phn}_{\phm{-}\phn\phd\phn\phn\phn}$ & $<\phn0.312^{\phm{+}\phn\phd\phn\phn\phn}_{\phm{-}\phn\phd\phn\phn\phn}\phm{\tablenotemark{a}}$ & $\phn2$ & $0.166$ & $128\phd\phn$\\
AD Leo & $11260\phd\phn\pm \phn530\phd\phn$ & $\phm{<}\ 0.112\phn^{+0.011}_{-0.011}$ & $\phm{<}\ \phn0.388^{+0.019}_{-0.019}\tablenotemark{g}$ & $14$ & $0.110$ & $\phn70.4$\\
IL Aqr & $\phn\phn167\phd\phn\pm \phn\phn99\phd\phn$ & $\phm{<}\ 0.177\phn^{+0.091}_{-0.097}$ & $\phm{<}\ \phn0.40\phn^{+0.10\phn}_{-0.11\phn}\tablenotemark{g}$ & $\phn1$ & $0.375$ & $\phn83.6$\\
HO Lib & $\phn\phn\phn\phn1.8\pm \phn\phn\phn8.1$ & $<0.592\phn^{\phm{+}\phn\phd\phn\phn\phn}_{\phm{-}\phn\phd\phn\phn\phn}$ & $<\phn0.588^{\phm{+}\phn\phd\phn\phn\phn}_{\phm{-}\phn\phd\phn\phn\phn}\phm{\tablenotemark{a}}$ & $\phn0$ & $0.000$ & $252\phd\phn$\\
Prox Cen & $\phn\phn188\phd\phn\pm \phn\phn87\phd\phn$ & $<0.132\phn^{\phm{+}\phn\phd\phn\phn\phn}_{\phm{-}\phn\phd\phn\phn\phn}$ & $<\phn0.131^{\phm{+}\phn\phd\phn\phn\phn}_{\phm{-}\phn\phd\phn\phn\phn}\phm{\tablenotemark{a}}$ & $\phn7$ & $0.082$ & $250\phd\phn$\\
GJ3877 & \nodata & \nodata & \nodata & \nodata & \nodata & \nodata\\
GJ3517 & \nodata & \nodata & \nodata & \nodata & \nodata & \nodata\\
\enddata

\statstablefootnotes
\end{deluxetable*}

\begin{deluxetable*}{lrrcrrr}
\tabletypesize{\scriptsize}
\tablecaption{Variability Statistics in \ion{Si}{4} \label{tbl:varsiv}}
\tablewidth{0pt}
\tablecolumns{7}
\tablehead{\colhead{Star} & \colhead{$\mf$} & \colhead{$\s_x$\tablenotemark{a}} & \colhead{$R_{\sigma_x}\tablenotemark{b}$} & \colhead{$N_{f}\tablenotemark{c}$} & \colhead{Duty\tablenotemark{d}} & \colhead{$\nfe\tablenotemark{e}$}\\
& \colhead{$10^{-17}$ erg s$^{-1}$ cm$^{-2}$} &  & \colhead{$R_J$} &  & \colhead{Cycle} & \colhead{s}}
\startdata
$\beta$ Cas\tablenotemark{f} & $81500\phd\phn\pm 1700\phd\phn$ & $\phm{<}\ 0.0089^{+0.0032}_{-0.0029}$ & $\phm{<}\ \phn0.84\phn^{+0.15\phn}_{-0.13\phn}\tablenotemark{g}$ & $\phn0$ & $0.000$ & $\phn\phn3.16$\\
$\delta$ Cep & $\phn\phn231\phd\phn\pm \phn\phn36\phd\phn$ & $<0.0259^{\phm{+}\phn\phd\phn\phn\phn\phn}_{\phm{-}\phn\phd\phn\phn\phn\phn}$ & $<18.5\phn\phn^{\phm{+}\phn\phd\phn\phn\phn}_{\phm{-}\phn\phd\phn\phn\phn}\phm{\tablenotemark{a}}$ & $\phn0$ & $0.000$ & $\phn29.6\phn$\\
$\alpha$ Per & $\phn1890\phd\phn\pm \phn240\phd\phn$ & $<0.0452^{\phm{+}\phn\phd\phn\phn\phn\phn}_{\phm{-}\phn\phd\phn\phn\phn\phn}$ & $<30.6\phn\phn^{\phm{+}\phn\phd\phn\phn\phn}_{\phm{-}\phn\phd\phn\phn\phn}\phm{\tablenotemark{a}}$ & $\phn0$ & $0.000$ & $\phn14.1\phn$\\
$\beta$ Dor & $\phn2840\phd\phn\pm \phn140\phd\phn$ & $<0.0107^{\phm{+}\phn\phd\phn\phn\phn\phn}_{\phm{-}\phn\phd\phn\phn\phn\phn}$ & $<18.9\phn\phn^{\phm{+}\phn\phd\phn\phn\phn}_{\phm{-}\phn\phd\phn\phn\phn}\phm{\tablenotemark{a}}$ & $\phn0$ & $0.000$ & $\phn11.9\phn$\\
Polaris & $\phn1800\phd\phn\pm \phn120\phd\phn$ & $<0.0214^{\phm{+}\phn\phd\phn\phn\phn\phn}_{\phm{-}\phn\phd\phn\phn\phn\phn}$ & $<\phn7.47\phn^{\phm{+}\phn\phd\phn\phn\phn}_{\phm{-}\phn\phd\phn\phn\phn}\phm{\tablenotemark{a}}$ & $\phn0$ & $0.000$ & $\phn20.8\phn$\\
HD25825 & $\phn\phn690\phd\phn\pm \phn160\phd\phn$ & $\phm{<}\ 0.050\phn^{+0.027\phn}_{-0.037\phn}$ & $\phm{<}\ \phn0.62\phn^{+0.17\phn}_{-0.23\phn}\phm{\tablenotemark{a}}$ & $\phn0$ & $0.000$ & $\phn24.5\phn$\\
HD209458/G130M & $\phn\phn\phn73\phd\phn\pm \phn\phn11\phd\phn$ & $<0.0634^{\phm{+}\phn\phd\phn\phn\phn\phn}_{\phm{-}\phn\phd\phn\phn\phn\phn}$ & $<\phn0.743^{\phm{+}\phn\phd\phn\phn\phn}_{\phm{-}\phn\phd\phn\phn\phn}\phm{\tablenotemark{a}}$ & $\phn1$ & $0.044$ & $\phn77.9\phn$\\
HD209458/E140M & $\phn\phn\phn16\phd\phn\pm \phn\phn28\phd\phn$ & $<0.277\phn^{\phm{+}\phn\phd\phn\phn\phn\phn}_{\phm{-}\phn\phd\phn\phn\phn\phn}$ & $<\phn1.55\phn^{\phm{+}\phn\phd\phn\phn\phn}_{\phm{-}\phn\phd\phn\phn\phn}\phm{\tablenotemark{a}}$ & $\phn0$ & $0.000$ & $344\phd\phn\phn$\\
$\chi^1$ Ori & $20160\phd\phn\pm \phn920\phd\phn$ & $\phm{<}\ 0.099\phn^{+0.020\phn}_{-0.015\phn}$ & $\phm{<}\ \phn0.844^{+0.086}_{-0.066}\phm{\tablenotemark{a}}$ & $\phn0$ & $0.000$ & $\phn25.5\phn$\\
HII314 & $\phn\phn255\phd\phn\pm \phn\phn55\phd\phn$ & $<0.0660^{\phm{+}\phn\phd\phn\phn\phn\phn}_{\phm{-}\phn\phd\phn\phn\phn\phn}$ & $<\phn0.414^{\phm{+}\phn\phd\phn\phn\phn}_{\phm{-}\phn\phd\phn\phn\phn}\phm{\tablenotemark{a}}$ & $\phn0$ & $0.000$ & $\phn40.5\phn$\\
EK Dra & $\phn4280\phd\phn\pm \phn190\phd\phn$ & $\phm{<}\ 0.0741^{+0.0066}_{-0.0061}$ & $\phm{<}\ \phn0.675^{+0.035}_{-0.033}\phm{\tablenotemark{a}}$ & $\phn0$ & $0.000$ & $\phn35.2\phn$\\
$\pi^1$ UMa & $\phn7030\phd\phn\pm \phn480\phd\phn$ & $\phm{<}\ 0.058\phn^{+0.014\phn}_{-0.011\phn}$ & $\phm{<}\ \phn0.580^{+0.070}_{-0.055}\phm{\tablenotemark{a}}$ & $\phn0$ & $0.000$ & $\phn16.3\phn$\\
HD90508 & $\phn\phn\phn37.1\pm \phn\phn\phn6.9$ & $\phm{<}\ 0.116\phn^{+0.054\phn}_{-0.092\phn}$ & $\phm{<}\ \phn0.59\phn^{+0.14\phn}_{-0.23\phn}\tablenotemark{g}$ & $\phn0$ & $0.000$ & $111\phd\phn\phn$\\
HD199288/G140L & $\phn\phn\phn32.6\pm \phn\phn\phn6.8$ & $<0.142\phn^{\phm{+}\phn\phd\phn\phn\phn\phn}_{\phm{-}\phn\phd\phn\phn\phn\phn}$ & $<\phn0.934^{\phm{+}\phn\phd\phn\phn\phn}_{\phm{-}\phn\phd\phn\phn\phn}\phm{\tablenotemark{a}}$ & $\phn0$ & $0.000$ & $104\phd\phn\phn$\\
HD199288/G130M & $\phn\phn\phn93\phd\phn\pm \phn\phn24\phd\phn$ & $<0.0696^{\phm{+}\phn\phd\phn\phn\phn\phn}_{\phm{-}\phn\phd\phn\phn\phn\phn}$ & $<\phn0.653^{\phm{+}\phn\phd\phn\phn\phn}_{\phm{-}\phn\phd\phn\phn\phn}\phm{\tablenotemark{a}}$ & $\phn0$ & $0.000$ & $\phn63.2\phn$\\
18 Sco/G140L & $\phn\phn370\phd\phn\pm \phn\phn44\phd\phn$ & $<0.131\phn^{\phm{+}\phn\phd\phn\phn\phn\phn}_{\phm{-}\phn\phd\phn\phn\phn\phn}$ & $<\phn0.963^{\phm{+}\phn\phd\phn\phn\phn}_{\phm{-}\phn\phd\phn\phn\phn}\phm{\tablenotemark{a}}$ & $\phn0$ & $0.000$ & $\phn40.5\phn$\\
18 Sco/G130M & $\phn\phn506\phd\phn\pm \phn\phn84\phd\phn$ & $<0.0650^{\phm{+}\phn\phd\phn\phn\phn\phn}_{\phm{-}\phn\phd\phn\phn\phn\phn}$ & $<\phn0.678^{\phm{+}\phn\phd\phn\phn\phn}_{\phm{-}\phn\phd\phn\phn\phn}\phm{\tablenotemark{a}}$ & $\phn0$ & $0.000$ & $\phn27.7\phn$\\
FK Com\tablenotemark{f} & $\phn8110\phd\phn\pm \phn120\phd\phn$ & $\phm{<}\ 0.0147^{+0.0020}_{-0.0022}$ & $\phm{<}\ \phn2.64\phn^{+0.18\phn}_{-0.19\phn}\tablenotemark{g}$ & $\phn2$ & $0.110$ & $\phn10.6\phn$\\
HD65583 & $\phn\phn\phn\phn8.6\pm \phn\phn\phn3.8$ & $<0.249\phn^{\phm{+}\phn\phd\phn\phn\phn\phn}_{\phm{-}\phn\phd\phn\phn\phn\phn}$ & $<\phn0.985^{\phm{+}\phn\phd\phn\phn\phn}_{\phm{-}\phn\phd\phn\phn\phn}\phm{\tablenotemark{a}}$ & $\phn0$ & $0.000$ & $226\phd\phn\phn$\\
HD103095/G140L & $\phn\phn\phn20.2\pm \phn\phn\phn8.8$ & $<0.383\phn^{\phm{+}\phn\phd\phn\phn\phn\phn}_{\phm{-}\phn\phd\phn\phn\phn\phn}$ & $<\phn1.04\phn^{\phm{+}\phn\phd\phn\phn\phn}_{\phm{-}\phn\phd\phn\phn\phn}\phm{\tablenotemark{a}}$ & $\phn0$ & $0.000$ & $153\phd\phn\phn$\\
HD103095/G130M & $\phn\phn\phn37\phd\phn\pm \phn\phn21\phd\phn$ & $<0.201\phn^{\phm{+}\phn\phd\phn\phn\phn\phn}_{\phm{-}\phn\phd\phn\phn\phn\phn}$ & $<\phn0.756^{\phm{+}\phn\phd\phn\phn\phn}_{\phm{-}\phn\phd\phn\phn\phn}\phm{\tablenotemark{a}}$ & $\phn0$ & $0.000$ & $101\phd\phn\phn$\\
HD282630 & $\phn\phn333\phd\phn\pm \phn\phn83\phd\phn$ & $<0.0999^{\phm{+}\phn\phd\phn\phn\phn\phn}_{\phm{-}\phn\phd\phn\phn\phn\phn}$ & \nodata & $\phn0$ & $0.000$ & $\phn34.2\phn$\\
HD189733 & $\phn\phn470\phd\phn\pm \phn\phn63\phd\phn$ & $<0.0430^{\phm{+}\phn\phd\phn\phn\phn\phn}_{\phm{-}\phn\phd\phn\phn\phn\phn}$ & $<\phn0.401^{\phm{+}\phn\phd\phn\phn\phn}_{\phm{-}\phn\phd\phn\phn\phn}\phm{\tablenotemark{a}}$ & $\phn0$ & $0.000$ & $\phn33.5\phn$\\
HD145417 & $\phn\phn\phn\phn4.8\pm \phn\phn\phn2.9$ & $<0.322\phn^{\phm{+}\phn\phd\phn\phn\phn\phn}_{\phm{-}\phn\phd\phn\phn\phn\phn}$ & $<\phn0.887^{\phm{+}\phn\phd\phn\phn\phn}_{\phm{-}\phn\phd\phn\phn\phn}\phm{\tablenotemark{a}}$ & $\phn0$ & $0.000$ & $324\phd\phn\phn$\\
V410-$\tau$ & $\phn\phn\phn33\phd\phn\pm \phn\phn47\phd\phn$ & $<0.175\phn^{\phm{+}\phn\phd\phn\phn\phn\phn}_{\phm{-}\phn\phd\phn\phn\phn\phn}$ & $<\phn2.89\phn^{\phm{+}\phn\phd\phn\phn\phn}_{\phm{-}\phn\phd\phn\phn\phn}\phm{\tablenotemark{a}}$ & $\phn0$ & $0.000$ & $257\phd\phn\phn$\\
EG Cha & $\phn\phn587\phd\phn\pm \phn\phn95\phd\phn$ & $\phm{<}\ 0.076\phn^{+0.030\phn}_{-0.031\phn}$ & $\phm{<}\ \phn1.27\phn^{+0.25\phn}_{-0.26\phn}\tablenotemark{g}$ & $\phn1$ & $0.354$ & $\phn34.0\phn$\\
HBC427 & $\phn\phn\phn64\phd\phn\pm \phn\phn33\phd\phn$ & $\phm{<}\ 0.192\phn^{+0.065\phn}_{-0.061\phn}$ & $\phm{<}\ \phn2.05\phn^{+0.35\phn}_{-0.33\phn}\tablenotemark{g}$ & $\phn0$ & $0.000$ & $\phn80.6\phn$\\
61 Cyg A & $\phn1510\phd\phn\pm \phn220\phd\phn$ & $\phm{<}\ 0.070\phn^{+0.024\phn}_{-0.021\phn}$ & $\phm{<}\ \phn0.184^{+0.031}_{-0.028}\phm{\tablenotemark{a}}$ & $\phn0$ & $0.000$ & $\phn24.7\phn$\\
LkCa 4 & $\phn\phn\phn59\phd\phn\pm \phn\phn34\phd\phn$ & $<0.226\phn^{\phm{+}\phn\phd\phn\phn\phn\phn}_{\phm{-}\phn\phd\phn\phn\phn\phn}$ & $<\phn2.27\phn^{\phm{+}\phn\phd\phn\phn\phn}_{\phm{-}\phn\phd\phn\phn\phn}\phm{\tablenotemark{a}}$ & $\phn0$ & $0.000$ & $\phn86.2\phn$\\
GJ832 & $\phn\phn\phn88\phd\phn\pm \phn\phn40\phd\phn$ & $\phm{<}\ 0.139\phn^{+0.067\phn}_{-0.089\phn}$ & $\phm{<}\ \phn0.48\phn^{+0.12\phn}_{-0.15\phn}\phm{\tablenotemark{a}}$ & $\phn0$ & $0.000$ & $\phn80.1\phn$\\
TWA13B & $\phn\phn267\phd\phn\pm \phn\phn67\phd\phn$ & $\phm{<}\ 0.104\phn^{+0.042\phn}_{-0.045\phn}$ & $\phm{<}\ \phn0.83\phn^{+0.16\phn}_{-0.18\phn}\tablenotemark{g}$ & $\phn1$ & $0.100$ & $\phn49.6\phn$\\
TWA13A & $\phn\phn264\phd\phn\pm \phn\phn74\phd\phn$ & $\phm{<}\ 0.077\phn^{+0.038\phn}_{-0.054\phn}$ & $\phm{<}\ \phn0.78\phn^{+0.19\phn}_{-0.27\phn}\tablenotemark{g}$ & $\phn0$ & $0.000$ & $\phn44.9\phn$\\
AU Mic & $\phn3320\phd\phn\pm \phn340\phd\phn$ & $<0.0596^{\phm{+}\phn\phd\phn\phn\phn\phn}_{\phm{-}\phn\phd\phn\phn\phn\phn}$ & $<\phn0.523^{\phm{+}\phn\phd\phn\phn\phn}_{\phm{-}\phn\phd\phn\phn\phn}\phm{\tablenotemark{a}}$ & $\phn4$ & $0.174$ & $\phn50.0\phn$\\
CE Ant & $\phn\phn560\phd\phn\pm \phn120\phd\phn$ & $\phm{<}\ 0.207\phn^{+0.044\phn}_{-0.035\phn}$ & $\phm{<}\ \phn1.16\phn^{+0.12\phn}_{-0.10\phn}\tablenotemark{g}$ & $\phn1$ & $0.143$ & $\phn78.9\phn$\\
GJ436 & $\phn\phn\phn\phn1.8\pm \phn\phn\phn4.3$ & $<0.386\phn^{\phm{+}\phn\phd\phn\phn\phn\phn}_{\phm{-}\phn\phd\phn\phn\phn\phn}$ & $<\phn0.703^{\phm{+}\phn\phd\phn\phn\phn}_{\phm{-}\phn\phd\phn\phn\phn}\phm{\tablenotemark{a}}$ & $\phn1$ & $0.091$ & $215\phd\phn\phn$\\
EV Lac & $\phn1160\phd\phn\pm \phn210\phd\phn$ & $<0.116\phn^{\phm{+}\phn\phd\phn\phn\phn\phn}_{\phm{-}\phn\phd\phn\phn\phn\phn}$ & $<\phn0.322^{\phm{+}\phn\phd\phn\phn\phn}_{\phm{-}\phn\phd\phn\phn\phn}\phm{\tablenotemark{a}}$ & $\phn2$ & $0.265$ & $\phn94.3\phn$\\
AD Leo & $\phn5870\phd\phn\pm \phn170\phd\phn$ & $\phm{<}\ 0.1296^{+0.0063}_{-0.0063}$ & $\phm{<}\ \phn0.417^{+0.010}_{-0.010}\tablenotemark{g}$ & $28$ & $0.303$ & $\phn54.3\phn$\\
IL Aqr & $\phn\phn\phn88\phd\phn\pm \phn\phn53\phd\phn$ & $\phm{<}\ 0.21\phn\phn^{+0.10\phn\phn}_{-0.10\phn\phn}$ & $\phm{<}\ \phn0.44\phn^{+0.11\phn}_{-0.10\phn}\tablenotemark{g}$ & $\phn1$ & $0.375$ & $\phn97.4\phn$\\
HO Lib & $\phn\phn\phn\phn1.8\pm \phn\phn\phn6.7$ & $<0.450\phn^{\phm{+}\phn\phd\phn\phn\phn\phn}_{\phm{-}\phn\phd\phn\phn\phn\phn}$ & $<\phn0.512^{\phm{+}\phn\phd\phn\phn\phn}_{\phm{-}\phn\phd\phn\phn\phn}\phm{\tablenotemark{a}}$ & $\phn0$ & $0.000$ & $208\phd\phn\phn$\\
Prox Cen & $\phn\phn524\phd\phn\pm \phn\phn42\phd\phn$ & $\phm{<}\ 0.264\phn^{+0.022\phn}_{-0.021\phn}$ & $\phm{<}\ \phn0.185^{+0.012}_{-0.012}\phm{\tablenotemark{a}}$ & $\phn9$ & $0.114$ & $157\phd\phn\phn$\\
GJ3877 & $\phn\phn\phn\phn1.6\pm \phn\phn\phn2.3$ & $<0.887\phn^{\phm{+}\phn\phd\phn\phn\phn\phn}_{\phm{-}\phn\phd\phn\phn\phn\phn}$ & $<\phn0.289^{\phm{+}\phn\phd\phn\phn\phn}_{\phm{-}\phn\phd\phn\phn\phn}\phm{\tablenotemark{a}}$ & $\phn0$ & $0.000$ & $691\phd\phn\phn$\\
GJ3517 & $\phn\phn\phn\phn1.3\pm \phn\phn\phn1.5$ & $<0.690\phn^{\phm{+}\phn\phd\phn\phn\phn\phn}_{\phm{-}\phn\phd\phn\phn\phn\phn}$ & $<\phn0.214^{\phm{+}\phn\phd\phn\phn\phn}_{\phm{-}\phn\phd\phn\phn\phn}\phm{\tablenotemark{a}}$ & $\phn0$ & $0.000$ & $938\phd\phn\phn$\\
\enddata

\statstablefootnotes
\end{deluxetable*}

\begin{deluxetable*}{lrrcrrr}
\tabletypesize{\scriptsize}
\tablecaption{Variability Statistics in Continuum \label{tbl:varcont}}
\tablewidth{0pt}
\tablecolumns{5}
\tablehead{\colhead{Star} & \colhead{$\mf$} & \colhead{$\s_x$\tablenotemark{a}} & \colhead{$R_{\sigma_x}\tablenotemark{b}$} & \colhead{$N_{f}\tablenotemark{c}$} & \colhead{Duty\tablenotemark{d}} & \colhead{$\nfe\tablenotemark{e}$}\\
& \colhead{$10^{-17}$ erg s$^{-1}$ cm$^{-2}$} &  & \colhead{$R_J$} &  & \colhead{Cycle} & \colhead{s}}
\startdata
$\beta$ Cas\tablenotemark{f} & $194000\phd\phn\pm 2900\phd\phn$ & $\phm{<}\ 0.0321\phn^{+0.0063}_{-0.0048}$ & $\phm{<}\ \phn1.59\phn^{+0.16\phn}_{-0.12\phn}\tablenotemark{g}$ & $0$ & $0.000$ & $\phn\phn8.21$\\
$\delta$ Cep & $\phn\phn\phn668\phd\phn\pm \phn\phn61\phd\phn$ & $\phm{<}\ 0.0427\phn^{+0.0055}_{-0.0051}$ & $\phm{<}\ 23.8\phn\phn^{+4.6\phn\phn}_{-4.6\phn\phn}\phm{\tablenotemark{a}}$ & $0$ & $0.000$ & $\phn23.8\phn$\\
$\alpha$ Per & $\phn13840\phd\phn\pm \phn640\phd\phn$ & $<0.0130\phn^{\phm{+}\phn\phd\phn\phn\phn\phn}_{\phm{-}\phn\phd\phn\phn\phn\phn}$ & $<16.4\phn\phn^{\phm{+}\phn\phd\phn\phn\phn}_{\phm{-}\phn\phd\phn\phn\phn}\phm{\tablenotemark{a}}$ & $0$ & $0.000$ & $\phn\phn4.27$\\
$\beta$ Dor & $\phn\phn2640\phd\phn\pm \phn160\phd\phn$ & $<0.00810^{\phm{+}\phn\phd\phn\phn\phn\phn}_{\phm{-}\phn\phd\phn\phn\phn\phn}$ & $<16.4\phn\phn^{\phm{+}\phn\phd\phn\phn\phn}_{\phm{-}\phn\phd\phn\phn\phn}\phm{\tablenotemark{a}}$ & $1$ & $0.132$ & $\phn11.0\phn$\\
Polaris & $\phn\phn6310\phd\phn\pm \phn220\phd\phn$ & $<0.00698^{\phm{+}\phn\phd\phn\phn\phn\phn}_{\phm{-}\phn\phd\phn\phn\phn\phn}$ & $<\phn4.27\phn^{\phm{+}\phn\phd\phn\phn\phn}_{\phm{-}\phn\phd\phn\phn\phn}\phm{\tablenotemark{a}}$ & $0$ & $0.000$ & $\phn\phn6.22$\\
HD25825 & $\phn\phn\phn186\phd\phn\pm \phn\phn81\phd\phn$ & $<0.193\phn\phn^{\phm{+}\phn\phd\phn\phn\phn\phn}_{\phm{-}\phn\phd\phn\phn\phn\phn}$ & $<\phn1.22\phn^{\phm{+}\phn\phd\phn\phn\phn}_{\phm{-}\phn\phd\phn\phn\phn}\phm{\tablenotemark{a}}$ & $0$ & $0.000$ & $\phn48.7\phn$\\
HD209458/G130M & $\phn\phn\phn\phn91\phd\phn\pm \phn\phn13\phd\phn$ & $<0.0427\phn^{\phm{+}\phn\phd\phn\phn\phn\phn}_{\phm{-}\phn\phd\phn\phn\phn\phn}$ & $<\phn0.610^{\phm{+}\phn\phd\phn\phn\phn}_{\phm{-}\phn\phd\phn\phn\phn}\phm{\tablenotemark{a}}$ & $0$ & $0.000$ & $\phn63.9\phn$\\
HD209458/E140M & $\phn\phn\phn\phn33\phd\phn\pm \phn\phn36\phd\phn$ & $<0.173\phn\phn^{\phm{+}\phn\phd\phn\phn\phn\phn}_{\phm{-}\phn\phd\phn\phn\phn\phn}$ & $<\phn1.23\phn^{\phm{+}\phn\phd\phn\phn\phn}_{\phm{-}\phn\phd\phn\phn\phn}\phm{\tablenotemark{a}}$ & $0$ & $0.000$ & $225\phd\phn\phn$\\
$\chi^1$ Ori & $\phn\phn5110\phd\phn\pm \phn390\phd\phn$ & $\phm{<}\ 0.014\phn\phn^{+0.007\phn}_{-0.011\phn}$ & $\phm{<}\ \phn0.32\phn^{+0.08\phn}_{-0.13\phn}\phm{\tablenotemark{a}}$ & $0$ & $0.000$ & $\phn\phn7.72$\\
HII314 & $\phn\phn\phn\phn49\phd\phn\pm \phn\phn24\phd\phn$ & $<0.124\phn\phn^{\phm{+}\phn\phd\phn\phn\phn\phn}_{\phm{-}\phn\phd\phn\phn\phn\phn}$ & $<\phn0.567^{\phm{+}\phn\phd\phn\phn\phn}_{\phm{-}\phn\phd\phn\phn\phn}\phm{\tablenotemark{a}}$ & $0$ & $0.000$ & $\phn76.6\phn$\\
EK Dra & $\phn\phn\phn850\phd\phn\pm \phn130\phd\phn$ & $<0.0407\phn^{\phm{+}\phn\phd\phn\phn\phn\phn}_{\phm{-}\phn\phd\phn\phn\phn\phn}$ & $<\phn0.500^{\phm{+}\phn\phd\phn\phn\phn}_{\phm{-}\phn\phd\phn\phn\phn}\phm{\tablenotemark{a}}$ & $1$ & $0.100$ & $\phn37.3\phn$\\
$\pi^1$ UMa & $\phn\phn1670\phd\phn\pm \phn220\phd\phn$ & $<0.0341\phn^{\phm{+}\phn\phd\phn\phn\phn\phn}_{\phm{-}\phn\phd\phn\phn\phn\phn}$ & $<\phn0.446^{\phm{+}\phn\phd\phn\phn\phn}_{\phm{-}\phn\phd\phn\phn\phn}\phm{\tablenotemark{a}}$ & $0$ & $0.000$ & $\phn11.9\phn$\\
HD90508 & $\phn\phn\phn211\phd\phn\pm \phn\phn17\phd\phn$ & $<0.0656\phn^{\phm{+}\phn\phd\phn\phn\phn\phn}_{\phm{-}\phn\phd\phn\phn\phn\phn}$ & $<\phn0.445^{\phm{+}\phn\phd\phn\phn\phn}_{\phm{-}\phn\phd\phn\phn\phn}\phm{\tablenotemark{a}}$ & $0$ & $0.000$ & $\phn44.6\phn$\\
HD199288/G140L & $\phn\phn\phn228\phd\phn\pm \phn\phn18\phd\phn$ & $<0.0725\phn^{\phm{+}\phn\phd\phn\phn\phn\phn}_{\phm{-}\phn\phd\phn\phn\phn\phn}$ & $<\phn0.667^{\phm{+}\phn\phd\phn\phn\phn}_{\phm{-}\phn\phd\phn\phn\phn}\phm{\tablenotemark{a}}$ & $0$ & $0.000$ & $\phn54.6\phn$\\
HD199288/G130M & $\phn\phn\phn233\phd\phn\pm \phn\phn38\phd\phn$ & $<0.0425\phn^{\phm{+}\phn\phd\phn\phn\phn\phn}_{\phm{-}\phn\phd\phn\phn\phn\phn}$ & $<\phn0.510^{\phm{+}\phn\phd\phn\phn\phn}_{\phm{-}\phn\phd\phn\phn\phn}\phm{\tablenotemark{a}}$ & $0$ & $0.000$ & $\phn37.0\phn$\\
18 Sco/G140L & $\phn\phn\phn401\phd\phn\pm \phn\phn44\phd\phn$ & $<0.0935\phn^{\phm{+}\phn\phd\phn\phn\phn\phn}_{\phm{-}\phn\phd\phn\phn\phn\phn}$ & $<\phn0.813^{\phm{+}\phn\phd\phn\phn\phn}_{\phm{-}\phn\phd\phn\phn\phn}\phm{\tablenotemark{a}}$ & $0$ & $0.000$ & $\phn32.2\phn$\\
18 Sco/G130M & $\phn\phn\phn374\phd\phn\pm \phn\phn72\phd\phn$ & $<0.0708\phn^{\phm{+}\phn\phd\phn\phn\phn\phn}_{\phm{-}\phn\phd\phn\phn\phn\phn}$ & $<\phn0.707^{\phm{+}\phn\phd\phn\phn\phn}_{\phm{-}\phn\phd\phn\phn\phn}\phm{\tablenotemark{a}}$ & $0$ & $0.000$ & $\phn29.2\phn$\\
FK Com\tablenotemark{f} & $\phn\phn2332\phd\phn\pm \phn\phn97\phd\phn$ & $<0.00812^{\phm{+}\phn\phd\phn\phn\phn\phn}_{\phm{-}\phn\phd\phn\phn\phn\phn}$ & $<\phn1.96\phn^{\phm{+}\phn\phd\phn\phn\phn}_{\phm{-}\phn\phd\phn\phn\phn}\phm{\tablenotemark{a}}$ & $2$ & $0.101$ & $\phn12.2\phn$\\
HD65583 & $\phn\phn\phn\phn49.9\pm \phn\phn\phn7.9$ & $<0.126\phn\phn^{\phm{+}\phn\phd\phn\phn\phn\phn}_{\phm{-}\phn\phd\phn\phn\phn\phn}$ & $<\phn0.700^{\phm{+}\phn\phd\phn\phn\phn}_{\phm{-}\phn\phd\phn\phn\phn}\phm{\tablenotemark{a}}$ & $0$ & $0.000$ & $\phn89.2\phn$\\
HD103095/G140L & $\phn\phn\phn\phn87\phd\phn\pm \phn\phn18\phd\phn$ & $<0.173\phn\phn^{\phm{+}\phn\phd\phn\phn\phn\phn}_{\phm{-}\phn\phd\phn\phn\phn\phn}$ & $<\phn0.701^{\phm{+}\phn\phd\phn\phn\phn}_{\phm{-}\phn\phd\phn\phn\phn}\phm{\tablenotemark{a}}$ & $0$ & $0.000$ & $\phn71.4\phn$\\
HD103095/G130M & $\phn\phn\phn\phn80\phd\phn\pm \phn\phn33\phd\phn$ & $\phm{<}\ 0.176\phn\phn^{+0.063\phn}_{-0.068\phn}$ & $\phm{<}\ \phn0.71\phn^{+0.13\phn}_{-0.14\phn}\phm{\tablenotemark{a}}$ & $0$ & $0.000$ & $\phn80.4\phn$\\
HD282630 & $\phn\phn\phn\phn32\phd\phn\pm \phn\phn27\phd\phn$ & $<0.234\phn\phn^{\phm{+}\phn\phd\phn\phn\phn\phn}_{\phm{-}\phn\phd\phn\phn\phn\phn}$ & \nodata & $0$ & $0.000$ & $112\phd\phn\phn$\\
HD189733 & $\phn\phn\phn\phn99\phd\phn\pm \phn\phn29\phd\phn$ & $<0.0749\phn^{\phm{+}\phn\phd\phn\phn\phn\phn}_{\phm{-}\phn\phd\phn\phn\phn\phn}$ & $<\phn0.529^{\phm{+}\phn\phd\phn\phn\phn}_{\phm{-}\phn\phd\phn\phn\phn}\phm{\tablenotemark{a}}$ & $0$ & $0.000$ & $\phn58.6\phn$\\
HD145417 & $\phn\phn\phn\phn29.3\pm \phn\phn\phn5.6$ & $<0.157\phn\phn^{\phm{+}\phn\phd\phn\phn\phn\phn}_{\phm{-}\phn\phd\phn\phn\phn\phn}$ & $<\phn0.618^{\phm{+}\phn\phd\phn\phn\phn}_{\phm{-}\phn\phd\phn\phn\phn}\phm{\tablenotemark{a}}$ & $0$ & $0.000$ & $109\phd\phn\phn$\\
V410-$\tau$ & $\phn\phn\phn\phn31\phd\phn\pm \phn\phn43\phd\phn$ & $<0.149\phn\phn^{\phm{+}\phn\phd\phn\phn\phn\phn}_{\phm{-}\phn\phd\phn\phn\phn\phn}$ & $<\phn2.66\phn^{\phm{+}\phn\phd\phn\phn\phn}_{\phm{-}\phn\phd\phn\phn\phn}\phm{\tablenotemark{a}}$ & $0$ & $0.000$ & $240\phd\phn\phn$\\
EG Cha & $\phn\phn\phn147\phd\phn\pm \phn\phn42\phd\phn$ & $\phm{<}\ 0.219\phn\phn^{+0.041\phn}_{-0.037\phn}$ & $\phm{<}\ \phn2.15\phn^{+0.20\phn}_{-0.18\phn}\tablenotemark{g}$ & $0$ & $0.000$ & $\phn87.3\phn$\\
HBC427 & $\phn\phn\phn\phn12\phd\phn\pm \phn\phn17\phd\phn$ & $<0.324\phn\phn^{\phm{+}\phn\phd\phn\phn\phn\phn}_{\phm{-}\phn\phd\phn\phn\phn\phn}$ & $<\phn2.67\phn^{\phm{+}\phn\phd\phn\phn\phn}_{\phm{-}\phn\phd\phn\phn\phn}\phm{\tablenotemark{a}}$ & $0$ & $0.000$ & $158\phd\phn\phn$\\
61 Cyg A & $\phn\phn\phn620\phd\phn\pm \phn140\phd\phn$ & $\phm{<}\ 0.053\phn\phn^{+0.028\phn}_{-0.034\phn}$ & $\phm{<}\ \phn0.160^{+0.042}_{-0.051}\phm{\tablenotemark{a}}$ & $0$ & $0.000$ & $\phn25.2\phn$\\
LkCa 4 & $\phn\phn\phn\phn\phn2.1\pm \phn\phn\phn8.0$ & $<0.481\phn\phn^{\phm{+}\phn\phd\phn\phn\phn\phn}_{\phm{-}\phn\phd\phn\phn\phn\phn}$ & $<\phn3.31\phn^{\phm{+}\phn\phd\phn\phn\phn}_{\phm{-}\phn\phd\phn\phn\phn}\phm{\tablenotemark{a}}$ & $0$ & $0.000$ & $239\phd\phn\phn$\\
GJ832 & $\phn\phn\phn\phn16\phd\phn\pm \phn\phn16\phd\phn$ & $<0.311\phn\phn^{\phm{+}\phn\phd\phn\phn\phn\phn}_{\phm{-}\phn\phd\phn\phn\phn\phn}$ & $<\phn0.712^{\phm{+}\phn\phd\phn\phn\phn}_{\phm{-}\phn\phd\phn\phn\phn}\phm{\tablenotemark{a}}$ & $0$ & $0.000$ & $134\phd\phn\phn$\\
TWA13B & $\phn\phn\phn\phn45\phd\phn\pm \phn\phn27\phd\phn$ & $<0.174\phn\phn^{\phm{+}\phn\phd\phn\phn\phn\phn}_{\phm{-}\phn\phd\phn\phn\phn\phn}$ & $<\phn1.07\phn^{\phm{+}\phn\phd\phn\phn\phn}_{\phm{-}\phn\phd\phn\phn\phn}\phm{\tablenotemark{a}}$ & $0$ & $0.000$ & $\phn92.8\phn$\\
TWA13A & $\phn\phn\phn\phn56\phd\phn\pm \phn\phn37\phd\phn$ & $<0.254\phn\phn^{\phm{+}\phn\phd\phn\phn\phn\phn}_{\phm{-}\phn\phd\phn\phn\phn\phn}$ & $<\phn1.42\phn^{\phm{+}\phn\phd\phn\phn\phn}_{\phm{-}\phn\phd\phn\phn\phn}\phm{\tablenotemark{a}}$ & $0$ & $0.000$ & $104\phd\phn\phn$\\
AU Mic & $\phn\phn\phn269\phd\phn\pm \phn\phn84\phd\phn$ & $<0.0821\phn^{\phm{+}\phn\phd\phn\phn\phn\phn}_{\phm{-}\phn\phd\phn\phn\phn\phn}$ & $<\phn0.613^{\phm{+}\phn\phd\phn\phn\phn}_{\phm{-}\phn\phd\phn\phn\phn}\phm{\tablenotemark{a}}$ & $1$ & $0.048$ & $\phn91.2\phn$\\
CE Ant & $\phn\phn\phn149\phd\phn\pm \phn\phn61\phd\phn$ & $<0.168\phn\phn^{\phm{+}\phn\phd\phn\phn\phn\phn}_{\phm{-}\phn\phd\phn\phn\phn\phn}$ & $<\phn1.05\phn^{\phm{+}\phn\phd\phn\phn\phn}_{\phm{-}\phn\phd\phn\phn\phn}\phm{\tablenotemark{a}}$ & $1$ & $0.071$ & $\phn54.5\phn$\\
GJ436 & $\phn\phn\phn\phn\phn1.6\pm \phn\phn\phn7.8$ & $<0.398\phn\phn^{\phm{+}\phn\phd\phn\phn\phn\phn}_{\phm{-}\phn\phd\phn\phn\phn\phn}$ & $<\phn0.714^{\phm{+}\phn\phd\phn\phn\phn}_{\phm{-}\phn\phd\phn\phn\phn}\phm{\tablenotemark{a}}$ & $0$ & $0.000$ & $296\phd\phn\phn$\\
EV Lac & $\phn\phn\phn156\phd\phn\pm \phn\phn70\phd\phn$ & $<0.121\phn\phn^{\phm{+}\phn\phd\phn\phn\phn\phn}_{\phm{-}\phn\phd\phn\phn\phn\phn}$ & $<\phn0.328^{\phm{+}\phn\phd\phn\phn\phn}_{\phm{-}\phn\phd\phn\phn\phn}\phm{\tablenotemark{a}}$ & $0$ & $0.000$ & $164\phd\phn\phn$\\
AD Leo & $\phn\phn\phn355\phd\phn\pm \phn\phn33\phd\phn$ & $<0.0512\phn^{\phm{+}\phn\phd\phn\phn\phn\phn}_{\phm{-}\phn\phd\phn\phn\phn\phn}$ & $<\phn0.262^{\phm{+}\phn\phd\phn\phn\phn}_{\phm{-}\phn\phd\phn\phn\phn}\phm{\tablenotemark{a}}$ & $9$ & $0.056$ & $\phn99.9\phn$\\
IL Aqr & $\phn\phn\phn\phn\phn2\phd\phn\pm \phn\phn13\phd\phn$ & $<0.593\phn\phn^{\phm{+}\phn\phd\phn\phn\phn\phn}_{\phm{-}\phn\phd\phn\phn\phn\phn}$ & $<\phn0.736^{\phm{+}\phn\phd\phn\phn\phn}_{\phm{-}\phn\phd\phn\phn\phn}\phm{\tablenotemark{a}}$ & $1$ & $0.344$ & $307\phd\phn\phn$\\
HO Lib & $\phn\phn\phn\phn\phn2\phd\phn\pm \phn\phn11\phd\phn$ & $<0.631\phn\phn^{\phm{+}\phn\phd\phn\phn\phn\phn}_{\phm{-}\phn\phd\phn\phn\phn\phn}$ & $<\phn0.607^{\phm{+}\phn\phd\phn\phn\phn}_{\phm{-}\phn\phd\phn\phn\phn}\phm{\tablenotemark{a}}$ & $0$ & $0.000$ & $291\phd\phn\phn$\\
Prox Cen & $\phn\phn\phn\phn77\phd\phn\pm \phn\phn24\phd\phn$ & $<0.0801\phn^{\phm{+}\phn\phd\phn\phn\phn\phn}_{\phm{-}\phn\phd\phn\phn\phn\phn}$ & $<\phn0.102^{\phm{+}\phn\phd\phn\phn\phn}_{\phm{-}\phn\phd\phn\phn\phn}\phm{\tablenotemark{a}}$ & $1$ & $0.022$ & $208\phd\phn\phn$\\
GJ3877 & $\phn\phn\phn\phn\phn7.0\pm \phn\phn\phn3.3$ & $<0.263\phn\phn^{\phm{+}\phn\phd\phn\phn\phn\phn}_{\phm{-}\phn\phd\phn\phn\phn\phn}$ & $<\phn0.157^{\phm{+}\phn\phd\phn\phn\phn}_{\phm{-}\phn\phd\phn\phn\phn}\phm{\tablenotemark{a}}$ & $0$ & $0.000$ & $248\phd\phn\phn$\\
GJ3517 & $\phn\phn\phn\phn\phn1.5\pm \phn\phn\phn1.3$ & $<0.249\phn\phn^{\phm{+}\phn\phd\phn\phn\phn\phn}_{\phm{-}\phn\phd\phn\phn\phn\phn}$ & $<\phn0.129^{\phm{+}\phn\phd\phn\phn\phn}_{\phm{-}\phn\phd\phn\phn\phn}\phm{\tablenotemark{a}}$ & $0$ & $0.000$ & $356\phd\phn\phn$\\
\enddata

\statstablefootnotes
\end{deluxetable*}

\section{Discussion}
\label{sec:discussion}

\subsection{The Range of Flare Behavior}
\label{sec:flarebehave}
\subsubsection{Strong Flares}
Several flares appear in the data that peak at tens of times the quiescent flux. The strongest of these is a flare on Prox Cen, displayed in Figure~\ref{fig:flarepalette}a  (see also \citealt{christian04}). This flare raises the continuum-subtracted flux in each band, normalized by the quiescent mean, to values at the peak data point of $27.5\pm0.2$ in \cii, $66.7\pm1.2$ in \siii, $91.6\pm0.3$ in \siv, and $13.5\pm0.6$ in the continuum. The $\nfe$ and duration of this flare in each band can be compared with the rest of the flare sample in Figure~\ref{fig:flaredist}, where these data points are labeled with 39, Prox Cen's dataset number from Table~\ref{tbl:obs}. Very strong flares also appear in the AD Leo (dataset no. 36; see also \citealt{hawley03}) and IL Aqr (dataset no. 37; see also \citealt{france12}) data and are also labeled in Figure~\ref{fig:flaredist}. The AD Leo flare peaks at $8.2\pm0.1$ in \cii, $14.0\pm0.2$ in \siii, $33.7\pm0.1$ in \siv, and $45.9\pm0.3$ in the continuum (Figure~\ref{fig:flarepalette}b). The IL Aqr flare peaks at 15.4$\pm0.3$ in \cii, $36.9\pm0.3$ in \siii, $46.5\pm0.3$ in \siv, and $20.1\pm1.4$ in the continuum (not included in Figure~\ref{fig:flarepalette}).

\subsubsection{Symmetric Flares}
Some of the flares in the sample show roughly equal rise and decay times, in contrast to the impulse-decay shape of many flares. The clearest example of such a flare is that which appears in the EG Cha data, plotted in Figure~\ref{fig:flarepalette}c. Note that this event was not flagged in the continuum or \cii\ data because the short span of the data, 0.82 hours, resulted in a poor determination of the quiescent scatter. The continuum and \cii\ did not sufficiently exceed the estimated scatter to result in an identification. Other flares that show a relatively clear symmetric photometric profile (though not nearly as clear as that of EG Cha) appear in the \siv\ data of EV Lac at mean Julian Date 52172.709 and the AD Leo data at mean Julian Date 51616.119. Many flares in the data appear as though they might be symmetric, but their rise and fall are not adequately resolved. 

\subsubsection{F Star Anomalies}
The flare identification algorithm flagged two events on F stars, one in \cii\ on $\delta$ Cep and another in the continuum on $\beta$ Dor. The latter is displayed in Figure~\ref{fig:flarepalette}d. Both events are gradual elevations of the flux in a single band relative to the other three. The $\beta$ Dor flare is particularly curious, as the other three bands clearly do not show the same consistent decline in flux that the elevated continuum flux does. These events seem likely to be true anomalies rather than spurious detections.

\subsubsection{Multi-peak Flares}
Many of the flares in the data exhibit complicated shapes, and might be a superposition of nearby small flares. Figure~\ref{fig:flarepalette}e depicts a flare on AD Leo that exhibits two peaks, separated by 240 s. This was the clearest specimen of a multi-peak flare. In most such flares, the data are not as clearly resolved above the surrounding scatter and/or different bands show different behavior. 

\subsubsection{Response in Different Lines}
\begin{figure}
\includegraphics[width=\columnwidth]{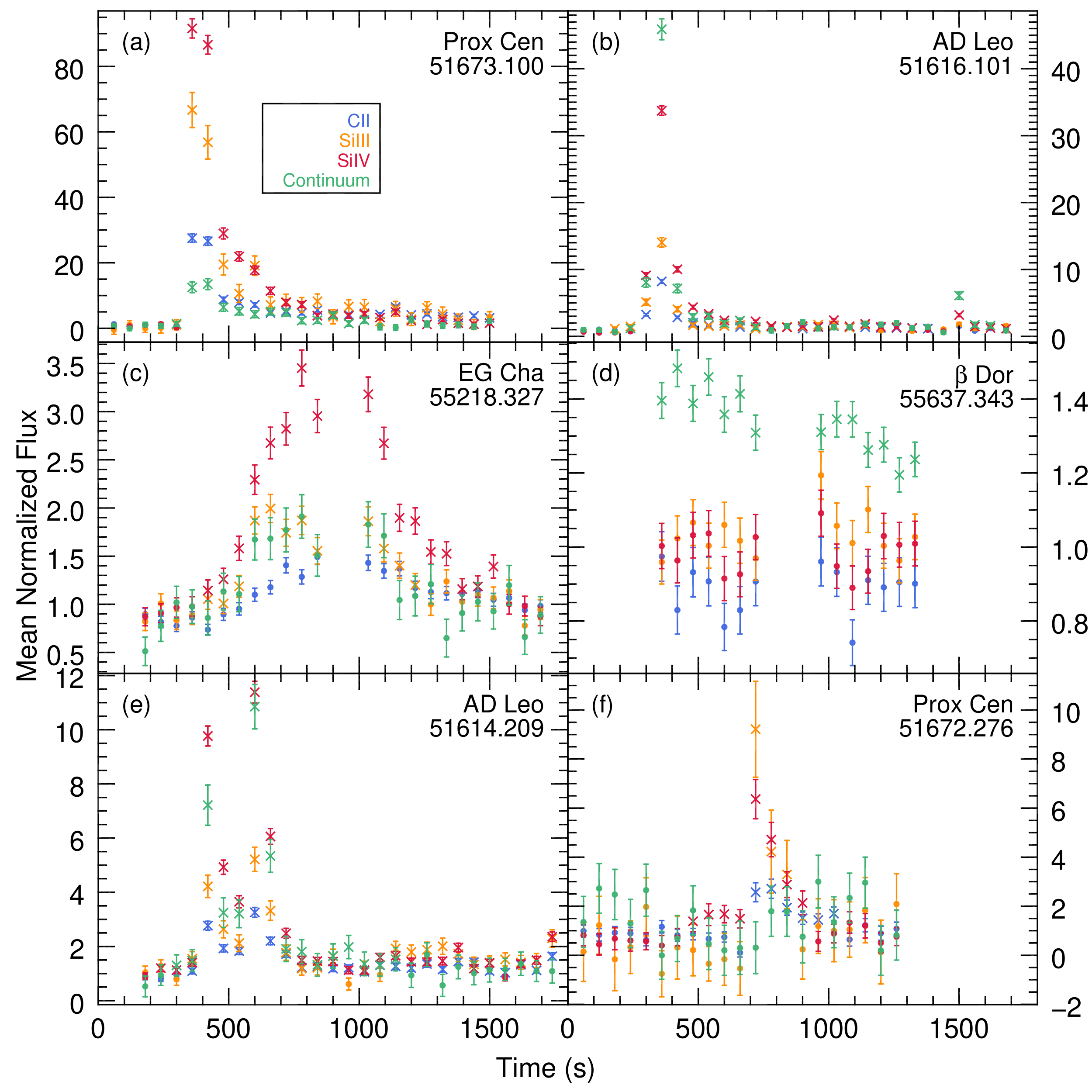}
\caption{A palette of flare behaviors, discussed in Section~\ref{sec:flarebehave}. Each subplot is labeled with the star and the mean Julian Date of $t = 0$. The fluxes are continuum-subtracted, high-pass filtered, and normalized to the mean of the quiescent points. The x symbols denote points flagged for removal before computing excess noise values (\emph{not} equivalent to the flare duration, see Section~\ref{sec:flareresults}. Note that the y-axis scales differ.}
\label{fig:flarepalette}
\end{figure}

We found that flare signals generally vary significantly between bands. In addition, these variations are not consistent from flare to flare. The difference in the response between bands is clear in Figure~\ref{fig:flarepalette}. For example, the Prox Cen, EG Cha, and AD Leo flares of subplots a, c, and e peak highest in \siv, with a roughly similar shape in each band. Alternatively, the AD Leo flare depicted in subplot b is the strongest in the continuum, and that of Prox Cen depicted in subplot f is the strongest in \siii. For the latter, the continuum shows essentially no response, though better signal to noise might reveal otherwise. (Note in Figure~\ref{fig:flarepalette}f that some \siii\ points are negative because the subtracted signal from the Ly$\alpha$ wing sometimes exceeds the low \siii\ signal.) Differences between lines in emitted flux during a flare have also been observed on the Sun (e.g. \citealt{brekke96}) and other low mass stars (e.g. GJ876, \citealt{ayres10,france12}).

\subsection{Risks Flares Pose to Transit Measurements}
\label{sec:flarerisks}
All of the detected flares are short lived, with one lasting 27 min, the others lasting less than 20 min, and over half lasting 4 min or less. These could easily be hidden by longer cadence data. Typical cadences in UV exoplanet transit observations are $\dt \approx$ 30-60 min, based on recent literature. While hour or longer cadences might often be unavoidable due to instrument or signal-to-noise limitations, a flare hidden in such data could bias a measurement of transit depth (see Section~\ref{sec:varintro}).

Of the 116 flares we identified (again, excluding FK Com), 57 (roughly one event per 5 h of data) would boost a 60 min integrated flux measurement by $\gtrsim10\%$, and all flares (roughly one event per 2.5 h of data) would boost a 60 min flux measurement by $\gtrsim\ $1\%. Flares produced the largest $\nfe$ values in the \siv\ band in roughly 2/3 of the 60~separate events and 7/9 of the events that registered a detection in all four bands. Thus, it appears that transit observations in \siv\ are somewhat likely to be more strongly affected by flaring. 

The flare stars (namely AD Leo and Prox Cen) account for most of the flares we detected, 92 of 116.  However, 24 flares were identified on objects not classified as ``flare stars.'' Such flares are of particular interest because these stars are more likely to be targeted in exoplanet search programs. These occurred on 5/8 M, 1/8 K, 3/11 G, and 2/5 F stars, and occurred roughly once per 5 h in the time all 32 stars were observed. Of these 24 flares, 9 (roughly one event per 13 h) would boost a 60 min integrated flux measurement by $\gtrsim10\%$. Most such flares occurred on M stars (7/9), while the remaining 2/9 are the same event observed in two different bands on the K star EG Cha.

A 10\% boost in flux exceeds the transit depth in \cii\ or \siii\ of HD209458b \citep{madjar04,linsky10}. However, not all stars in the sample are similar in size to HD209458, and it is the relative size of the star and planet that determines the transit depth and thus the impact of a flare. In other words, a strong flare will have less of an impact on a measurement of a Jupiter transiting and M star than a Jupiter transiting an F star. We used the stellar radii from the literature to determine the size of an object that would produce a transit signal of the same amplitude as the boost in flux from a flare in a 60 min integration. These range from 0.02 $R_J$ to 12.5 $R_J$ for the 116 flares identified. Of these, 90 (one flare per 3 h of data) would boost a 60 min flux integration by an amount larger than the signal of an Earth transiting the flaring star, 23 (one flare per 13 h) would boost flux by an amount exceeding a Neptune transit, and 7 (one flare per 42 h) would boost flux by an amount exceeding a Jupiter transit. Limiting the sample to the 24 events on non-flare stars only, 5 flares (one per 31 h) boost flux beyond an Earth signal, 2 (one per 76 h) beyond a Neptune signal, and 1 (one per 153 h) beyond a Jupiter signal.

We did not explore trends in flare rates with respect to stellar properties because the rates were only well constrained on Prox Cen and AD Leo. In addition, the small size, high diversity, and, most importantly, range of flare detection limits in the sample pose problems to such an analysis. Our general conclusion is that all low-mass stars likely pose a risk of flaring near or within a transit observation. Because flares might be so easily hidden in long-cadence data, we recommend that, when possible, transit observers employ minute-scale cadences to inspect their data prior to employing longer cadences for noise suppression. This would enable a sweep of the lightcurve for obvious flare events below the transit timescale. Flares could then be excised and the data, if desired, binned to a longer cadence, correcting for the ``dead time." Cadences of arbitrary length are possible with data from the photon-counting detectors common in UV work, provided the data are recorded as a time-tagged event list, rather than time-integrated counts.

\subsection{Flare Statistics on AD Leo}
\begin{figure}
\includegraphics[width=\columnwidth]{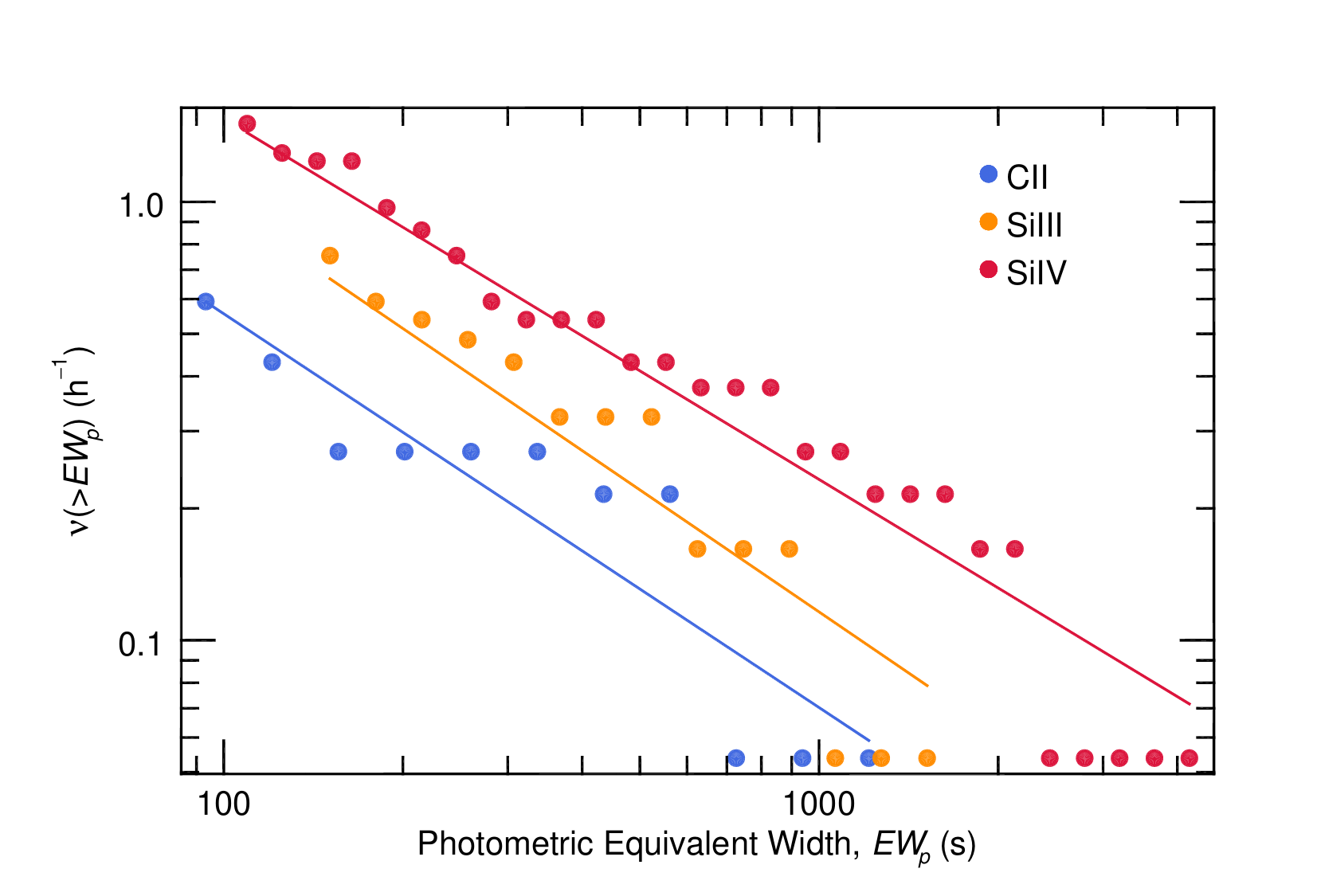}
\caption{A cumulative flare-frequency distribution for the flares identified in data for AD Leo. Power-law fits are overplotted. While the data are binned for display, following the methodology of \citet{crawford70} they were not binned when fitting the power laws.}
\label{fig:adleoflares}
\end{figure}

We detected enough flares on AD Leo to examine their distribution in $\nfe$. As such, Figure~\ref{fig:adleoflares} plots the frequency of flares, $\nu$, with $>\nfe$ versus $\nfe$. We fit a power law to these distributions of the form
\begin{equation}
\nu = \alpha\nfe^\beta
\end{equation}
using the maximum likelihood method of \citet{crawford70}. The resulting values of $\beta$ are $-0.90\pm0.29$ for \cii, $-0.92\pm0.27$ for \siii, and $-0.82\pm0.17$ for \siv. In comparison, previous values include
\begin{itemize}
\item{$-0.82\pm0.27$ from 21 h of visible and near-ultraviolet observations \citep{lacy76}}
\item{$-0.62\pm0.09$ from 111.5 h (spread over $>5$ years) of {\it U} band observations \citep{pettersen84}}
\item{$-1.01\pm0.28$ from $<72$ h of extreme-ultraviolet observations \citep{audard00}}
\item{$-0.68\pm0.16$ from 139.7 h of visible observations \citep{hunt12}.}
\end{itemize}
The above values all agree with those we computed for each band. This agreement is consistent with the response of each band tracing common energy deposition events, even for emission resulting from different regions of the stellar atmosphere. To explore further how impulsive energy deposition affects differing regions of a stellar atmosphere, simultaneous, panchromatic flare observations would be desirable.

\subsection{Size of Detectable Transiting Objects}
This work aims to explore the boundaries placed on transit observations by stellar variability in the ultraviolet. Stochastic fluctuations in the host star determine the minimum transit depth that will stand out from these fluctuations, consequently limiting the minimum detectable size of transiting objects. As a metric for this limitation set by each star + band's stochastic fluctuations, we computed the radius of an occulting disk that would produce a transit signal equivalent to $\s_x$ projected to 3.5~h, $R_{\s_x}$ -- in essence, the object size needed for a 1-$\s$ detection of a single transit in the absence of photometric noise. Explicitly, we compute $R_{\s_x}$ from 
\begin{equation}
R_{\s_x}^2 = \s_x R_\star^2,
\end{equation}
where $R_\star$ is the radius of the host star. This $R_{\s_x}$ does not represent an actual detection limit. The true minimum detectable object size depends on the instrument and observing time available. Instead of a true limit, $R_{\s_x}$ is an instrument independent means of comparing the suitability of stars for transit measurements. This metric is thus free from any assumptions about the number of photons an instrument will collect from the star or what other noise the instrument will add.

\begin{figure}
\plotone{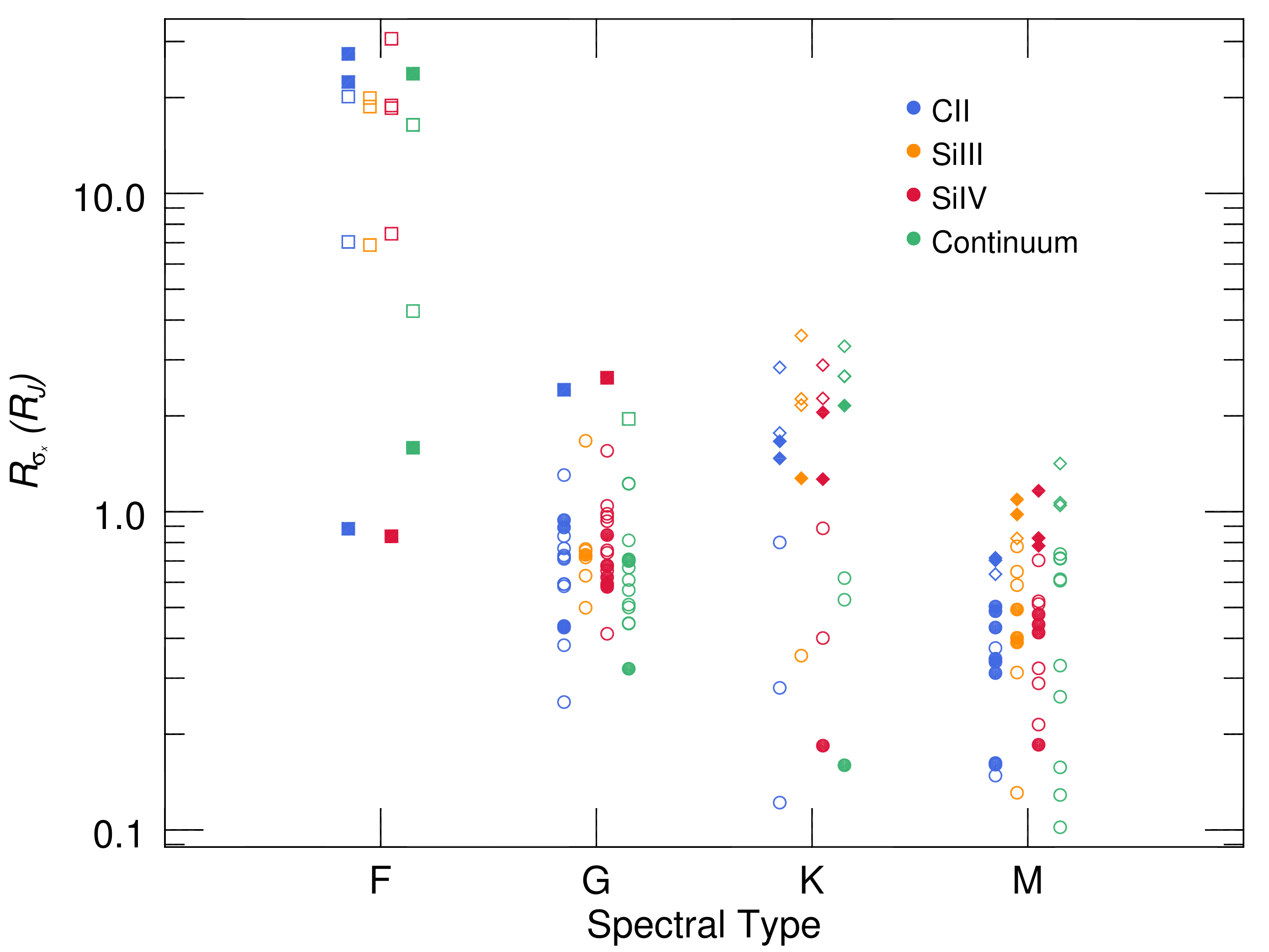}
\caption{The size in Jupiter radii of an occulting disk that would cause a dip equivalent to the 1-$\s$ scatter in the stellar flux at $\dt = 3.5$ h. Filled circles represent confident measurements of $\s_x$ whereas open circles represent upper limits. Squares represent post-main-sequence stars, circles represent main sequence stars, and diamonds represent WTTS. We group values on the x-axis by spectral type (labels below axis), then by band (symbol color).}
\label{fig:disksizes}
\end{figure}

Figure~\ref{fig:disksizes} shows the results, grouped by spectral type. These suggest that, in the absence of photometric uncertainties, roughly Jovian-size disks would produce the smallest detectable transit signal (for a reasonable quantity of data) in a typical system. Indeed, HD209458b, HD189733b, and WASP-12b (see Section~\ref{sec:intro}) are all of Jovian dimensions. However, $R_{\s_x}$ spans around an order of magnitude within each spectral type, and more than two orders of magnitude overall.

The results in Figure~\ref{fig:disksizes} are grouped by spectral type to examine the tradeoff between the smaller $R_\star$ but higher $\s_x$ (see Section~\ref{sec:varcor}) of less massive stars. Smaller stellar disks imply deeper transit signals, but higher levels of stochastic fluctuations better hide these signals. From Figure~\ref{fig:disksizes}, it appears stellar size trumps $\s_x$: the smallest stars also permit the smallest objects to produce detectable transits. However, this apparent trend is significantly weakened when the F stars, all post main-sequence giants or sub-giants, are removed.

Of the stars with data robustly sampling their stochastic fluctuations on transit timescales  (mean flux over $5\sn{-16}$ erg s$^{-1}$ cm$^{-2}$ and total accumulated observations $\geq3.5$ h), Prox Cen has the most generous $R_{\s_x}$ limits in each band. These are $1.8\pm0.1~ R_\earth$ in \cii, $<1.5~R_\earth$ in \siii, $2.1\pm0.1~R_\earth$ in \siv\, and $<1.1~R_\earth$ in the FUV continuum. Interestingly, Prox Cen is classified as a flare star, and the chances that it could flare near or within a transit are significant. 

Like Prox Cen, most of the stars in this sample have received flare, variable, or WTTS classifications. The sample is biased: The targets were preferentially selected for variability by the various individuals that commissioned the observations. It seems probable that a less-biased sample would produce detection limits clustered lower than those in Figure~\ref{fig:disksizes}. Therefore, the prospects for FUV transit measurements of planets the size of Neptune or super-Earths seem promising. Furthermore, if the observed planet hosts an atmosphere inflated to several times the area of the solid disk, like HD209458b \citep{madjar03}, then in an ideal case even an Earth-size planet might produce a detectable FUV signal within a few folded transits. These prospects are exciting, but it bears remembering that, to attain such limits, photometric noise must be suppressed to below the level of stochastic fluctuations.

\subsection{Correlations with Stellar Properties}
\label{sec:varcor}
\begin{figure*}
\plotone{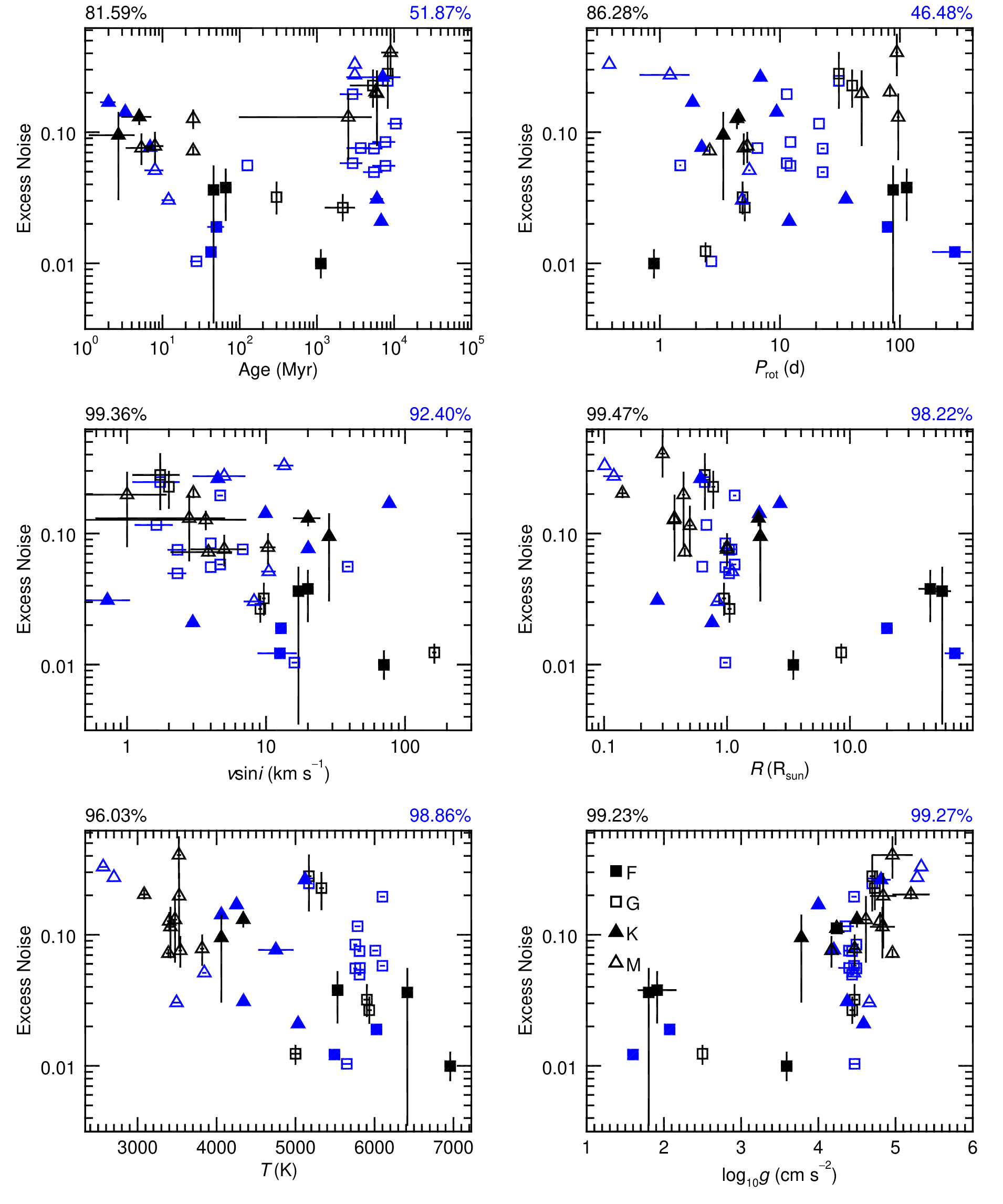}
\caption{Excess noise versus stellar properties in the \cii\ band. Symbols differentiate between spectral types (see legend in bottom right plot). Black symbols represent excess noise detections, while blue symbols represent upper limits. The black numbers above each plot give the significance of the correlation (probability it is not produced by uncorrelated points) using only excess noise detections while the blue number gives the significance including upper limits (see Section~\ref{sec:varcor}).}
\label{fig:varcorCII}
\end{figure*}

\begin{figure*}
\plotone{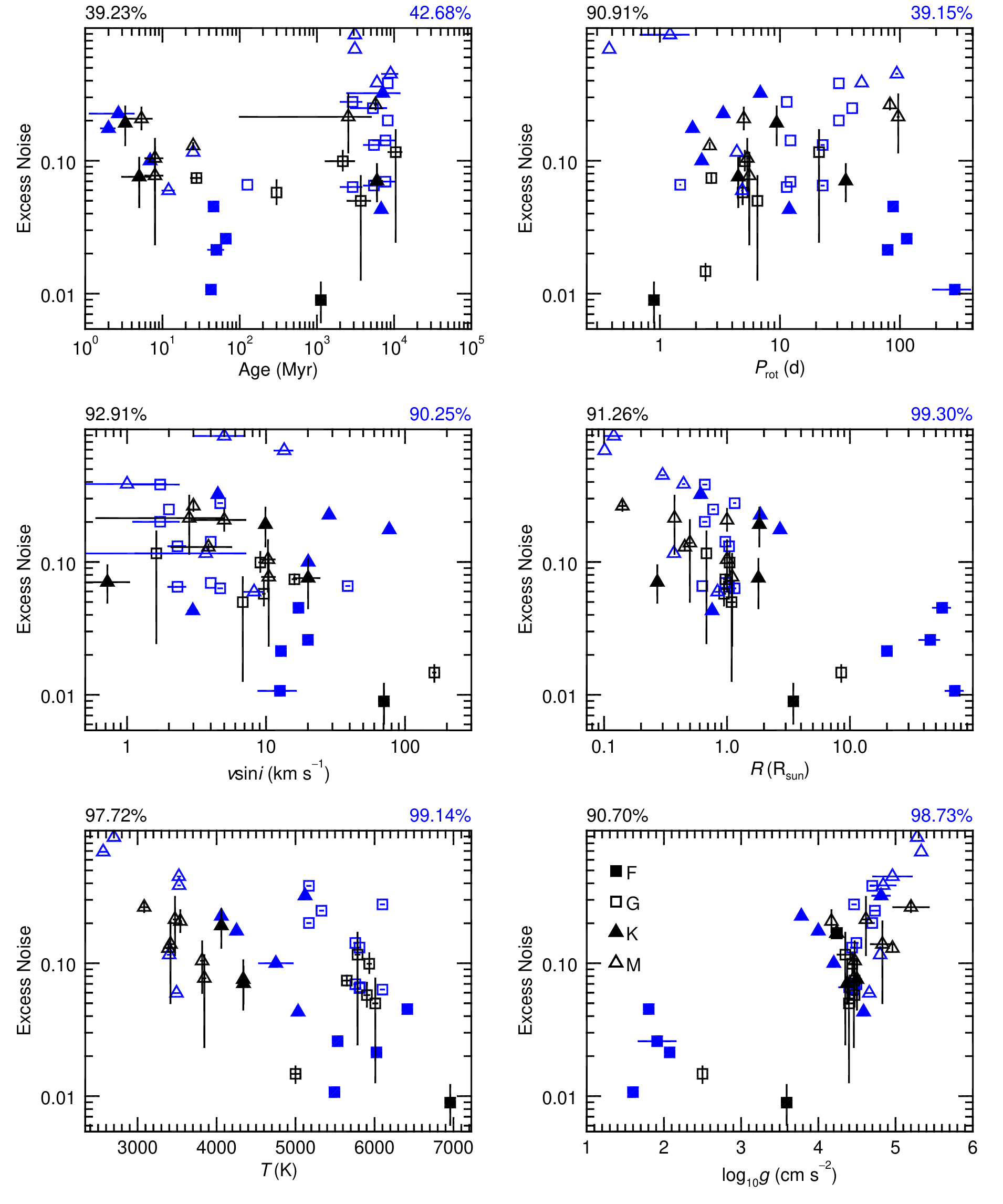}
\caption{Excess noise versus stellar properties in the \ion{Si}{4} band, following the same format as Figure~\ref{fig:varcorCII}.}
\label{fig:varcorSiIV}
\end{figure*}

We explored correlations between the mean-normalized excess noise measurements in \cii\ and \siv\ with properties of the sample stars (see Table~\ref{tbl:props}). Excess noise values in the remaining bands, \siii\ and the FUV continuum, were too poorly constrained (essentially, there were too few detections and too many upper limits) to support correlations with any value. As a means of visualizing potential correlations, Figures~\ref{fig:varcorCII} and~\ref{fig:varcorSiIV} graph $\s_x$ values (both detections, black, and upper limits, blue) against each of the stellar properties, excluding mass and luminosity, for the \cii\ and \siv\ bands respectively. 

These figures show that the error bars of the mean-normalized excess noise values and stellar properties often overlap. Thus, changes in the point values even under the 1-$\s$ error bars could change the Spearman Rank-Order correlation coefficient for the data and, more importantly, confidence that the correlation cannot be explained by randomly scattered points. To account for the uncertainty in how the true values of the points fall, we constructed a Monte-Carlo simulation, generating $10^4$ possible arrangements of the data points given the uncertainties. For each trial in the Monte-Carlo simulation, we randomly drew stellar parameter (e.g., age, $M$, $P_{rot}$) values from Gaussians matching the means and 1-$\s$ uncertainties. When there was no uncertainty accompanying the measurement we found in the literature, we assigned an uncertainty of 10\%. For the mean-normalized excess noise values, we did not use Gaussians. Instead we randomly drew values from the actual likelihood distributions of $\s_x$ for each star/band (Appendix~\ref{sec:likelihood}). 

For each of the $10^4$ such realizations of the data, we recorded probability to exceed (PTE) for the null hypothesis of randomly distributed points (i.e. no correlation) using the Spearman Rank-Order test. We multiplied the PTE by $10^{-4}$ to compute the joint probability of the data producing the correlation coefficient \emph{and} such a correlation coefficient resulting from random point scatter. The integral of these values over the range of possible correlation coefficients (-1 to 1), provided our overall estimate of the PTE for random scatter in light of the data and uncertainties. We then recorded the significance of the correlation as $1-$PTE. 

We carried out this process with and without the $\s_x$ upper limits and, for both cases, quote the significance of the correlation in Figures~\ref{fig:varcorCII} and~\ref{fig:varcorSiIV}. The $\s_x$ upper limits provide useful constraints on correlations when they fall below the surrounding $\s_x$ detections. As with the $\s_x$ detections, when we included upper limits we generated $\s_x$ values in the Monte-Carlo simulation from the likelihood distribution that produced each upper limit. For instance, an upper limit of 0.01 on $\s_x$ for a point meant that we randomly drew a value between 0 and 0.01 with roughly uniform probability for each trial in the Monte-Carlo simulation. 

The results computed without including the upper limits suggest weak correlations in all cases except $\s_x$-age and, in \cii\, $\s_x$-$P_{rot}$. The $\s_x$ upper limits further constrain these correlations, both quantitatively and by eye, and bring \cii\ and \siv\ into closer agreement. The subsections below address each of these in light of previous literature.

\subsubsection{Temperature, Radius, Mass, and Luminosity}
\label{sec:sptcorr}
Possible correlations of excess noise with temperature and radius (as well as mass and luminosity, though these are left out of Figures~\ref{fig:varcorCII} and~\ref{fig:varcorSiIV}) become strong when upper limits are included. This indicates a more general correlation between excess noise and spectral type: Stars of later spectral type typically exhibit higher $\s_x$. Correlations between stellar variability and spectral type were explored by \citet{mcquillan12} for all {\it Kepler} stars save those with known eclipsing companions (stellar or planetary) and lightcurve discontinuities. They found that cooler, later type stars exhibited high levels of stochastic variability by their metric, in agreement with our findings. Similarly, \citet{ciardi11} found a relationship between variability and temperature in their analysis of {\it Kepler} field stars, in both dwarfs and giants. We note, however, that {\it Kepler} utilizes a broad optical bandpass ($\sim$4000-9000 \AA). Therefore, variability in the {\it Kepler} data reflects processes occurring in the photosphere, rather than the transition region, and there is no guarantee that the two are directly related. 

\subsubsection{Rotation Period, Equatorial Velocity, and Age}
Excess noise does not correlate with age, and the hint of a correlation between $\s_x$ and rotation period is all but eliminated with the inclusion of $\s_x$ upper limits. A possible anticorrelation seems to exist between $\s_x$ and \vsini; however, when $\s_x$ upper limits are included, the anticorrelation is probably weak enough to dismiss outright. Furthermore, because $v = 2\pi R/P_{rot}$, the combination of a lack of a $\s_x$-$P_{rot}$ correlation and the presence of a strong $\s_x$-$R$ anticorrelation are capable of producing the $\s_x$-\vsini\ trend. Thus, it is the $\s_x$-$R$ anticorrelation (or, rather, the relationship of $\s_x$ to spectral type) that drives the $\s_x$-\vsini\ trend. Indeed, $\s_x$ and the quotient $R/P_{rot}$ (excluding the 8 stars for which we computed $P_{rot}/\sin{i}$ from \vsini\ and $R$) are anticorrelated to roughly the same confidences as $\s_x$-\vsini. 

Interpreting the lack of correlations with age and rotation in light of previous work is difficult, as we did not find any studies exploring correlations between these properties and stellar stochastic fluctuations. Both age and rotation, however, have been tied to stellar activity as quantified by chromospheric emission (e.g., \citealt{wilson63,skumanich72,noyes84,mamajek08}). In turn, chromospheric emission likely has a direct relationship with stochastic fluctuations at visible wavelengths \citep{hall09}. This suggests that younger, faster rotating stars might exhibit higher levels of UV emission line variability.

Given these past results, the lack of $\s_x$-age and $\s_x$-$P_{rot}$ correlations in this sample could be explained if either the age-rotation-activity or the variability-activity relationships do not hold for this sample. That either might not hold would not be particularly alarming given that this sample includes stars approaching, dwelling on, and departing the main sequence, whereas the above-mentioned studies only analyze stars on the main sequence. The pre and post main-sequence stars in this sample might exhibit magnetic behavior not in line with that of main-sequence stars, such that relationships between magnetic activity (chromospheric emission) and rotation, age, and/or variability for main-sequence stars are not extensible to this broader sample. 

\subsubsection{Surface Gravity}
A strong correlation exists between excess noise and surface gravity in the sample, particularly in \cii. This is in conflict with the recent results of \citet{bastien13}, who find an inverse relationship between variability and surface gravity in {\it Kepler} field stars. However, compared to this work, the stellar sample employed by \citet{bastien13} is far more restrictive. They limit their study to stars with $4500$ K $< T_{eff}<6500$ K, $2.5< \log_{10}{g} <4.5$ (cgs units), and relative brightness variations (measured using a 30 min cadence) of $< 0.003$. Applying the same cuts in $T_{eff}$ and $\log_{10}{g}$ to this stellar sample eliminates any correlation (null hypothesis can produce the observed or stronger correlation with $\sim$50\% probability).  The strong correlation present in this data when all stars are included is driven primarily by the F giants, as these have both very low $g$ and $\s_x$. However, high $\s_x$, high $g$ M stars also promote the correlation.

\subsection{Star-Planet Interactions}
Close-in planets might interact with their host star magnetically and tidally, producing enhanced activity in the stellar upper atmosphere \citep{cuntz00}. Evidence for these interactions could be found by monitoring a single star with a transiting planet to search for signs of elevated activity near transit, when the portion of the star interacting with the planet is in view. Indeed, this has been attempted for two systems with transiting planets in this sample, HD209458 \citep{shkolnik04} and HD189733 \citep{shkolnik08,fares10}. However, these studies found no clear evidence of enhanced activity correlated with orbital phase for either of these systems. 

Alternatively, evidence for star-planet interactions might result from comparing the overall activity or variability of the stars with close-in planets to a sample of control stars. \citet{shkolnik13} recommended investigations into star-planet interactions in the form of time-resolved observations of UV flux variability rather than activity (i.e. mean line flux), such as those presented in this paper. The small size and high diversity of the 32 stars in this sample that do not host planets make for a poor control. Nevertheless, we looked for increased $\s_x$ in the six known planet hosts. These hosts are identified in the $\s_x$ detections and upper limits presented in Figures~\ref{fig:longvar} and~\ref{fig:longvar} as open points. 

The hosts with $\s_x$ detections in Figure~\ref{fig:longvar} do appear clustered at higher $\s_x$. However, these are all M dwarfs, shown in the previous section to have the highest $\s_x$ values in the sample. The M dwarf planet hosts do not stand out when grouped with the other M dwarfs in the sample. The G and K planet hosts (HD209458 and HD189733) have insufficient data for a $\s_x$ detection, allowing only 95\% upper limits. Because only upper limits are possible, these stars are not constrained to be more variable than other sample stars of the same spectral type. Thus, star-planet interactions, manifested as increased variability, are not supported by the limited volume of this data. 

\section{Summary}
\label{sec:summary}

We have examined stellar luminosity fluctuations and stellar flares in a sample of 38 cool stars using 60~s cadence lightcurves constructed from narrow spectroscopic bands containing the \ion{Si}{3} $\lm$1206, \ion{C}{2} $\lm\lm$1334,1335, and \ion{Si}{4} $\lm\lm$1393,1402 resonance lines and a combined 36.5 \AA\ of interspersed FUV continuum bands. In the high pass filtered lightcurves, we detected 116 flares, occurring roughly once per 2.5 h. Flares commonly radiated more energy in \siv, relative to quiescent levels, than the other bands. Shorter flares are more prevalent, with over half lasting 4 min or less. Most (90 of 116) of the detected flares could annihilate the signal of an Earth transit, while 7 of the 116 could annihilate the signal of a Jupiter transit. These results highlight the usefulness of minute-scale cadences for finding and removing flares prior to estimating transit depths.

To quantify stochastic fluctuations, we found the maximum-likelihood ``excess noise" that, in addition to the photometric noise, accounts for the scatter of the high-pass filtered, mean-normalized lightcurves excluding flares. Values of the excess noise, relative to the mean flux, range from about 1-41\% in \cii, 8-18\% in \siii, 0.9-26\% in \siv, and 1-22\% in the FUV continuum. Where the likelihood distribution of the excess noise was one-sided, we instead quote a 95\% upper limit on the excess noise. These upper limits on excess noise are often strong enough to be lower than the values for stars where excess noise was detected.

We found significant anticorrelations of excess noise with mass, radius, temperature, and luminosity in \cii\ and \siv. An additional, weaker anticorrelation of excess noise with \vsini\ could be an artifact of the strong underlying correlation between radius and rotation rate in the sample. There was no correlation with age or rotation period.

The median levels of stochastic stellar fluctuations we estimated, integrated over a typical transit timescale of 3.5~h, would impose a rough 1-$\s$ transit detection limit of $\sim1~R_J$ occulting disks for many of the stars in the sample. However, the range in these limits is broad, spanning from tenths of $R_J$ to tens of $R_J$.  M dwarfs might permit the FUV observation of transiting objects as small as Neptunes or even super-Earths in the absence of photometric noise. While the large fluctuations of some stars might stymie transit spectroscopy in the FUV for any but the largest planets, these results suggest that many planetary host stars might be found for which the limits of FUV transit spectroscopy can be pushed well below the hot Jupiters observed thus far. 

{\it Acknowledgments:} The authors wish to acknowledge the anonymous referee, whose familiarity with the field and resulting thoughtful and constructive comments greatly improved this work. The authors also wish to thank Tom Ayres for his careful reading and helpful suggestions that similarly improved this work. Kevin France acknowledges support through a NASA Nancy Grace Roman Fellowship during a portion of this work. This research has made use of the Exoplanet Orbit Database and the Exoplanet Data Explorer at exoplanets.org \citep{wright11} as well as extensive use of the SIMBAD database and VizieR catalog access tool, operated at CDS, Strasbourg, France (\citet{wenger00}, \citet{ochsenbein00}). The authors wish in particular to acknowledge \citet{glebocki05} for the excellent stellar rotation catalog that we used in this work. Some of the data presented in this paper were obtained from the Mikulski Archive for Space Telescopes (MAST). STScI is operated by the Association of Universities for Research in Astronomy, Inc., under NASA contract NAS5-26555. This work was supported by NASA grant NNX08AC146 to the University of Colorado at Boulder.

{\it Facility:} \facility{{\it HST} (COS, STIS)}

\appendix
\section{Target Confusion}
\label{sec:doubles}
For some of the targets, other FUV sources were present within the instrument aperture. Targets with known or suspected companions within the field of view of the instrument, but that should not contribute more than 10\% of the measured FUV flux, are as follows.
\begin{itemize}
\item{EK Dra has low-mass (0.5$\pm$0.1 M$_\sun$) binary companion GJ559.1B at 0.7" angular separation as of 2002 Oct and an orbital period of 45$\pm$5 years with a {\it V} band magnitude difference of 6 \citep{konig05}.}
\item{Polaris Aa has binary companion Polaris Ab of spectral type F6V at 0.170$\pm$0.003" separation as of 2006 Aug and an orbital period of 29.59$\pm$0.02 years with a {\it V} band magnitude difference of 7.2 \citep{kamper96,evans08}.}
\item{HD103095 might have companion CF UMa, but its existence is uncertain \citep{heintz84}.}
\end{itemize}

\section{Sensitivity to Excess Noise}
\label{sec:sensitivity}
A brief look at the sample variance statistic, $\hat{\s}^2$, provides insight into the driving factors concerning the sensitivity of excess noise measurements. Since $\hat{\s}$ monotonically increases with $\hat{\s}^2$, greater sensitivity to one translates to the other. When $\hat{\s}^2$ is computed from a set of $N$ points drawn from a Gaussian distribution with true variance $\s^2$, the statistic $(N-1)\hat{\s}^2/\s^2$ follows a $\chi^2$ distribution of $N-1$ degrees of freedom. Thus, the variance of this statistic is
\begin{equation}\label{eq:chi2var}
\textup{V}\left [\frac{(N-1)\hat{\s}^2}{\s^2} \right ] = 2(N-1).
\end{equation}

However, when examining the stochastic fluctuations of a star, we are concerned with the magnitude of the sample variance relative to the mean, $\hat{\s}^2/\mu^2$. Assuming the null hypothesis that the data are described by a Poisson distribution such that $\s^2 = \mu$,
\begin{equation}
\textup{V}\left [\frac{\hat{\s}^2}{\mu^2}\right ] = \frac{2}{(N-1)\mu^2}.
\end{equation}
Thus, for a given measurement of $\hat{\s}^2$, the null hypothesis can be rejected with greater confidence as $\mu$ and $N$ increase. In other words, sensitivity to excess variance ($\hat{\s}^2$ above $\mu$) is better for targets with stronger signal and longer durations of observations, as one might intuitively expect. Although we employ a maximum-likelihood method when estimating excess noise in the lightcurves rather than computing $\hat{\s}^2$, these drivers of sensitivity are apparent in the results.

Of further interest would be a cadence that minimizes the variance on the cadence-independent quantity
$$\frac{\hat{\s}^2}{\mu^2}\dt.$$
The mean is directly proportional to the cadence length, $\mu\propto\dt$, and, for a fixed quantity of data, $N\propto\dt^{-1}$. Additionally, if each lightcurve point is uncorrelated to surrounding points no matter how short of a cadence is employed, then $\s^2$ scales as $\dt$. For large $N$ such that the 1 in the $(N-1)$ terms of Equation (\ref{eq:chi2var}) can be neglected,
\begin{equation}
\textup{V}\left [\frac{\hat{\s}^2}{\mu^2}\dt\right ]\propto\dt.
\end{equation}
Hence, by this analysis it is always beneficial to choose the shortest possible cadence. However, $\s^2$ does not scale as $\dt$ for all $\dt$ as we have assumed in this analysis. This factor is part of what drives the choice of $\dt$, as discussed in Section~\ref{sec:lcextract}.

\section{Maximum-Likelihood Estimation of Excess Noise}
\label{sec:likelihood}
To explain in detail how we estimated $\s_x$, we must first discuss again the high-pass filtering to which we subjected the data. This filtering alters a series of points even if they exhibit pure white noise (i.e. no periodic signals). Yet the white noise in the data is exactly what we wished to preserve and measure. Thus, we attempted a simple correction. To formulate our correction, we assumed the high-pass filtering scales the white-noise scatter in a lightcurve by a constant factor, $\alpha$. However, this factor was unknown for the actual data because the portion of the data represented by white noise was also unknown. For a set of simulated data (i.e. data known to be purely white noise), the factor could be determined by computing the standard deviation before filtering, $\s$, and after filtering, $\s'$. Then $\alpha = \s/\s'$. Thus, each simulated dataset would produce a different value for $\alpha$. After simulating many datasets and recording $\alpha$ for each, the results could be histogrammed to estimate the probability density function (PDF) of $\alpha$. The shape of the PDF depends on the spacing of the points and the cutoff frequency of the high-pass filter, but the PDF is independent of the amplitude of $\s$. For our purposes, we estimated the PDF of $\alpha$ using $10^4$ white-noise simulations of each lightcurve separately. The PDFs were often noticeably asymmetric, with means generally within a few percent of unity.

With PDFs for $\al$ in hand, we computed maximum-likelihood values of $\s_x$ as follows. First, we modeled each high-pass filtered lightcurve as points randomly drawn from independent Gaussians (one for each point). The variance of the Gaussian for point $i$ we assumed to be
\begin{equation}
\s_i^2 = \s_{p,i}^2 + \s_x^2 = F_{signal,i} + F_{bkgnd,i} + \s_x^2,
\end{equation}
where $F_{signal,i}$ and $F_{bkgnd,i}$ are the signal counts and expected background counts in the extraction region for lightcurve point $i$. Then we assumed this white-noise lightcurve (more specifically the mean-normalized, mean-subtracted lightcurve) was scaled by $\alpha$, the unknown factor discussed above, due to the high pass filtering. Thus, the set of $\s_{p,i}$, $\s_x$, and $\alpha$ fully specified a given lightcurve in our model. The $\s_{p,i}$ we estimated as the quadratic sum the signal and background Poisson noise for each point. The values of $\s_x$ and $\alpha$ we sampled over a grid and computed the likelihood of the data, $\mathcal{L}(\mathbf{F}\ |\s_x,\alpha)$, at every grid point. Here $\mathbf{F}$ represents the vector of $N$ lightcurve points.

We found $\mathcal{L}(\mathbf{F}\ |\s_x,\alpha)$ as follows. First, we scaled the data back by $\alpha^{-1}$. In other words, we computed $\al^{-1}(\mathbf{F}- \mf)$, where $\mf$ is the mean value of the lightcurve. If the sampled $\alpha$ were the true $\alpha$, this would have rescaled the white noise in the data to its original state (without the filtered low-frequency signals). Next, we computed the likelihood that the lightcurve points were randomly drawn from the Gaussians with variances augmented by $\s_x$. Finally, we multiplied this by the probability, $p(\al)$, of the $\al$ we assumed. This determined the likelihood that the original data were randomly drawn from the specified Gaussians \emph{and} were then scaled by $\alpha$. Altogether, this gives
\begin{equation}
\mathcal{L}(\mathbf{F}\ |\s_x,\alpha) =  p(\al)\prod_{i=1}^{N}\frac{1}{\s_i\sqrt{2\pi}}e^{-\al^{-2}(F_i-\mf)^2/2\s_i^2}.
\end{equation}
We computed likelihood values in log space, such that
\begin{equation}
\ln{\left [\mathcal{L}(\mathbf{F}\ |\s_x,\alpha) \right ]} = \ln{[p(\al)]} - \sum_{i=1}^N \left[\ln{(\sigma_i\sqrt{2\pi})}+\frac{\al^{-2}(F_i-\mf)^2}{2\sigma_i^2}\right ] + A,
\end{equation}
where $A$ is a normalization factor, constant for all sampled values of $\s_x$ for which we computed $\mathcal{L}$. We chose it to avoid arithmetic underflow when we returned to $\mathcal{L}$ from $\ln{\mathcal{L}}$. After returning from log-space, for each sampled value of $\s_x$ we marginalized over all sampled values of $\al$ to estimate
\begin{equation}
\mathcal{L}(\mathbf{F}\ |\s_x) = \int_{-\infty}^{\infty} \mathcal{L}(\mathbf{F}\ |\s_x,\alpha)d\alpha.
\end{equation}

The quantity of interest was the likelihood of $\s_x$ given the data $\mathbf{F}$ rather than the likelihood of the data given a value of $\s_x$. Thus, we turned to Bayes Theorem and enforced the prior probability distribution that all $\s_x \geq 0$ were uniformly probable while all $\s_x < 0$ were impossible. Values of $\s_x < 0$ are nonsensical. Bayes Theorem simply yields
\begin{equation}
	\mathcal{L}(\s_x|\mathbf{F}\ )= \left \{ 
	\begin{array}{lr} 
		\mathcal{L}(\mathbf{F}\ |\s_x) & : \s_x \geq 0 \\
		 0 & : \s_x < 0 
	\end{array}
	\right .
\end{equation}
We then used the likelihood distribution to compute the maximum-likelihood value and 68.3\% error bars or the 95\% upper limit.


\end{document}